\documentclass[tighten,twocolumn,appendixfloats,apj]{likeapj}
\pdfoutput=1

\usepackage{graphicx,bm,mathptmx,amsmath}

\usepackage{natbib}
\usepackage{array}

\hypersetup{linkcolor=blue,citecolor=blue,filecolor=cyan,urlcolor=blue,bookmarksnumbered=true,bookmarksopen=true}
\usepackage[all]{hypcap}

\usepackage{float}
\usepackage{enumitem}

\newcolumntype{C}[1]{>{\centering}m{#1}}

\newcommand\fauxsc[1]{\fauxschelper#1 \relax\relax}
\def\fauxschelper#1 #2\relax{%
  \fauxschelphelp#1\relax\relax%
  \if\relax#2\relax\else\ \fauxschelper#2\relax\fi%
}
\def\Hscale{.85}\def\Vscale{.74}\def\Cscale{1.12}
\def\fauxschelphelp#1#2\relax{%
  \ifnum`#1>``\ifnum`#1<`\{\scalebox{\Hscale}[\Vscale]{\uppercase{#1}}\else%
    \scalebox{\Cscale}[1]{#1}\fi\else\scalebox{\Cscale}[1]{#1}\fi%
  \ifx\relax#2\relax\else\fauxschelphelp#2\relax\fi}

\usepackage{textgreek} 
\usepackage{upgreek} 

\usepackage{catchfile}
\newcommand{\sh}[2][]{%
  \CatchFileEdef{\temp}{"|kpsewhich --var-value #2"}{\endlinechar=-1}%
  \if\relax\detokenize{#1}\relax\temp\else\let#1\temp\fi}

\newcommand{\cena}{Centaurus~A}

\newcommand{\meee}{MCG$+$8$-$11$-$11}

\newcommand{\gc}{NGC~}

\newcommand{\cunits}{cm$^{-2}$}
\newcommand{\funits}{erg~cm$^{-2}$~s$^{-1}$}

\newcommand{\kmps}{km s$^{-1}$}

\newcommand{\lunits}{erg~s$^{-1}$}

\newcommand{\msun}{$M_{\odot}$}

\newcommand{\phunits}{photons~cm$^{-2}$~s$^{-1}$}

\newcommand{\reunits}{cts~s$^{-1}$~keV$^{-1}$}

\newcommand{\cs}{$\chi^2$}

\newcommand{\eshift}{$E_{\rm shift}$}

\newcommand{\gammasoft}{$\Gamma_{\rm soft}$}

\newcommand{\lxht}{$L_{\rm X,2.0-10.0 \ keV}$}

\newcommand{\nh}{$N_{\rm H}$}

\newcommand{\nhz}{$N_{\rm H,Z}$}

\newcommand{\nhs}{$N_{\rm H,S}$} 
 
\newcommand{\nhsoft}{$N_{\rm H,soft}$}

\newcommand{\nhGal}{$N_{\rm H}^{\rm Gal}$}

\newcommand{\Nrel}{$\cal{N}_{\rm rel}$}

\newcommand{\sigl}{$\sigma_L$}

\newcommand{\x}{X-ray}

\newcommand{\ifeka}{$I_{\rm Fe K\alpha}$}

\newcommand{\lcra}{$L_{\rm 2-10, c, rest, abso}$}
\newcommand{\lcru}{$L_{\rm 2-10, c, rest, unabso}$}
\newcommand{\lcraH}{$L_{\rm 10-30, c, rest, abso}$}
\newcommand{\lcruH}{$L_{\rm 10-30, c, rest, unabso}$}

\newcommand{\fcobs}{$f_{\rm 2-10, c, obs}$}
\newcommand{\fcobsH}{$f_{\rm 10-30, c, obs}$}

\newcommand{\afe}{$A_{\rm Fe}$}

\newcommand{\afesun}{$A_{\rm Fe,\odot}$}

\newcommand{\fek}{Fe~K}
\newcommand{\feka}{Fe~K$\alpha$}

\newcommand{\fekb}{Fe~K$\beta$}

\newcommand{\fetwofive}{Fe{\sc \,xxv}}
\newcommand{\fetwosix}{Fe{\sc \,xxvi}}

\newcommand{\hone}{H{\sc \,i}}

\newcommand{\chandra}{\textit{Chandra}}

\newcommand{\cpx}{$C_{\rm PIN:XIS}$}

\newcommand{\nustar}{\textit{NuSTAR}}
\newcommand{\suzaku}{\textit{Suzaku}}

\newcommand{\xmm}{\textit{XMM-Newton}}
\newcommand{\xispin}{XIS$+$PIN}

\newcommand{\myt}{{\textsc{mytorus}}}

\newcommand{\myts}{{\sc myt}orus{\sc s}}
\newcommand{\mytl}{{\sc myt}orus{\sc l}}
\newcommand{\mytz}{{\sc myt}orus{\sc z}}

\newcommand{\xspec}{{\textsc{xspec}}}

\newcommand{\cthin}{Compton-thin}
\newcommand{\cthick}{Compton-thick}

\newcommand{\scr}{Sec.~\ref}
\newcommand{\tr}{Table~\ref}
\newcommand{\exi}{\begin{equation}}
\newcommand{\exo}{\end{equation}}
\newcommand{\los}{line-of-sight}

\newcommand{\aer}[3]{$#1^{#2}_{#3}$} 
\newcommand{\ten}[2]{$#1\times 10^{#2}$} 
\newcommand{\up}[1]{$^{#1}$}

\newcommand{\zphabs}{{\tt zphabs}}

\newcommand{\zgauss}{{\tt zgauss}}

\def\spose#1{\hbox to 0pt{#1\hss}} 
\def\approxlt{\mathrel{\spose{\lower 3pt\hbox{$\sim$}}
        \raise 2.0pt\hbox{$<$}}}
\def\approxgt{\mathrel{\spose{\lower 3pt\hbox{$\sim$}}
        \raise 2.0pt\hbox{$>$}}}

\newcommand{\simgt}{$\approxgt$}

\definecolor{aqua}{rgb}{0.0, 1.0, 1.0}

\newcommand{\sss}{$\sim$}

\newcommand{\lt}{~$<$~}

\usepackage{tikz}

\usepackage[normalem]{ulem}

\defcitealias{torrejon2010}{T10} 

\received{2020}
\revised{}
\accepted{}
\published{}

\shorttitle{None}
\shortauthors{Tzanavaris et al.}

\begin{document}
\newcommand\redsout{\bgroup\markoverwith{\textcolor{red}{\rule[0.5ex]{2pt}{1pt}}}\ULon}

\title{Are Compton-thin AGNs Globally Compton Thin?} 

\author{P.~Tzanavaris}
\affil{University of Maryland, Baltimore County, 1000
  Hilltop Circle, Baltimore, MD 21250, USA}
\affil{Laboratory for X-ray Astrophysics, NASA/Goddard
  Spaceflight Center, Mail Code 662, Greenbelt, MD 20771, USA}

\author{T.~Yaqoob}
\affil{University of Maryland, Baltimore County, 1000
  Hilltop Circle, Baltimore, MD 21250, USA}
\affil{Laboratory for X-ray Astrophysics, NASA/Goddard
  Spaceflight Center, Mail Code 662, Greenbelt, MD 20771, USA}
\affil{Department of Physics and Astronomy, The Johns
  Hopkins University, Baltimore, MD 21218, USA}

\author{S.~LaMassa}
\affil{Space Telescope Science Institute, 3700 San
  Martin Drive, Baltimore, MD 21218, USA}

\author{A.~Ptak}
\affil{Laboratory for X-ray Astrophysics, NASA/Goddard
  Spaceflight Center, Mail Code 662, Greenbelt, MD 20771, USA}
\affil{Department of Physics and Astronomy, The Johns
  Hopkins University, Baltimore, MD 21218, USA}

\author{M.~Yukita}
\affil{Laboratory for X-ray Astrophysics, NASA/Goddard
  Spaceflight Center, Mail Code 662, Greenbelt, MD 20771, USA}
\affil{Department of Physics and Astronomy, The Johns
  Hopkins University, Baltimore, MD 21218, USA}

\begin{abstract}
  We select eight nearby AGNs which, based on previous work, appear to be
  \cthin\ in the line of sight. We model with \myt\
  their broadband \x\ spectra from 20 individual observations with
  \suzaku, accounting
  self-consistently for \feka\ line emission, as well as direct and scattered
  continuum from matter with finite column density and solar Fe abundance.
  Our model configuration allows us to measure the global, out of the line of sight, equivalent hydrogen
  column density separately from that in the line of sight.  For 5 out
  of 20 observations (in 3 AGNs) we find that the global column density is
  in fact \simgt\ten{1.5}{24} \cunits,
  consistent with the distant scattering matter being \cthick. 
  For a fourth AGN, 2 out of 5 observations are also
  consistent with being \cthick, although with large errors.
  Some of these AGNs have been reported to
  host relativistically broadened \feka\ emission. Based on our
  modeling, the \feka\ emission line is not resolved in all but
  two \suzaku\ observations, and the data can be fitted
  well with models that only include a narrow \feka\ emission line.

\end{abstract}

\keywords{black hole physics -- radiation mechanisms: general -- scattering -- galaxies: active }

\section{Introduction}\label{sec-intro}
It is now widely accepted that bulged galaxies harbor nuclear
supermassive black holes (SMBHs, \sss$10^6$--$10^9$~\msun; for a
review see, e.g.,~\citealt{graham2016}). Gas accretion onto the SMBH
leads to substantial energy release. A population of hot electrons,
perhaps residing in a hot corona, induce inverse Compton upscattering
of thermal photons in the accretion disk, thus producing a primary
\x\ power-law continuum, which constitutes a ubiquitous,
characteristic observational spectral signature of Active Galactic
Nuclei (AGNs).  Primary continuum emission may also be
Compton-scattered either at the accretion disk or further out by
optically thick, cold, neutral or mildly ionized material, and give
rise to a series of fluorescent emission lines and an associated
secondary, scattered (also known as ``reflected'') continuum.
The most
prominent and frequently observed such line is the \feka\ emission
line at 6.4~keV \citep{george1991}.

 \renewcommand{\baselinestretch}{0.7} 
 \begin{deluxetable*}{ccccc ccc c}
     \tablecaption{\footnotesize \vspace{3pt} Exposure Times and Count Rates for \suzaku\  Spectra. \vspace{-5pt} \label{tab-suz}}
\tabletypesize{\footnotesize}
\tablehead{
  [-0pt]
\colhead{\#} &
\colhead{Name} &
\colhead{ObsID} &
\colhead{Date/Time} &
\colhead{Detector} &
\colhead{Exposure} &
\colhead{Energy ranges} &
\colhead{Count Rate} &
\colhead{Percentage of} 
\\[-.15cm]
\colhead{} &
\colhead{} & 
\colhead{} & 
\colhead{} &
\colhead{} &
\colhead{(ks)} &
\colhead{(keV)} &
\colhead{(count s\up{-1})} &
\colhead{on-source Rate} 
\\[-.6cm]}
\colnumbers
\startdata 
\vspace{-15pt}\\
  1 &         CenA &    100005010 &  2005-08-19T03:39:19 & XIS & 254.4 &       2.3--10.0 & 4.3072$\pm$0.0042 & 99.5 \\
    &              &              &                      & PIN &  62.7 &      12.0--45.0 & 1.3082$\pm$0.0055 & 72.0 \\
  2 &         CenA &    708036010 &  2013-08-15T04:22:42 & XIS &  32.0 &      2.3--10.00 & 7.8154$\pm$0.0162 & 98.7 \\
    &              &              &                      & PIN &   9.5 &    12.80--40.80 & 1.8326$\pm$0.0148 & 88.7 \\
  3 &         CenA &    708036020 &  2014-01-06T11:18:21 & XIS &  22.1 &      2.3--10.00 & 4.1996$\pm$0.0146 & 98.5 \\
    &              &              &                      & PIN &   6.2 &    13.20--32.00 & 0.8706$\pm$0.0132 & 81.3 \\
  4 &         CenA &    704018010 &  2009-07-20T08:55:29 & XIS & 187.8 &      2.3--10.00 & 6.0583$\pm$0.0058 & 98.5 \\
    &              &              &                      & PIN &  31.9 &    12.00--51.00 & 2.1287$\pm$0.0091 & 82.6 \\
  5 &         CenA &    704018020 &  2009-08-05T07:23:27 & XIS & 153.9 &      2.3--10.00 & 5.4177$\pm$0.0061 & 98.4 \\
    &              &              &                      & PIN &  40.7 &    12.00--51.00 & 2.0889$\pm$0.0079 & 83.0 \\
  6 &         CenA &    704018030 &  2009-08-14T09:06:56 & XIS & 167.8 &      2.3--10.00 & 6.1118$\pm$0.0062 & 98.5 \\
    &              &              &                      & PIN &  43.0 &    12.00--52.00 & 2.3040$\pm$0.0081 & 83.5 \\
  7 &  MCG+8-11-11 &    702112010 &  2007-09-17T01:46:02 & XIS & 296.3 &       2.3--10.0 & 1.3904$\pm$0.0022 & 98.0 \\
    &              &              &                      & PIN &  82.9 &     12.0--35.00 & 0.2592$\pm$0.0029 & 38.4 \\
  8 &      NGC526A &    705044010 &  2011-01-17T18:13:23 & XIS &  69.8 &       2.3--10.0 & 1.0197$\pm$0.0041 & 98.5 \\
    &              &              &                      & PIN &  58.4 &     15.0--30.75 & 0.1271$\pm$0.0026 & 34.0 \\
  9 &      NGC2110 &    100024010 &  2005-09-16T09:06:40 & XIS & 369.2 &       2.3--10.0 & 2.7912$\pm$0.0028 & 98.9 \\
    &              &              &                      & PIN &  80.6 &      12.0--40.0 & 0.5508$\pm$0.0036 & 54.7 \\
 10 &      NGC2110 &    707034010 &  2012-08-31T02:31:38 & XIS & 309.8 &      2.3--10.00 & 3.0121$\pm$0.0032 & 98.5 \\
    &              &              &                      & PIN &  95.8 &    12.80--52.10 & 0.5687$\pm$0.0031 & 65.8 \\
 11 &      NGC3227 &    703022020 &  2008-11-04T03:36:31 & XIS & 161.1 &       2.3--10.0 & 0.3631$\pm$0.0016 & 96.8 \\
    &              &              &                      & PIN &  46.7 &      13.0--30.7 & 0.1428$\pm$0.0032 & 31.0 \\
 12 &      NGC3227 &    703022030 &  2008-11-12T02:48:55 & XIS & 169.7 &      2.3--10.00 & 0.5520$\pm$0.0019 & 97.6 \\
    &              &              &                      & PIN &  46.7 &    12.00--28.85 & 0.1469$\pm$0.0035 & 28.1 \\
 13 &      NGC3227 &    703022040 &  2008-11-20T17:00:00 & XIS & 193.7 &       2.3--9.00 & 0.1514$\pm$0.0010 & 97.2 \\
    &              &              &                      & PIN &  43.4 &    12.00--26.60 & 0.0623$\pm$0.0026 & 22.4 \\
 14 &      NGC3227 &    703022050 &  2008-11-27T21:29:20 & XIS & 238.3 &      2.3--10.00 & 0.4539$\pm$0.0015 & 97.4 \\
    &              &              &                      & PIN &  37.4 &    12.00--27.00 & 0.0859$\pm$0.0029 & 28.9 \\
 15 &      NGC3227 &    703022060 &  2008-12-02T14:28:03 & XIS & 154.2 &       2.3--9.70 & 0.2598$\pm$0.0014 & 98.0 \\
    &              &              &                      & PIN &  36.9 &    12.00--28.85 & 0.0974$\pm$0.0037 & 21.1 \\
 16 &      NGC4258 &    701095010 &  2006-06-10T12:59:29 & XIS & 368.0 &      2.3--10.00 & 0.1682$\pm$0.0008 & 91.0 \\
    &              &              &                      & PIN &  91.6 &    12.00--28.00 & 0.0314$\pm$0.0022 &  7.5 \\
 17 &      NGC4507 &    702048010 &  2007-12-20T12:20:59 & XIS & 310.9 &       2.3--9.70 & 0.0895$\pm$0.0006 & 94.4 \\
    &              &              &                      & PIN & 103.2 &    12.00--35.50 & 0.1289$\pm$0.0024 & 23.2 \\
 18 &      NGC5506 &    701030010 &  2006-08-08T16:30:05 & XIS & 191.0 &      2.3--10.00 & 2.6782$\pm$0.0039 & 98.7 \\
    &              &              &                      & PIN &  38.6 &    12.00--36.00 & 0.4364$\pm$0.0049 & 49.0 \\
 19 &      NGC5506 &    701030020 &  2006-08-11T02:26:18 & XIS & 213.2 &      2.3--10.00 & 2.8307$\pm$0.0037 & 98.7 \\
    &              &              &                      & PIN &  44.8 &    12.00--40.50 & 0.4707$\pm$0.0047 & 49.5 \\
 20 &      NGC5506 &    701030030 &  2007-01-31T02:12:12 & XIS & 172.2 &      2.3--10.00 & 2.5013$\pm$0.0039 & 98.4 \\
    &              &              &                      & PIN &  44.7 &    12.00--39.30 & 0.4374$\pm$0.0046 & 48.9 \\

\enddata
\tablecomments{Column (4) is the observation start-date (header keyword {\sc date-obs}).
  Column (6) is the exposure time (header keyword {\sc exposure}).
  Column (8) is the background-subtracted count rate
  in the energy bands specified. For the XIS, this is the rate per XIS
  unit, averaged over XIS0, XIS1, and XIS3. Column (9) is the
  background-subtracted source count rate as a percentage of the total
  on-source count rate, in the energy intervals shown.
}
  \renewcommand{\baselinestretch}{1} 
\vspace{-1cm}
\end{deluxetable*}

When the primary \x\ continuum is scattered by distant matter at
several thousands of gravitational radii, the line
is narrow (FWHM\lt10,000~\kmps) and is unlikely to carry any
information about conditions in the accretion disk and the
strong-gravity regime near the SMBH. Narrow lines have been detected
in the great majority of AGNs with luminosities \lxht\lt$10^{45}$
\lunits.  The mean FWHM reported from \chandra\ High Energy
Transmission Grating (HETG) spectra is \sss2000~\kmps\
(\citealt{yaqoob2004,shu2010,shu2011}; see
  also \citealt{nandra2006}).  Low
equivalent width values from \suzaku\ observations
\citep[tens of eV,][]{fukazawa2011a} have also been interpreted as
evidence of narrow lines originating in distant material
\citep{ricci2014}. In contrast, when the primary \x\ continuum is
scattered close to the SMBH, Doppler and general relativistic effects
combined may give rise to a significantly broader line (FWHM tens of
thousands \kmps), reported in at least \sss36\%\ of AGNs
(\citealt{delacalleperez2010}; see also, e.g.,
\citealt{porquet2004,jimenez-bailon2005,guainazzi2006,nandra2007,brenneman2009,patrick2012,liu2015,mantovani2016,baronchelli2018}). In
this case, contributions to the line's width become increasingly
stronger as the primary continuum is scattered closer, and up to, the
innermost stable circular orbit (ISCO), where the accretion disk's
inner edge is located.  Since the ISCO location depends directly on
black hole spin, the latter leaves an imprint on the broad line's
profile and can in principle be measured by means of relativistic modeling of
the line via the \x\ reflection method of spin determination. As a
result, there is a significant number of SMBH spin measurements
\citep[e.g.][and references therein]{brenneman2013}. SMBH spin
constraints have very significant implications for understanding both
SMBHs and the way they affect their enviroment. Apart from potentially
constituting a Kerr-metric-based test of general relativity in the
strong-field regime, spin measurements can inform on jet-driving
mechanisms, e.g.~via extraction of SMBH rotational energy
\citep{blandford1977}, which in turn critically affect galactic
environments and evolution.

While relativistic modeling of broad lines in reflection spectra is an
established method of measuring black hole spin, the validity of at
least some results remains unclear;
overall, spin measurements via the reflection
method span the allowed spin values, sometimes for the same
object and data. For instance, spin is degenerate
with Fe abundance, \afe, \citep{patrick2012}, and extreme \afe\ values
obtained via the reflection method have frequently been reported
\citep[e.g.~$>$8 \afesun\ for Fairall~9,][]{lohfink2012}, without
satisfactory explanation as to their origin. While newer high-density
relativistic modeling appears to be making significant progress on
this issue \citep{garcia2016}, the effect of spin on the line profile
still suffers from degeneracies both with the radial line emissivity
profile and the ISCO value. To complicate matters more, \x\ spectra
for some AGNs which were well-known for showing broad \feka\ lines
have been modeled successfuly with narrow lines.  As explained in
\citet[][see their Figure~5]{yaqoob2016} for Fairall~9 \citep[see
  also][for MCG$+$8$-$11$-$11]{murphy2014}, reflection continuum
complexities could lead to a broad-line interpretation. Spectral
fitting with a model such as \myt\ \citep{murphy2009} that, unlike
earlier models, has finite
column density and models the \feka\ line and its associated continuum
in tandem can preclude the need for a broad line. It is also worth
noting that the earlier relativistic interpretation of time delays in
the Fe K band has recently not been confirmed in NGC 4151, following
an extensive \xmm\ campaign \citep{zoghbi2019}.

Whereas modeling the narrow line with infinite column-density material
excludes the possibility of probing the actual global column density,
the \myt\ model enables measurement of the column density of distant
obscuring matter both in and out of the line of sight. This is of
great significance for studies of the Cosmic \x\ Background
\citep[CXB,][]{giacconi1962}, which is thought to be produced by the
integrated emission of all accreting SMBHs in the universe. AGNs are
labeled Compton-thick or Compton-thin, depending on whether their
obscuring equivalent hydrogen column density exceeds
$10^{24}$\cunits\ or not.  CXB modeling relies on the number density
of Compton-thick, highly obscured AGNs, which according to some
estimates might make up to 50\%\ or more of the obscured AGN
population that contributes to the CXB \citep[e.g.][and references
  therein]{gilli2007,ueda2014,ricci2015}.  Traditionally, studies of
AGN samples have generally only probed a single, line-of-sight, column
density. However, measuring column density out of the line of sight
can specifically constrain global obscuration characteristics relevant
for CXB studies, especially at energies higher than
\sss10~keV. Several studies have now shown that the two column
densities may differ significantly
\citep[e.g.][]{lamassa2014,tzanavaris2019b,turner2020}, even
suggesting reclassification of an AGN from Compton-thick to
Compton-thin \citep[][for Mrk 3]{yaqoob2015}.

The main goals of this paper are (1) to measure equivalent hydrogen
column density both in and out of the line of sight, and assess the
global Compton-thinness for a sample of AGNs that are traditionally
thought to be bona fide Compton-thin based only on the
line-of-sight column density,
and
(2) to assess the nature of, and obtain parameters for,
neutral \feka\ emission.
We use a sample of AGNs in the local
universe that have 
line-of-sight equivalent hydrogen column densities $<$$10^{24}$ \cunits,
i.e.~are likely \cthin\ in the line of sight, and do
not show strong ionization absorption or emission features, based on
the available literature and preliminary inspections of available
spectra.
Traditionally, AGN Compton thickness classifications
have been based on line-of-sight equivalent hydrogen
column densities, which may not
be at all representative of the global equivalent hydrogen column density.
In addition, we choose a sample with
available \suzaku\ \x\ broadband spectra, whose
coverage beyond \sss10~keV is
critical for optimally fitting simultaneously both
the direct and reprocessed continua
with physically motivated models.
This is not a statistical study as
our sample is too small to draw any
statistically robust conclusions about global
properties of bona fide \cthin\ AGNs.
However, we expect that AGNs
which appear to be \cthin\ in the line of sight and have relatively
simple \x\ spectra can in principle best constrain the ubiquitous
narrow \feka\ line emission from material distant from the central
SMBH, which would not be the case with Type I or ``bare'' Seyferts.
While bare Seyferts have traditionally been thought best
for studying broad \fek\ emission lines, it is
particularly challenging for such studies to measure
the column density of the narrow line component, making
deconvolution of the two components ambiguous.
On the contrary, Compton-thin AGNs can successfully
constrain the narrow line, and at the same time the
line-of-sight column density is not so large as to
obscure the broad-line component.

This is of interest, as some of the AGNs in our sample
have reported broad lines in the literature.  We note up front that in this
paper we will use three symbols to refer to column densities. The
symbol \nh\ will refer to the general equivalent hydrogen column
density concept, while \nhz\ and \nhs\ will be used to distinguish
between column densities in and out of the line of sight
(\scr{sec-mytorus}).

The structure of the paper is as follows. \scr{sec-obs} describes the
observations and data reduction. \scr{sec-modeling} presents our
spectral modeling methodology. Results are presented
and discussed in \scr{sec-res}. We summarize in
\scr{sec-summ}. Details on individual fits are presented
in Appendix~\ref{app-indi} and spectral plots for all
data are shown in Appendix~\ref{app-plots}.

\section{Observations and Data Reduction}\label{sec-obs}

\subsection{Sample Selection}

Our goal is to use specifically \cthin\ systems which would provide
miminal constraints to \los\ column density but still allow signal due
to any broad line component to get through. We first use the
\chandra\ sample of \citet{yaqoob2004} and \citet{shu2010} in
conjunction with the \suzaku\ archive to identify relatively nearby
AGNs that have broadband \x\ coverage. The \chandra\ results allow us
to have a baseline width for the narrow line thus constraining any
broad component. The choice of \suzaku\ is motivated by the broadband
\x\ coverage, allowing us to constrain the reflected continuum and the
Compton hump at $\gtrsim$10~keV, while providing the highest combined
spectral resolution and sensitivity among all archival \x\ mission
data in the region of \feka\ emission.
\footnote{We note that simultaneous \xmm-\nustar\ observations would also be very useful for this work, however only a few were available at the start of this project.}

Based on this initial AGN set,
our final sample selection is a two-tiered process. First, we
perform a literature as well as visual search to identify AGNs that
the evidence suggests are \cthin\ in the line of sight, i.e.~with
equivalent hydrogen column densities $<$10\up{24}\cunits.  We then
further exclude systems that are known to show pronounced ionized
outflows which would significantly complicate modeling and increase
model uncertainties.
Finally,
we assess the observed spectral shape for each observation for evidence
that there is sufficient column density in the \los\ to ensure robust
measurements of this quantity. To this end we
measure the ratio of normalized
\reunits\ in the immediate redward vicinity of the
\feka\ line $\Delta E_{\rm 6 keV} \equiv (5.8 - 6.0) / (1+z)$ to those in the
spectral range $\Delta E_{\rm soft} \equiv (0.5 - 5.0) / (1+z)$.  This
empirically based test suggests that if the minimum number of counts
in $\Delta E_{\rm soft}$ are lower than the maximum number of counts
in $\Delta E_{\rm 6 keV}$, the turnover in the
spectral data would be sufficient to robustly measure the \los\ column
density. In addition, lower energy components due to, e.g., additional
soft emission could safely be ignored.
The final sample thus contains 20 individual observations for
eight \cthin\ AGNs, shown
in \tr{tab-suz}.

\subsection{Data Reduction}
We study 20 archival observations carried out by the joint
Japan/US \x\ astronomy satellite, \suzaku\ \citep{mitsuda2007}.
\suzaku\ had four \x\ imaging spectrometers
\citep[XIS,][]{koyama2007} and a collimated Hard \x\ Detector
\citep[HXD,][]{takahashi2007}.  Each XIS consisted of four CCD detectors
with a field of view of $17.8\times17.8$~arcmin\up{2}. Of the three front-side
illuminated (FI) CCDs, (XIS0, XIS2, and XIS3) XIS2 had ceased to operate
prior to the observations studied here.  We thus used FI CCDs XIS0 and
XIS3, as well as the back-side-illuminated (BI) XIS1.  The operational
bandpass is (0.2--12)~0.4--12~keV for (BI) FI. However, the useful
bandpass depends on the signal-to-noise ratio of the
background-subtracted source data since the effective area strongly
diminishes at the ends of the operational bandpass.  The HXD consisted
of two non-imaging instruments (the PIN and GSO) with combined
bandpass of $\sim$10--600~keV. Both instruments were background-limited,
with the GSO having the smaller effective area. We only use the PIN
data because the GSO data have too low signal-to-noise to provide a reliable spectrum.

The principal data selection and screening criteria for the XIS are
the selection of only ASCA grades 0, 2, 3, 4, and 6, the removal of
flickering pixels with the FTOOL {\tt cleansis}, and exclusion of data
taken during satellite passages through the South Atlantic Anomaly (SAA), as
well as for time intervals less than 256~s after passages through the
SAA, using the T\_SAA\_HXD house-keeping parameter. Data are also
rejected for Earth elevation angles (ELV) less than 5\degr, Earth
daytime elevation ang-
\global\pdfpageattr\expandafter{\the\pdfpageattr/Rotate 0}
\clearpage

\begin{minipage}{0.9\textwidth}
\vspace{100pt} 
  \begin{rotatetable}
     \renewcommand{\baselinestretch}{0.8} 
\begin{deluxetable*}{ccccc ccccc ccc} 
  \tablecaption{\footnotesize \vspace{3pt} Spectral Fitting Results for the Decoupled \myt\ Model. \vspace{-5pt} \label{tab-fit}}
  \tabletypesize{\footnotesize}
\tablehead{
  [-0pt]
\colhead{\#} &               
\colhead{Name} &             
\colhead{ObsID} &            
\colhead{\cs/d.o.f.} &       
\colhead{\cpx} &             
\colhead{\nhGal} &           
\colhead{\nhz} &             
\colhead{\nhs} &             
\colhead{$A_S$}&             
\colhead{$\Gamma$}&          
\colhead{$\Gamma_{\rm soft}$}&  
\colhead{\nhsoft} &          
\colhead{\Nrel}              
\\[-.15cm]                         
\colhead{} &                  
\colhead{} &                  
\colhead{} &                  
\colhead{} &                  
\colhead{} &                  
\colhead{($10^{22}$\cunits)} & 
\colhead{($10^{22}$\cunits)} & 
\colhead{($10^{22}$\cunits)} & 
\colhead{} &                  
\colhead{} &                  
\colhead{} &                  
\colhead{($10^{22}$\cunits)} & 
\colhead{}                    
\\[-.6cm]}
\colnumbers
\startdata
\vspace{-15pt}\\
  1  &         Cen A  &   100005010  &  552.7/342   &      \aer{0.99}{+0.01}{-0.01}  	&  0.0235  	&  \aer{11.292}{+0.174}{-0.152}  	&  \aer{30.317}{+13.466}{-10.657}  	&   \aer{0.826}{+0.226}{-0.159}  	&    \aer{1.728}{+0.014}{-0.014} & \ldots                      & \ldots                      & \aer{0.014}{+0.002}{-0.002}    \\ 
  2  &         Cen A  &   708036010  &  355.2/331   &                1.16$^{\rm f}$  	&  0.0235  	&  \aer{10.397}{+0.293}{-0.283}  	&  \aer{15.602}{+2.593}{-2.292}  	&               1.000$^{\rm f}$  	&    \aer{1.743}{+0.013}{-0.012} & \ldots                      & \ldots                      & \aer{0.014}{+0.005}{-0.005}    \\ 
  3  &         Cen A  &   708036020  &  299.7/305   &                1.16$^{\rm f}$  	&  0.0235  	&  \aer{10.960}{+0.493}{-0.426}  	&  \aer{40.854}{+35.036}{-25.427}  	&   \aer{1.176}{+0.923}{-0.268}  	&    \aer{1.799}{+0.044}{-0.035} & \ldots                      & \ldots                      & \aer{0.021}{+0.007}{-0.008}    \\ 
  4  &         Cen A  &   704018010  &  533.8/358   &      \aer{1.05}{+0.01}{-0.01}  	&  0.0235  	&  \aer{10.874}{+0.249}{-0.137}  	&  \aer{19.957}{+9.135}{-1.334}  	&   \aer{0.962}{+0.963}{-0.230}  	&    \aer{1.725}{+0.009}{-0.011} & \ldots                      & \ldots                      & \aer{0.008}{+0.002}{-0.002}    \\ 
  5  &         Cen A  &   704018020  &  463.1/359   &                1.18$^{\rm f}$  	&  0.0235  	&  \aer{11.313}{+0.150}{-0.133}  	&  \aer{31.141}{+12.850}{-6.509}  	&   \aer{0.878}{+0.270}{-0.150}  	&    \aer{1.754}{+0.011}{-0.007} & \ldots                      & \ldots                      & \aer{0.013}{+0.002}{-0.002}    \\ 
  6  &         Cen A  &   704018030  &  544.6/361   &      \aer{1.15}{+0.01}{-0.01}  	&  0.0235  	&  \aer{11.290}{+0.203}{-0.157}  	&  \aer{25.557}{+16.145}{-12.596}  	&   \aer{0.790}{+0.497}{-0.210}  	&    \aer{1.743}{+0.012}{-0.014} & \ldots                      & \ldots                      & \aer{0.014}{+0.002}{-0.002}    \\ 
  7  &   MCG+8-11-11  &   702112010  &  350.0/316   &      \aer{1.01}{+0.06}{-0.06}  	&  0.176  	&   \aer{0.125}{+0.086}{-0.090}  	&  \aer{146.350}{+63.720}{-53.177}  	&   \aer{0.943}{+0.100}{-0.097}  	&    \aer{1.740}{+0.019}{-0.016} & \ldots                      & \ldots                      &$\ldots$                        \\ 
  8  &       NGC526A  &   705044010  &  302.8/298   &                1.16$^{\rm f}$  	&  0.0213  	&   \aer{2.248}{+0.785}{-0.498}  	&  \aer{235.000}{+125.910}{-89.350}  	&               1.000$^{\rm f}$  	&    \aer{1.784}{+0.039}{-0.043} & \ldots                      & \ldots                      & \aer{0.318}{+0.334}{-0.247}    \\ 
  9  &       NGC2110  &   100024010  &  423.6/330   &      \aer{1.04}{+0.02}{-0.02}  	&  0.163  	&   \aer{3.913}{+0.061}{-0.053}  	&  \aer{60.195}{+37.733}{-17.334}  	&   \aer{0.421}{+0.065}{-0.044}  	&    \aer{1.668}{+0.011}{-0.010} & \ldots                      & \ldots                      &$\ldots$                        \\ 
 10  &       NGC2110  &   707034010  &  438.5/361   &                1.16$^{\rm f}$  	&  0.163  	&   \aer{4.557}{+0.050}{-0.066}  	&  \aer{40.013}{+12.587}{-15.063}  	&   \aer{0.559}{+0.188}{-0.073}  	&    \aer{1.687}{+0.008}{-0.010} & \ldots                      & \ldots                      &$\ldots$                        \\ 
 11  &       NGC3227  &   703022020  &  306.7/303   &                1.16$^{\rm f}$  	&  0.0186  	&  \aer{16.227}{+1.197}{-1.168}  	&  \aer{62.362}{+23.420}{-19.297}  	&   \aer{1.579}{+0.269}{-0.204}  	&                1.400$^{\rm f}$  & \aer{3.579}{+2.042}{-0.444} & \aer{5.112}{+3.769}{-1.016} & \aer{5.700}{+3.881}{-2.475}    \\ 
 12  &       NGC3227  &   703022030  &  389.9/301   &      \aer{1.04}{+0.04}{-0.04}  	&  0.0186  	&   \aer{5.436}{+0.824}{-0.295}  	&  \aer{10.031}{+10.723}{-7.866}  	&   \aer{4.518}{+8.910}{-2.100}  	&                1.400$^{\rm f}$  & \aer{2.704}{+0.560}{-0.398} & \aer{4.938}{+0.930}{-0.624} & \aer{4.838}{+0.890}{-1.553}    \\ 
 13  &       NGC3227  &   703022040  &  240.1/250   &      \aer{1.11}{+0.16}{-0.20}  	&  0.0186  	&  \aer{24.251}{+2.384}{-2.675}  	&  \aer{99.881}{+96.329}{-43.553}  	&   \aer{2.279}{+0.503}{-0.279}  	&                1.400$^{\rm f}$  & \aer{5.137}{+0.587}{-0.697} & \aer{7.378}{+7.348}{-2.323} & \aer{40.598}{+39.964}{-15.637} \\ 
 14  &       NGC3227  &   703022050  &  275.3/287   &                1.16$^{\rm f}$  	&  0.0186  	&   \aer{9.567}{+0.413}{-0.640}  	&  \aer{50.339}{+24.103}{-27.373}  	&   \aer{1.389}{+0.318}{-0.233}  	&    \aer{1.658}{+0.039}{-0.060} & \ldots                      & \ldots                      & \aer{0.156}{+0.018}{-0.010}    \\ 
 15  &       NGC3227  &   703022060  &  267.9/290   &      \aer{1.01}{+0.08}{-0.06}  	&  0.0186  	&  \aer{13.534}{+1.245}{-0.797}  	&  \aer{99.739}{+30.551}{-29.922}  	&               1.000$^{\rm f}$  	&    \aer{1.572}{+0.086}{-0.057} & \ldots                      & \ldots                      & \aer{0.182}{+0.009}{-0.020}    \\ 
 16  &       NGC4258  &   701095010  &  166.9/146   &      \aer{0.84}{+0.12}{-0.11}  	&  0.0419  	&  \aer{10.801}{+0.924}{-0.937}  	&  \aer{11.179}{+5.175}{-4.028}  	&               1.000$^{\rm f}$  	&    \aer{1.871}{+0.077}{-0.076} & \ldots                      & \ldots                      & \aer{0.088}{+0.011}{-0.012}    \\ 
 17  &       NGC4507  &   702048010  &  271.3/301   &      \aer{1.21}{+0.12}{-0.11}  	&  0.068  	&  \aer{80.560}{+7.486}{-7.137}  	&  \aer{27.993}{+4.609}{-4.837}  	&   \aer{1.367}{+0.258}{-0.236}  	&    \aer{1.531}{+0.111}{-0.103} & \ldots                      & \ldots                      & \aer{0.009}{+0.004}{-0.003}    \\ 
 18  &       NGC5506  &   701030010  &  342.8/316   &      \aer{1.06}{+0.03}{-0.03}  	&  0.0423  	&   \aer{3.204}{+0.079}{-0.079}  	&            1000.000$^{\rm f}$  	&   \aer{1.546}{+0.100}{-0.103}  	&    \aer{1.935}{+0.018}{-0.018} & \ldots                      & \ldots                      &$\ldots$                        \\ 
 19  &       NGC5506  &   701030020  &  393.1/332   &      \aer{1.07}{+0.03}{-0.03}  	&  0.0423  	&   \aer{3.148}{+0.074}{-0.072}  	&  \aer{380.560}{+73.110}{-51.120}  	&               1.000$^{\rm f}$  	&    \aer{1.895}{+0.016}{-0.015} & \ldots                      & \ldots                      &$\ldots$                        \\ 
 20  &       NGC5506  &   701030030  &  402.3/325   &      \aer{1.16}{+0.03}{-0.03}  	&  0.0423  	&   \aer{3.401}{+0.086}{-0.077}  	&  \aer{397.100}{+183.880}{-77.680}  	&               1.000$^{\rm f}$  	&    \aer{1.953}{+0.015}{-0.016} & \ldots                      & \ldots                      &$\ldots$                        \\

\enddata
\vspace{10pt}

\begin{minipage}{1.29\textwidth} 
\tablecomments{  Column
  clarifications: (4): fit \cs\ and degrees of freedom (d.o.f.); (5):
  cross-normalization between PIN and XIS data; (6): tabulated
  Galactic column density \citep{hi4pi2016};
  (7): equivalent hydrogen column density
  associated with the single zeroth-order (direct) continuum; (8):
  equivalent hydrogen column density associated with the scattered
  (reflected) continuum and the fluorescent line emission component;
  (9): relative normalization between direct and scattered continuum;
  (10): power-law slope for \myt\ components;
  (11): power-law slope for soft power law;
  (12): equivalent hydrogen column density associated with the soft power law;
  (13): relative normalizations of distant or scattered power-law component
  with respect to main \myt\ power law
  at 1~keV.
  $^{\rm f}$ indicates a frozen parameter.
  }
\end{minipage}

\renewcommand{\baselinestretch}{1}
\vspace{-.7cm}
\end{deluxetable*}

  \end{rotatetable}
\vspace{-25cm}
 \end{minipage}

\global\pdfpageattr\expandafter{\the\pdfpageattr/Rotate 90}

\clearpage

\global\pdfpageattr\expandafter{\the\pdfpageattr/Rotate 0}

\noindent les (DYE\_ELV) less than 20\degr, 
and values of
the magnetic cutoff rigidity less than 6~GeV/$c^2$.  Residual
uncertainties in the XIS energy scale are of the order of 0.2\%\ or
less (or \sss13~eV~at 6.4~keV). The cleaning and data selection
results in the net exposure times shown in \tr{tab-suz}.
We extract XIS source spectra in a circular extraction region with a
radius of 3\hspace{1pt}\arcmin\hspace{-3pt}.\hspace{1pt}5.  
We
construct background XIS spectra from off-source areas of the detector,
after removing a circular region with a radius of
4\hspace{1pt}\arcmin\hspace{-3pt}.\hspace{1pt}5 centered on the
source, as well as the calibration sources (using rectangular
masks). The background-subtraction method for the HXD/PIN uses
files corresponding to the ``tuned'' version of the background model.

Spectral response matrix files (RMFs) and telescope effective area
files (ARFs) for the XIS data are made using the mission-specific
FTOOLS {\tt xisrmfgen} and {\tt xisimarfgen}, respectively. The XIS spectra from
XIS0, XIS1, and XIS3 are combined into a single spectrum for spectral
fitting. The three RMFs and ARFs are all combined, using the
appropriate weighting (according to the count rates and exposure times
for each XIS), into a single response file for the combined XIS
background-subtracted spectrum. For the HXD/PIN spectrum, the supplied
spectral response matrices appropriate for the times and nominal
pointing mode of the observations
are used for spectral fitting.

\renewcommand{\baselinestretch}{0.8} 
\begin{deluxetable*}{ccccccccc} 
\tablecaption{\footnotesize \vspace{3pt} Properties of Narrow \feka\ line. \vspace{-5pt} \label{tab-line}}
\tabletypesize{\footnotesize}
\tablehead{
  [-0pt]
\colhead{\#}                             
&\colhead{Name}                          
&\colhead{ObsID}                         
&\colhead{\eshift}                       
&\colhead{\sigl}                         
&\colhead{FWHM}                          
&\colhead{Residuals}                     
&\colhead{Flux}                          
&\colhead{EW}                            
\\[-.15cm]
\colhead{}                               
&\colhead{}                              
&\colhead{}                              
&\colhead{(eV)}                          
&\colhead{(eV)}                          
&\colhead{\kmps}                         
&\colhead{(\%)}                          
&\colhead{($10^{-5}$\phunits)}           
&\colhead{(eV)}                          
\\[-.6cm]}
\colnumbers
\startdata
\vspace{-15pt}\\
  1  &         Cen A  &   100005010  &  \aer{14.0}{+3.5}{-3.3}  	&      $20.0^{\rm f}$  	&  $2353^{\rm f}$  	&                     3.6   	&  \aer{24.21}{+6.64}{-4.65}  	&        \aer{52}{+14}{-10} \\ 
  2  &         Cen A  &   708036010  &  \aer{16.1}{+11.9}{-12.0}  	&      $20.0^{\rm f}$  	&  $2353^{\rm f}$  	&                    10.5   	&  \aer{29.94}{+37.09}{-14.52}  	&        \aer{42}{+52}{-20} \\ 
  3  &         Cen A  &   708036020  &  \aer{20.2}{+10.8}{-13.2}  	&      $20.0^{\rm f}$  	&  $2353^{\rm f}$  	&                    11.6   	&  \aer{28.92}{+22.72}{-6.60}  	&        \aer{62}{+49}{-14} \\ 
  4  &         Cen A  &   704018010  &  \aer{12.1}{+4.2}{-4.1}  	&      $20.0^{\rm f}$  	&  $2353^{\rm f}$  	&                     5.3   	&  \aer{32.47}{+32.49}{-7.76}  	&        \aer{47}{+47}{-11} \\ 
  5  &         Cen A  &   704018020  &  \aer{14.1}{+3.8}{-4.2}  	&      $20.0^{\rm f}$  	&  $2353^{\rm f}$  	&                     4.1   	&  \aer{34.99}{+10.75}{-6.00}  	&         \aer{55}{+17}{-9} \\ 
  6  &         Cen A  &   704018030  &  \aer{17.7}{+4.2}{-4.4}  	&      $20.0^{\rm f}$  	&  $2353^{\rm f}$  	&                     5.6   	&  \aer{31.54}{+19.82}{-8.39}  	&        \aer{46}{+29}{-12} \\ 
  7  &   MCG+8-11-11  &   702112010  &  \aer{20.6}{+7.5}{-7.3}  	&      $25.0^{\rm f}$  	&  $2941^{\rm f}$  	&                     2.3   	&  \aer{6.01}{+0.64}{-0.62}  	&          \aer{59}{+6}{-6} \\ 
  8  &       NGC526A  &   705044010  &  \aer{28.0}{+29.2}{-31.6}  	&       $0.8^{\rm f}$  	&  $100^{\rm f}$  	&                     3.4   	&  \aer{3.24}{+0.56}{-1.20}  	&         \aer{46}{+8}{-17} \\ 
  9  &       NGC2110  &   100024010  &  \aer{18.2}{+6.4}{-5.8}  	&      $20.0^{\rm f}$  	&  $2353^{\rm f}$  	&                     2.4   	&  \aer{6.39}{+0.99}{-0.67}  	&          \aer{36}{+6}{-4} \\ 
 10  &       NGC2110  &   707034010  &  \aer{10.2}{+8.1}{-6.1}  	&      $20.0^{\rm f}$  	&  $2353^{\rm f}$  	&                     2.4   	&  \aer{8.55}{+2.88}{-1.12}  	&         \aer{40}{+13}{-5} \\ 
 11  &       NGC3227  &   703022020  &  \aer{24.8}{+8.2}{-11.6}  	&  \aer{46.1}{+19.0}{-23.7}  	&  \aer{5426}{+2235}{-2785}  	&                     1.4   	&  \aer{4.43}{+0.76}{-0.57}  	&        \aer{90}{+15}{-12} \\ 
 12  &       NGC3227  &   703022030  &  \aer{7.1}{+11.6}{-12.6}  	&  \aer{60.7}{+20.3}{-21.8}  	&  \aer{7145}{+2390}{-2568}  	&                     1.4   	&  \aer{4.43}{+8.74}{-2.06}  	&       \aer{78}{+153}{-36} \\ 
 13  &       NGC3227  &   703022040  &  \aer{20.0}{+8.4}{-7.5}  	&       $0.8^{\rm f}$  	&  $100^{\rm f}$  	&                     0.6   	&  \aer{3.93}{+0.87}{-0.48}  	&        \aer{95}{+21}{-12} \\ 
 14  &       NGC3227  &   703022050  &  \aer{12.1}{+3.2}{-2.0}  	&       $0.8^{\rm f}$  	&  $100^{\rm f}$  	&                     1.0   	&  \aer{3.77}{+0.86}{-0.63}  	&        \aer{65}{+15}{-11} \\ 
 15  &       NGC3227  &   703022060  &  \aer{17.1}{+13.8}{-9.2}  	&       $0.8^{\rm f}$  	&  $100^{\rm f}$  	&                     0.8   	&  \aer{2.54}{+3.03}{-0.10}  	&         \aer{56}{+66}{-2} \\ 
 16  &       NGC4258  &   701095010  &  \aer{29.6}{+36.1}{-34.6}  	&      $26.0^{\rm f}$  	&  $3059^{\rm f}$  	&                     0.6   	&  \aer{0.43}{+1.48}{-0.33}  	&       \aer{31}{+106}{-24} \\ 
 17  &       NGC4507  &   702048010  &  \aer{15.3}{+3.7}{-4.0}  	&       $0.8^{\rm f}$  	&  $100^{\rm f}$  	&                     0.4   	&  \aer{5.39}{+1.02}{-0.93}  	&       \aer{146}{+27}{-25} \\ 
 18  &       NGC5506  &   701030010  &  \aer{34.7}{+8.2}{-8.2}  	&      $34.0^{\rm f}$  	&  $4000^{\rm f}$  	&                     3.2   	&  \aer{9.49}{+0.61}{-0.63}  	&          \aer{62}{+4}{-4} \\ 
 19  &       NGC5506  &   701030020  &  \aer{14.1}{+8.0}{-8.5}  	&      $34.0^{\rm f}$  	&  $4000^{\rm f}$  	&                     3.6   	&  \aer{8.43}{+1.16}{-0.14}  	&          \aer{49}{+7}{-1} \\ 
 20  &       NGC5506  &   701030030  &  \aer{-8.2}{+13.8}{-17.6}  	&      $34.0^{\rm f}$  	&  $4000^{\rm f}$  	&                     3.8   	&  \aer{7.81}{+1.16}{-0.22}  	&          \aer{52}{+8}{-1} \\ 

\enddata
\tablecomments{Column information: (4) - line peak energy shift; (5) - energy width of line's Gaussian convolution kernel;  (6) - line velocity width calculated from \sigl\ value; (7) - maximum residuals as percentage of model value in the vicinty of \feka\ line (energy range 6.3--6.5 keV, rest frame); (8) - line flux; (9) - line equivalent width. $^{\rm f}$ indicates a frozen parameter. In such cases values other than 100~\kmps\ (0.8~eV) are taken from literature results with the \chandra/HETG.
}
\renewcommand{\baselinestretch}{1}
\vspace{-0.6cm}
\end{deluxetable*}

We determine useful energy bandpasses for the spectrum from each
instrument by first assessing background-subtraction systematics.  For
XIS we use spectra with a uniform binning, with 30 eV bin width, and
identify spectral ranges (Column (7) in \tr{tab-suz}) in which
there were $>$20 counts per bin
for the unscaled background, total source, and background-subtracted
source. In addition, in these regions, the background counts, scaled
by the relative areas of source vs.~background region, are $<$50\%\ of
the background-subtracted source counts. Since the counts per bin are
$>$20 in the stated energy bands, we are able to use the $\chi^{2}$
statistic for spectral fitting.
Note that we do not group spectral bins using a
signal-to-noise ratio threshold, which can wash out
weak features. We further exclude spectral regions that are subject to
calibration uncertainties in the effective area due to atomic
features.  Specifically, it is known that this calibration is poor in
the ranges $\sim$1.8--1.9 and $\sim$2.0--2.3~keV due to Si in the
detector and Au M edges in the telescope, respectively. The effective
area also has a steep change at $\sim$1.56~keV due to Al in the
telescope. Thus for the purposes of analysis and spectral fitting we
conservatively choose to exclude the energy range 1.5--2.3~keV. After
preliminary fitting of the XIS data, we further exclude the region
below 1.5~keV. As explained in \scr{sec-modeling} fitting this region
would increase model complexity without improving scientific
conclusions as the soft energy components leave higher-energy fitted
components largely unaffected. 

For HXD/PIN, we first perform background subtraction on the original
256-bin spectrum to identify the maximum continuous spectral range
with nonnegative background-subtracted counts, since negative
background-subtracted counts would indicate an obvious breakdown of
the background model.  This is then rebinned uniformly to bin widths
of 1.5~keV, leading to the final useful spectral ranges shown in
\tr{tab-suz}.

\section{Spectral Modeling: \feka\ line emission and Reflection Continuum}\label{sec-modeling}
\subsection{Overview}

Our primary modeling goal is to apply a model with strong physical
basis, which self-consistently produces the \feka\ line and its
associated reflection continuum with solar Fe abundance and finite
equivalent hydrogen column density. Some of the AGNs in our sample
have been modeled with relativistically broadened \feka\ lines.  On
the other hand, many also have results from \chandra-HETG spectral
fitting of a narrow \feka\ line.  Using the \myt\ model
\citep{murphy2009,yaqoob2012}, we specifically test whether the data
can be fitted with narrow \feka\ emission and its associated
continuum, both arising in distant scattering material with a patchy
geometry. At the same time, the decoupled implementation used here
(see \scr{sec-mytorus}) allows us to measure column densities both in
and out of the line of sight, thus testing whether AGNs traditionally
classified as \cthin\ are also Compton-thin globally. 

We use \xspec\ \citep[][version~12.10.0{\it c}]{arnaud1996} and the
\cs\ statistic for minimization.  We include Galactic absorption,
\nhGal, modeled with a {\tt phabs} component and fixed at the
tabulated Galactic column density values in \citet{hi4pi2016}. We use
photoelectric cross sections from \citet{verner1996} with element
abundances from \citet{anders1989}.

We calculate statistical errors at the 90\%\ confidence level for one
parameter of interest, corresponding to $\Delta\chi^2=2.706$, by
iteratively stepping away from the best-fit minimum. For line flux and
equivalent width we determine errors as explained in
\scr{sec-fxew}. We do not give statistical errors on continuum fluxes
and luminosities because absolute continuum fluxes are dominated by
systematic uncertainties that are not well quantified, typically of
the order of $\sim$$10\%$--$20\%$
\citep[e.g.][]{tsujimoto2011,madsen2017}.

\subsectionU{Suzaku \textit{XIS Preliminary Fits}}
\label{subsec-XIS}

We first carry out preliminary fits using only the combined XIS-data
over the XIS spectral ranges shown in \tr{tab-suz} and excluding the
1.5--2.3~keV range as mentioned.  Although the AGN sample had been
selected to be as free as possible of complicated absorption or
emission features, all spectra need additional \x\ components that dominate the
soft ($\la$2~keV) spectral range. These components model either
thermal emission (modeled with, e.g., {\tt apec}\footnote{\href{https://heasarc.gsfc.nasa.gov/xanadu/xspec/manual/node135.html}{https://heasarc.gsfc.nasa.gov/xanadu/xspec/manual/node135.html}})
or additional soft power laws, whose normalizations decrease by up to
orders of magnitude in the harder \x\ spectral range. We also perform
preliminary ``restricted'' XIS fits using the $>$2.3 keV data only,
excluding the components dominating in the $<$1.5 keV range.  Given
our primary science goals (\scr{sec-intro}), the key \myt\ model
parameters (\scr{sec-mytorus}) of interest are then column densities
in and out of the line of sight, as well as the direct and reflected
continuum power law index, and \feka\ emission line parameters.
%
With respect to the unrestricted XIS fits, these parameters change by
$<$10\%\ in the restricted fits. Since we are also not interested in the
details of the components dominating the soft \x\ region, we choose
to exclude the region below 2.3 keV in all final fits,
exchanging the uncertainty due to the complications from the soft
components with an uncertainty from a more limited fitting region
(see, e.g., also \citealt{fukazawa2011b} who also fit XIS spectra
between 2--10 keV for the \suzaku\ observations of \cena, and
\citealt{markowitz2007} who fit the hard band separately to
constrain the fit).
This latter uncertainty is ultimately incorporated in the final
parameter estimates. 
We then perform joint \xispin\ fits, treating the \cpx\
cross-normalization factor as explained in \scr{sec-xispin}.

\vspace{10pt}

\subsectionMb{}
\vspace{-16pt}
\hspace{0.3\columnwidth}{\textit{Decoupled \fauxsc{mytorus} Model}}
\label{sec-mytorus}

\myt\ provides for several possible configurations apart from the
default toroidal geometry. For in-depth descriptions see
\citet{murphy2009}, \citet{yaqoob2011}, \citet{yaqoob2012},
\citet{lamassa2014}, \citet{yaqoob2016}, and the
\myt\ manual\footnote{\vspace{-.1cm}\href{http://mytorus.com/manual}{http://mytorus.com/manual}}.
In this paper we use the ``decoupled'' implementation, which allows
for obscuring gas that has a patchy or clumpy geometry, unlike a
smooth torus of uniform column density. In this case the model's
equivalent neutral hydrogen column density associated with the
primary, unscattered (``zeroth order'') \x\ continuum (\nhz) is purely
a line-of-sight parameter, while the column density \nhs, associated
with the secondary, scattered or reflected continuum, probes
obscuration out of the line of sight. The primary and
secondary continua have (the same) power-law photon index $\Gamma$.  In some
cases an additional power-law continuum is needed (labeled ``second PL''
in spectral plots). In most cases, this continuum
has the same photon index $\Gamma$ as the primary
continuum.
Since we assume a patchy geometry, this component
is interpreted as AGN primary continuum that
does not intercept the obscuring medium.
However, in three \gc3227 obsIDs a very soft
power-law continuum is needed instead, attributed
to a ``soft-excess'' and also observed
before with \xmm\ \citep{markowitz2009b}.

The model uses solar abundances and self-consistently produces the
\feka\ and \fekb\ fluorescent emission-line spectrum, as well as
absorption and Compton scattering effects on continuum and line
emission.
As the reflected continuum and these emission lines
result from Compton scattering and fluorescence, respectively,
in the same global matter distribution, the column density associated
with the lines is by definition identical to \nhs, and the same applies
to continuum normalizations associated with the two columns.

An illustrative sketch of the assumed
configuration is shown in Figure~1 of \citet{tzanavaris2019b};
Table~2 in that paper also summarizes the main continua, associated
column densities, terminology, and symbols used.
%

The model includes a parameter for the relative normalization between
the direct and scattered continuum ($A_S$).  In decoupled mode a value
of 1.0 implies a covering factor of \sss0.5 in a steady state; other
values are a convolution of covering factor and time variability.
Regardless of the actual $A_S$ value, the reflection continuum and
\feka\ line are always self-consistent with each other.  Occasionally,
this parameter cannot be independently constrained during fitting. In
such cases we arbitratily fix $A_S$ to 1.0, as indicated in \tr{tab-fit}
and in what follows.

\subsection{\feka\ Line Energy, Velocity Width, and Equivalent Width}
\label{sec-fxew}
\myt\ models Fe ${\rm K}\alpha_1$, ${\rm
  K}\alpha_2$ and \fekb\ emission at rest
energies 6.404, 6.391, and 7.058~keV, respectively.
As in \suzaku\ data the
line peaks are likely to be offset due to instrumental calibration
systematics and/or mild ionization effects,
we used the best-fit redshift value 
to calculate a \feka\ line energy offset
in the observed frame, with positive shifts implying \feka\ centroid
energies higher than the \feka\ model mean centroid rest energy of 6.400~keV.

We used the Gaussian convolution
kernel {\tt gsmooth} in \xspec\ to implement line velocity
broadening, with width $\sigma_E =
\sigma_L \left( \frac{E_0}{6 {\rm keV}} \right)$, where
$E_0$ is the centroid energy,
and \sigl\ the fitting parameter.
To avoid fitting instabilities, we first
step through the line redshift parameter $z$ with
\sigl\ fixed at \ten{8.5}{-4} keV (100 \kmps, FWHM).
After determining a stable minimum, we fix $z$ and
step through \sigl\ instead. In practice, in most cases
the lower 90\%\ limit tends to \sigl~$=0$, indicating
the narrow line is not resolved. In this case,
we choose to freeze
\sigl\ to the values reported by \chandra\ HETG work,
or to 100 \kmps\ otherwise.

After the best fit is obtained, we isolate the model emission-line table
to measure the observed flux of the
\feka\ line, \ifeka, in an energy range excluding the \fekb\ line with
the \xspec\ {\tt flux} command. We measure the equivalent width (EW)
by means of the line flux and the total monochromatic continuum flux
at the observed line peak energy. \ifeka\ and EW in the AGN frame are
then obtained by multiplying observed values by $(1+z)$. As these are not
explicit model parameters, we estimate fractional errors by using the
fractional errors on $A_S$.

\subsection{Continuum Fluxes and Luminosities}
We calculate continuum fluxes and luminosities using the best-fit
model and the {\tt flux} and {\tt lumin} commands in \xspec. We
obtain absorbed fluxes in the observed frame (labeled ``obs'') and
both absorbed and unabsorbed luminosities in the AGN frame (``rest,
abso,'' ``rest, unabso,'' respectively).
For absorbed values, we use the total
best-fit model minus any additional Gaussian emission lines, and for
unabsorbed ones only the direct power-law component. The energy
ranges used are 2--10 and 10--30~keV.

\subsection{\suzaku\ PIN vs.~XIS cross-normalization}\label{sec-xispin}
The relative cross-normalization between PIN and XIS data involves
many factors \citep[see][for a detailed discussion]{yaqoob2012}.  The
recommended PIN:XIS ratios (hereafter \cpx) for ``HXD-nominal,''
vs.~``XIS-nominal'' observations are 1.18 and 1.16,
respectively.\footnote{\href{ftp://legacy.gsfc.nasa.gov/suzaku/doc/xrt/suzakumemo-2008-06.pdf}{ftp://legacy.gsfc.nasa.gov/suzaku/doc/xrt/suzakumemo-2008-06.pdf}}
These values do not take into account background-subtraction
systematics, sensitivity to spectral shape, and other factors that
could affect the actual ratio. Allowing \cpx\ to be a free parameter
does not optimally address this issue, because that could skew the
best-fitting model parameters at the expense of obtaining a
\cpx\ ``best-fit'' value, which in actuality is unrelated to the true
normalization ratio of the instruments. We thus apply the method
presented in \citet{tzanavaris2019b} and carry out preliminary
investigations for each dataset before deciding whether we could fix
\cpx.  In essence, we obtain a preliminary best fit with \cpx\ free,
and then explore \nhs$-$\cpx\ parameter space by means of
two-dimensional contours. If the contours overlap with the region
corresponding to the recommended PIN:XIS ratio, we fix \cpx\ to the
recommended value, and leave it free otherwise. For further details on
this method see \citet[][Section~3.8 and Figure~2]{tzanavaris2019b}.

\renewcommand{\baselinestretch}{0.8} 
\begin{deluxetable*}{ccccccccc} 
\tablecaption{\footnotesize \vspace{3pt} Continuum Fluxes and Luminosities\vspace{-5pt} \label{tab-cont}}
\tabletypesize{\footnotesize}
\tablehead{
  [-0pt]
\colhead{\#}                             
&\colhead{Name}                          
&\colhead{ObsID}                         
&\colhead{\fcobs}                        
&\colhead{\lcra}                         
&\colhead{\lcru}                         
&\colhead{\fcobsH}                       
&\colhead{\lcraH}                        
&\colhead{\lcruH}                        
\\[-.15cm]
\colhead{}                               
&\colhead{}                              
&\colhead{}                              
&\colhead{($10^{-11}$\funits)}              
&\colhead{($10^{42}$\lunits)}               
&\colhead{($10^{42}$\lunits)}               
&\colhead{($10^{-11}$\funits)}              
&\colhead{($10^{42}$\lunits)}               
&\colhead{($10^{42}$\lunits)}               
\\[-.6cm]}
\colnumbers
\startdata
\vspace{-15pt}\\
  1  &         Cen A  &   100005010  &       21.68  	&                1.61  	&      3.06  	&                    14.71  	&                     2.84  	&                      3.00 \\ 
  2  &         Cen A  &   708036010  &       36.04  	&                2.67  	&      4.94  	&                    23.30  	&                     5.22  	&                      4.76 \\ 
  3  &         Cen A  &   708036020  &       18.83  	&                1.40  	&      2.37  	&                    12.11  	&                     2.68  	&                      2.12 \\ 
  4  &         Cen A  &   704018010  &       33.22  	&                2.46  	&      4.66  	&                    22.22  	&                     4.54  	&                      4.59 \\ 
  5  &         Cen A  &   704018020  &       29.38  	&                2.18  	&      4.17  	&                    19.62  	&                     4.48  	&                      3.96 \\ 
  6  &         Cen A  &   704018030  &       32.85  	&                2.43  	&      4.68  	&                    21.98  	&                     4.91  	&                      4.50 \\ 
  7  &   MCG+8-11-11  &   702112010  &        6.53  	&               62.05  	&     63.24  	&                     2.82  	&                    72.44  	&                     61.20 \\ 
  8  &       NGC526A  &   705044010  &        4.46  	&               36.59  	&     32.28  	&                     2.02  	&                    51.64  	&                     29.46 \\ 
  9  &       NGC2110  &   100024010  &       10.93  	&               14.68  	&     19.97  	&                     5.96  	&                    22.12  	&                     21.23 \\ 
 10  &       NGC2110  &   707034010  &       12.78  	&               17.18  	&     24.31  	&                     7.04  	&                    28.91  	&                     25.20 \\ 
 11  &       NGC3227  &   703022020  &        1.83  	&                0.61  	&      0.74  	&                     1.53  	&                     1.76  	&                      1.11 \\ 
 12  &       NGC3227  &   703022030  &        2.61  	&                0.87  	&      0.85  	&                     1.73  	&                     1.71  	&                      1.28 \\ 
 13  &       NGC3227  &   703022040  &        0.98  	&                0.33  	&      0.44  	&                     0.95  	&                     1.10  	&                      0.66 \\ 
 14  &       NGC3227  &   703022050  &        2.16  	&                0.72  	&      0.97  	&                     1.41  	&                     1.45  	&                      1.04 \\ 
 15  &       NGC3227  &   703022060  &        1.57  	&                0.52  	&      0.76  	&                     1.19  	&                     1.13  	&                      0.92 \\ 
 16  &       NGC4258  &   701095010  &        0.82  	&                0.04  	&      0.07  	&                     0.46  	&                     0.05  	&                      0.06 \\ 
 17  &       NGC4507  &   702048010  &        0.57  	&                1.89  	&     15.71  	&                     0.99  	&                    12.57  	&                     19.99 \\ 
 18  &       NGC5506  &   701030010  &        9.93  	&                8.44  	&     10.89  	&                     4.45  	&                    11.41  	&                      8.12 \\ 
 19  &       NGC5506  &   701030020  &       10.58  	&                8.98  	&     11.62  	&                     4.87  	&                    12.00  	&                      9.13 \\ 
 20  &       NGC5506  &   701030030  &        9.86  	&                8.36  	&     11.18  	&                     4.37  	&                    11.54  	&                      8.13 \\ 

\enddata
\tablecomments{Column information: (4) 2-10 keV continuum flux, observed frame; (5) 2--10 keV continuum absorbed luminosity, AGN frame; (6) 2--10 keV continuum unabsorbed luminosity, AGN frame; (7) 10--30 keV continuum flux, observed frame; (8) 10--30 keV continuum absorbed luminosity, AGN frame; (9) 10--30 keV continuum unabsorbed luminosity, AGN frame.
}
\renewcommand{\baselinestretch}{1}
\vspace{-0.6cm}
\end{deluxetable*}

\section{Results and Discussion}\label{sec-res}
Spectral fitting results with the decoupled \myt\ model for 20
observations of the 8
\cthin\ AGNs are presented in \tr{tab-fit}. Fitting results and
deduced properties for the \feka\ emission line are shown separately
in \tr{tab-line}, and continuum fluxes and luminosities in
\tr{tab-cont}. Energy shifts and continuum fluxes are in the observed
frame in the interest of comparisons with literature
values. Luminosities and other parameters are in the AGN frame.
Details on individual objects and obsIDs are discussed in
Appendix~\ref{app-indi}, and detailed spectral plots for all obsIDs
are shown in Appendix~\ref{app-plots}.  Here, left-hand plots (labeled
(a)) show the total model fit (continuous red line) over the full
extent of the spectral data used, as well as data/model
ratios. Right-hand plots labeled (b) show total model (continuous
black line) and model components (as indicated in each plot legend),
while those labeled (c) show the total model and data in the region of
the \feka\ line.

\tr{tab-fit} presents equivalent hydrogen column densities in, and out
of, the line of sight (\nhz\ and \nhs) in Columns (7) and (8), respectively.
With the
exception of one observation, all values are constrained for both
quantities, showing a significant range of \sss2 orders of magnitude
for each.  While all \nhz\ values are in the \cthin\ regime, this is
not the case for \nhs. Instead, column densities out of the line of
sight for the single obsID of \meee, the single obsID of \gc526A, and
all three obsIDs of \gc5506 are \cthick.
For a further two obsIDs of
\gc3227 (703022040, 703022060) column densities
out of the line of sight are also
borderline \cthick\ at $\approxgt$\ten{0.99}{24}\cunits.  {\it Since all
objects were initially selected as likely \cthin, these results
highlight the model-dependence for Compton-thickness
classifications.} There is in fact a second level of complication in
such classifications that is related to individual observations of the
same object, as in the case of \gc3227. The \nhs\ column density range
is \sss\ten{0.1}{24} to $>$10\up{24} \cunits, and two out of five obsIDs are on the
threshold of being \cthick\ when errors are taken into account.
Thus models of the CXB that rely on the
fraction of \cthick\ AGNs in the universe
would benefit from being explicit about the precise meaning of the
column density associated with them. In particular, if only
line-of-sight column density is modeled, and Compton-thickness is
based only on that column density, there is an implied assumption that
the obscuring material in the entire population of AGN is spherically
symmetric.

Previous work has reported relativistically broadened emission for
some of the objects and/or obsIDs in this paper:

For \meee\
with \xmm\ data, \citet{matt2006} only find evidence for a narrow line
but \citet{nandra2007} report a broad line in the same data.
Using the same \suzaku\ obsID as in this paper,
\citet{bianchi2010} report a broad line specifically when the PIN
data are included; \citet{patrick2012} also report a broad
line. \citet{tortosa2018} report residuals in \nustar\ data first
fitted with a narrow line, that they subsequently model with an
additional broad line.

For \gc526A, \citet{landi2001} report a broad
line with {\it BeppoSAX} data and \citet{nandra2007} with \xmm.

For \gc2110, \citet{nandra2007} report a broad line with \xmm.

For \gc3227, broad lines are reported by \citet{patrick2012} and
\citet{noda2014} for the \suzaku\ obsIDs in this paper (in addition to
narrow ones). In contrast, \citet{markowitz2009b} with \xmm\ only report
the possibility that a feature in the \feka\ region
could be evidence for relativistic broadening in addition to
narrow \feka\ emission.

Finally, in the case of \gc5506 broad lines are
either reported or
absent in the last two decades right up to the present.
The analysis of \citet{bianchi2003} found no evidence
for broad lines in \xmm\ and \chandra\ HETG data.
Broad lines have been reported instead from \xmm\ data
by \citet{nandra2007} and
\citet{guainazzi2010}.
\citet{sun2018} analyze data from several
different missions, including the three obsIDs in this paper, and
report broad line results in all cases.

Recently, \citet{zoghbi2020} model \xmm\ data both with narrow-line and
with relativistic models, but caution on the difficulty of establishing
a broad line from spectroscopy alone.
\citet{laha2020} analyzed all
obsIDs and AGNs in our sample, except for \meee\ and \gc3227, and do
not report any broad lines from these \suzaku\ data via
non\myt\ modeling. They also do not report any broad lines after
analyzing further data from other missions for these and other
\cthin\ objects.  Our modeling is in almost all cases
entirely distinct from previous work and suggests that narrow emission
is sufficient for modeling these data,
although this does not conclusively \textit{prove}
  the absence of relativistic components.
There are only two cases where
\myt\ has been used before for these targets. In both
cases, results corroborate a narrow line.
In the case of \meee,
\citet{murphy2014}
fit the same \suzaku\ data with a narrow line, although they
use the coupled \myt\ model and thus do not report separate column
densities in and out of the line of sight. The
second case is \gc4507, for which the decoupled
narrow-line \myt\ modeling results of \citet{braito2013} are in
excellent agreement with our results.

\tr{tab-line} presents fitting results for the \feka\ emission line.
The line is mostly not resolved by \suzaku, whose XIS spectral resolution is
\sss6000~\kmps\ (\sss120~eV FWHM at 6
keV\footnote{\href{https://heasarc.gsfc.nasa.gov/docs/suzaku/about/overview.html}{https://heasarc.gsfc.nasa.gov/docs/suzaku/about/overview.html}},
also \citealt{mitsuda2007}). As during the fitting process in all but two
obsIDS the lower limit of \sigl\ goes to zero, we fix \sigl\ to
an arbitrary narrow value, either 100 \kmps\ or values reported by
previous work with \chandra-HETG data, as shown in Columns (5) and (6)
of \tr{tab-line}. It can be seen in plots (c),
Appendix~\ref{app-plots}, that these values fit the \feka\ line well,
and this is also particularly illustrated by the residuals between model
and data in the feka region (Column (7) in \tr{tab-line} and lower panel in
plots (c)).
It is possible that adding relativistic components might
  further improve the goodness-of-fit. However, this is beyond the scope of
  the present analysis, and we choose not to explore it further. Overall,
  our results show that
the physically motivated \myt\ model, where \fek\
emission and scattered continuum from matter with finite column density
and solar Fe abundance are produced self-consistently
in tandem, can model these data well
with no need for a broad
 / relativistic
component.

Clumpy-torus models
\citep[e.g.][]{elitzur2006,risaliti2007,nenkova2008,honig2013} propose
a more realistic picture, where \x\ reprocessing likely occurs in
nonhomogeneous, rather than uniform-density, structures. Variability
in line-of-sight \nh\ has also been observed on timescales of months
to years, with variations up to factors of \sss3 \citep{risaliti2002},
corroborating the clumpy-torus picture
\citep{markowitz2014,laha2020}. In the case of \gc3227, hardening of
the primary continuum has been interpreted as occultation from
broad-line region clouds \citep{turner2018}.
This is a particularly complex object situated in an overdense
  and disturbed group environment \citep{garcia1993,crook2007},
  strongly interacting with nearby systems
  \citep{mundell1995,mundell2004,davies2014}, and showing neutral
  hydrgoen streaming motions out to scales of several parsecs, and
  molecular inflows / outflows out to several hundreds of parsecs
  \citep{davies2014,alonso-herrero2019}. \x\ warm absorbers
  \citep{beuchert2015} are the signature of an AGN wind. These
  properties, coupled with \x\ variability and state changes, would
  lead one to expect global variations in column density.
Thus, overall, as observed for
the data in this paper, different column densities between the line of
sight and out of the line of sight, as well as differences between
observations of the same object, should not come as a surprise.

\section{Summary and Conclusions}\label{sec-summ}
We select eight nearby AGNs which, based on previous work and
preliminary data inspection, appeared to be \cthin.
For these objects, we fit twenty individual \suzaku\ broadband \x\
spectra with \myt, self-consistently modeling
the \feka\ line emission and associated reflected
continuum from material of finite column density with
solar Fe abundance, carefully assessing the optimal
value of the PIN~vs.~XIS cross-normalization.
Our main results are:
\vspace{-4pt}
\begin{enumerate}
  \itemsep-4pt
\item All data can be fitted well with narrow-only
  \feka\ line emission and associated reflected continuum.
\item The \feka\ line is not resolved in any but two
  \suzaku\ obsIDs. The line is consistent
  with being narrow as found in a number of previous analyses with
  \chandra-HETG, \xmm, or the same \suzaku\ data.
Some previous analyses alternatively reported evidence
    for broad / relativistic components. Exploring this is beyond
    the scope of this paper.
\item Fits model an \x\
  reprocessor with finite column density far from the central
  SMBH that does not require nonsolar Fe abundance.
\item We measure global column densities
  associated with Compton scattering out of the line of sight (\nhs)
  separately from line-of-sight column densities (\nhz).
  Along the line of sight all \nhz\ column densities are \cthin.
  In contrast, for four AGNs and seven obsIDs, global \nhs\
  column densities are consistent with being \cthick.
\end{enumerate}

These results show that Compton-thinness is necessarily an ambiguous
concept without additional qualifiers concerning the geometry of the
global matter distribution and its orientation with respect to the
observer.

\acknowledgments

\noindent
We thank the anonymous referee for their constructive
comments that helped improve this paper.
P.T. acknowledges support from
NASA grant 80NSSC18K0408 (solicitation NNH17ZDA001N-ADAP).
This research has made use of data obtained from the \suzaku\ satellite, a
collaborative mission between the space agencies of Japan (JAXA) and
the USA (NASA).
This work is supported by NASA under the CRESST Cooperative Agreement, award number 80GSFC21M0002.

\noindent\facilities{\suzaku}

\appendix
\numberwithin{figure}{section}

\section{\\Notes on Individual Objects}\label{app-indi}
We present more detailed information on fitting and results for all
objects and obsIDs in our sample.
We include comparisons with results from other work and
mostly different modeling, but only for results on narrow lines, which
are the focus of this paper.

We also present simple
estimates on the distance, $r$, from the nucleus of material where
narrow \feka\ emission may be produced based on the reported
line FWHM.  Assuming virialized material, $v$ is related to
FWHM velocity by $\langle v^2\rangle=\frac{3}{4} \ {\rm FWHM_{Fe K\alpha}^2}$ \citep{netzer1990}, so that $r=\frac{4c^2}{3
  \ \rm FWHM_{Fe K\alpha}^2} \ r_g$, where the gravitational radius is
$r_g\equiv GM_{\bullet} /c^2$.

\subsection{\cena}
As can be seen in \tr{tab-fit}, for all obsIDs both \nhs\ and
\nhz\ are in the \cthin\ regime.  All power law slopes $\Gamma$ are
between 1.7--1.8, and all $A_S$ are \sss1. In the case of obsID
708036010 $A_S$ could not be independently constrained and was fixed
to 1. For all obsIDs, we fix $\sigma=20$~eV, consistent with the
\chandra\ HETG results of \citet{evans2004}, as the lower
\sigl\ limit vanishes.
Fit residuals in the \feka\ region are between
\sss4--12\%. Although we fit a model that has not been applied
to these data before,
overall the $\Gamma$ values and column densities agree with those
reported earlier from \chandra\ and \xmm\ work
\citep{evans2004}. \citet{fukazawa2011a} also report \feka\ results
for obsID 100005010, consistent with a
narrow line.  This obsID is also analyzed by \citet{markowitz2007}. In
spite of the very different modeling, $\Gamma$, column density, and
\feka\ values essentially agree with our results.
The line-of-sight \nh\ values of \citet{laha2020} are also
in good agreement with our \nhz\ values. For obsID 100005010,
these authors also obtain a separate column density for a partially
covering absorber, which agrees well with our \nhs\
result for that observation.

The contribution of a leaking scattered power law has
a relative normalization of \sss1\%\ for all obsIDs.

For our quoted
line width, the line producing region's radius is estimated at
\ten{2}{4} $r_g$.

\subsection{\meee}
\tr{tab-fit} shows that while \nhs\ is \cthick, \nhz\ is \cthin, with
a change in column density of \sss3 orders of magnitude between the
two. In fact, the line-of-sight column density converges to a value
lower than the lower limit of the corresponding model table, so that
we replace it with \zphabs.  We fix $\sigma=25$~eV, consistent with
the \chandra\ HETG result of \citet{murphy2014}, as required by the
vanishing \sigl\ limit. We model \fetwosix\ emission at \sss6.97 keV 
with an additional Gaussian \citep{bianchi2010,patrick2012}.
\citet{fukazawa2011a} also report \feka\ results for the same
obsID. Their single column density is \sss\ten{0.6}{21} \cunits, of
the order of our line-of-sight result.  They also report a narrow
\feka\ line, with \sigl\ somewhat larger than the \chandra-HETG
result.  \citet{murphy2014} report a single column density from
nondecoupled \myt\ fitting of the same data together with the
\chandra-HETG data. In their model, that column density is out of the
line of sight, and their result is consistent with our \nhs\ result.

Otherwise, $A_S\sim1$ and $\Gamma\sim1.7$. 
For our quoted line width,
the line-producing region has a radius of \sss\ten{1.4}{4} $r_g$.

\subsection{\gc526A}
In \tr{tab-fit} \nhz\ is \cthin\ but \nhs\ clearly \cthick. To
constrain \nhs, we need to fix $A_S=1$.  We also fix \sigl\ to 100
\kmps\ as its lower limit tends to zero.  The \citet{laha2020}
\nh\ reported for this obsID is in reasonable agreement with our
\nhz\ \cthin\ value. $\Gamma\sim1.8$ for this obsID and a leaking
power-law scattered continuum contributes at a relative normalization
of \sss30\%.

\subsection{\gc2110}
For both obsIDs of this AGN \tr{tab-fit} shows that both \nhz\ and
\nhs\ are \cthin. In both cases the lower \sigl\ limit tends to zero
and is fixed to the value quoted by \citet{evans2007} for
\chandra-HETG data, i.e., 20~eV.  The \citet{laha2020} \nh\ values
reported for these obsIDs are in general agreement with our
\nhz\ \cthin\ values. In particular, their separate partial-coverage
\nh\ results are in very good agreement with our \nhz\ results for
each obsID. The \nh\ value for obsID 100024010 reported by
\citet{fukazawa2011a} is of the same order, although somewhat larger,
than our \nhz\ value.

\subsection{\gc3227}
All obsIDs of this AGN have \nhz\ values that are in the \cthin\ regime.
\nhs\ values for observations 703022040 and 703022060 are
borderline \cthick\ within their upper limit
90\%\ uncertainties. Although $10^{24}$ \cunits\ is commonly used as
the threshold for \cthick ness, strictly this is \ten{1.24}{24}
\cunits, and even this is not without ambiguities\footnote{See the
  \myt\ manual
  (\href{http://mytorus.com/mytorus-manual-v0p0.pdf}{http://mytorus.com/mytorus-manual-v0p0.pdf}),
  Section 2.1, for a detailed discussion.}.  The reported \nh\ values
in \citet{fukazawa2011a} for the same obsIDs are of the same order as
our results within factors of a few. Our
results show a varying column density both for \nhs\ and \nhz\ between
observations,
within factors of a few, although, when taking errors into account,
only one observation might be discrepant.
This is
consistent with the values reported by \citet{fukazawa2011a} but not
\citet{patrick2012}, since the latter tie column densities between
observations.
There is considerable evidence across wavebands showing that
  \gc3227 is located in an environment that can affect the large scale
  distribution of neutral gas.  It is in an overdense group
  environment \citep{garcia1993,crook2007} and is strongly interacting
  both with \gc3226 and a nearby gas-rich dwarf / \hone\ cloud
  \citep{mundell1995,mundell2004,davies2014}, with evidence of
  large-scale neutral hydrogen streaming motions over several to
  hundreds of parsecs.  It is reported to have a large-scale bar
  \citep{mulchaey1997,alonso-herrero2019}.  \citet{alonso-herrero2019}
  report a nuclear molecular outflow extending out to scales of
  several parsecs, while \citet{davies2014} detect inflows and ouflows
  over scales of hundreds of parsecs. In the X-rays, the variable
  ionization states and column densities reported in
  \citet{beuchert2015} for the warm absorbers in the clumpy wind
  outflow consider only the line of sight.  There is no a priori
  reason why within the context of a three-dimensional wind, coupled
  with X-ray continuum variability, the global column density would not be affected. One might expect that state changes would have a global effect, affecting
column density in all directions.

\tr{tab-fit} shows that for obsID 703022060
$A_S$ has to be fixed to 1, in order to constrain \nhs.
For other obsIDs $A_S$ is on the high side, possibly indicating
time delays between direct and scattered continua. For
obsIDs 703022020, 703022030, and 703022040 the power-law continuum
slope is unstable but tends to hard values and we fix it to the lowest
value of $\Gamma=1.4$ allowed by the model. This has also been noted
by \citet{noda2014} for the same observations, and is interpreted as a
possible ``low (luminosity) / hard'' state, perhaps associated with a
Radiativally Inefficient Accretion Flow (RIAF,
\citealt{narayan1994,esin1997}). \xmm\ data also indicate the presence
of a hard power law \citep{markowitz2009b}. Similar ``softer when
brighter'' behavior is reported from the \xmm+\nustar\ analysis of
\citet{lobban2020}, where low-flux spectra have
$\Gamma\sim1.4$. \citet{turner2018} also interpret spectral hardening
in those data as due to an occultation event due to broad line region
clouds.  \citet{lobban2020} also obtain modest \feka\ variability
relative to significant continuum variability, suggesting that
\feka\ emission originates far from the central \x\ source.
For three obsIDs, the lower limit of \sigl\ tends to zero, and we fix it
to 100 \kmps. This is not the case for the remaining two obsIDs but
the values derived are still quite narrow.
\citet{fukazawa2011a} and \citet{noda2014} also report only narrow
\feka\ emission for the same obsIDs, (all $\approxlt$10000 \kmps).
\citet{markowitz2009b} report similarly narrow \feka\ emission with
\xmm\ with FWHM \sss7000 \kmps. Our quoted values correspond
to an emitting region at
\simgt2300~$r_g$.

For obsIDs 703022020 and 703022030, as in \citet{fukazawa2011a} for
the same data, we add absorption due to \fetwofive~K$\alpha$ at
\sss6.7 keV and Ni K$\alpha$ emission at \sss7.4 keV. For obsID
703022040, gaussians are used to model emission from \fetwofive\ at
$\sim 6.68$ keV, \fetwosix\ at \sss6.97 keV, and Ni K$\alpha$ at
\sss7.47 keV.

For the three obsIDs 703022020, 703022030, and 703022040, a second
soft power law (\gammasoft~$\sim 3-5$), dominant
in the softer spectral region, is needed to minimize
residuals. This is similar to the behavior observed by
\citet{lamer2003} and \citet{markowitz2009b} with \xmm.
However, note that, in terms of emitted power,
  Figures~\ref{fig-n3227-20} -- \ref{fig-n3227-60} show that
  the direct power law (MYTorusZ) is dominant above \sss4~keV.
  
\subsection{\gc4258}
According to \tr{tab-fit}, both \nhs\ and \nhz\ are \cthin\ for this
object. $A_S$ needs to be fixed to 1 for \nhs\ to be constrained.  As
before, the lower \sigl\ limit tends to zero, so we fix it to 100
\kmps. The \nh\ reported by \citet{laha2020} for the same obsID is in
very good agreement with our \cthin\ values.

\subsection{\gc4507}
Both column densities are \cthin.  However, \nhz\ is \sss3 times
larger than \nhs. This bears a certain similarity to the \nh\ reported
by \citet{laha2020}, as their partial-coverage column density (which,
within the errors, is consistent with our \nhz\ value) is \sss3--5
times larger then the full coverage one.  The \citet{fukazawa2011a}
\nh\ value is also of the same order as our \cthin\ \nh\ values.
\sigl\ is fixed at 100 \kmps. \citet{fukazawa2011a} also report a
narrow line with \sigl\sss40 eV.  \citet{braito2013} also analyze this
obsID. Their results with the decoupled \myt\ models for \nhs\ and
\nhz\ are in excellent agreement with ours. They also report
\sigl\sss30 eV.  We add three \zgauss\ components to account for
emission due to Ca~K$\alpha$ at \sss3.7 keV, \fetwofive\ at \sss6.7
keV, and Ni~K$\alpha$ at \sss7.4 keV, consistent with
\citet{braito2013}.

\subsection{\gc5506}
All three observations are very \cthin\ in the line of sight, but
clearly \cthick\ out of the line of sight.  In obsID 701030010
\nhs\ reaches the upper limit of the model and is fixed at $10^{25}$
\cunits.  In the other two obsIDs, $A_S$ cannot be well constrained and is
set to 1.0.  We also add two \zgauss\ components to account for
\fetwofive\ and \fetwosix\ K$\alpha$ at \sss6.7 and \sss6.96 keV,
respectively, consistent with \citet{bianchi2003}.  As \sigl\ tends to
zero at its lower limit, we fix \sigl~$=4000$ \kmps\ based on the
reported upper limit for the line FWHM in \chandra-HETG spectra
\citep{bianchi2003}.  The \nhz\ values of all three obsIDs are in
excellent agreement with the \nh\ values reported for the same data by
\citet{laha2020} and \citet{fukazawa2011a}.  The \sigl\ values
reported by the latter authors are up to \sss13000 \kmps\ for the
line in these obsIDs.

\section{\\Comprehensive Spectral Plots}\label{app-plots} 

For each obsID, we present here three types of spectral
plots:
\vspace{-20pt}
\begin{enumerate}[label=(\alph*),itemsep=-4pt]
\item Total model (red continuous line) and data over full spectral
  extent fitted, with data/model ratio.
\item Total model (solid black line) and model components over
  fitted region. These include:
  \vspace{-6pt}
  \begin{itemize}[itemsep=-1pt,leftmargin=*]
    \item The direct, line-of-sight, or zeroth-order continuum
  (\mytz) diminished via absorption and removal of photons from the
      line of sight via Compton scattering (red dashed line).
      For the single obsID of \meee,
    \nhz\ drops below the lower limit of the \myt\ table, so that
    there is negligible Compton scattering, allowing this component
    to be implemented by \zphabs\ instead.
  \item The secondary, scattered, or reflected continuum (\myts,
    blue dot dashed line).
  \item The \feka\ and \fekb\ emission lines (\mytl, purple dotted line).
  \item An additional scattered power law (orange triple dot dashed
    line).  This is either the scattered primary power law leaking
    through a patchy medium, or a distinct, softer power-law
    \scr{sec-mytorus}.
  \end{itemize}
\item Total model (solid red line) and data, zoomed-in over the \fek\
  fitting region. The redshifted location of 6.4 keV is shown by the
  vertical dashed line.
  \end{enumerate}

\clearpage
\begin{figure*}[t!]
    \begin{minipage}[c]{0.5\textwidth}
      \includegraphics[trim=0 50 0 -200,clip,width=1.\textwidth,angle=0]{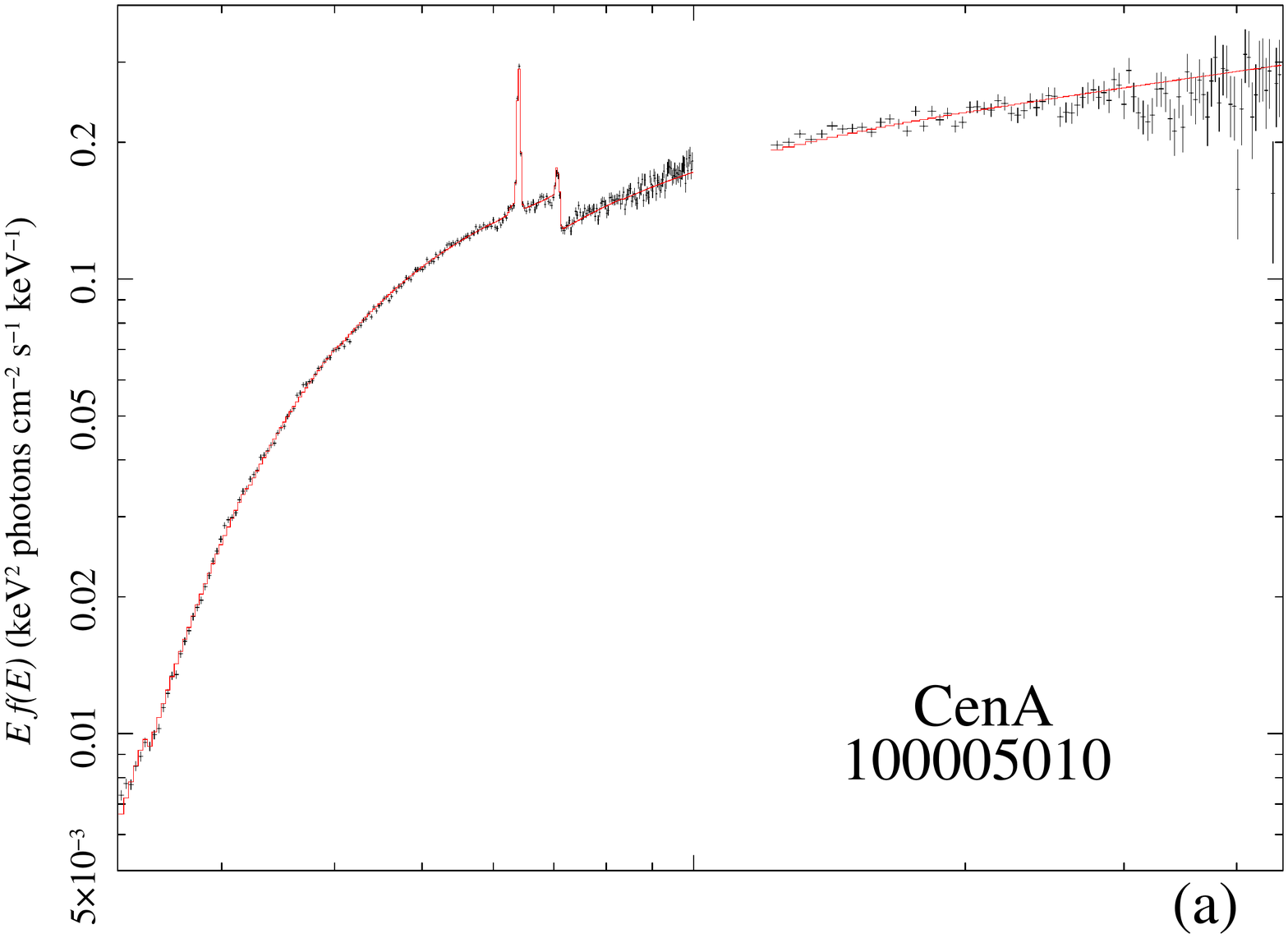}
    \end{minipage}
    \begin{minipage}[c]{0.5\textwidth}\vspace{-0pt}
      \includegraphics[trim=0 30 0 50,clip,width=1.\textwidth,angle=0]{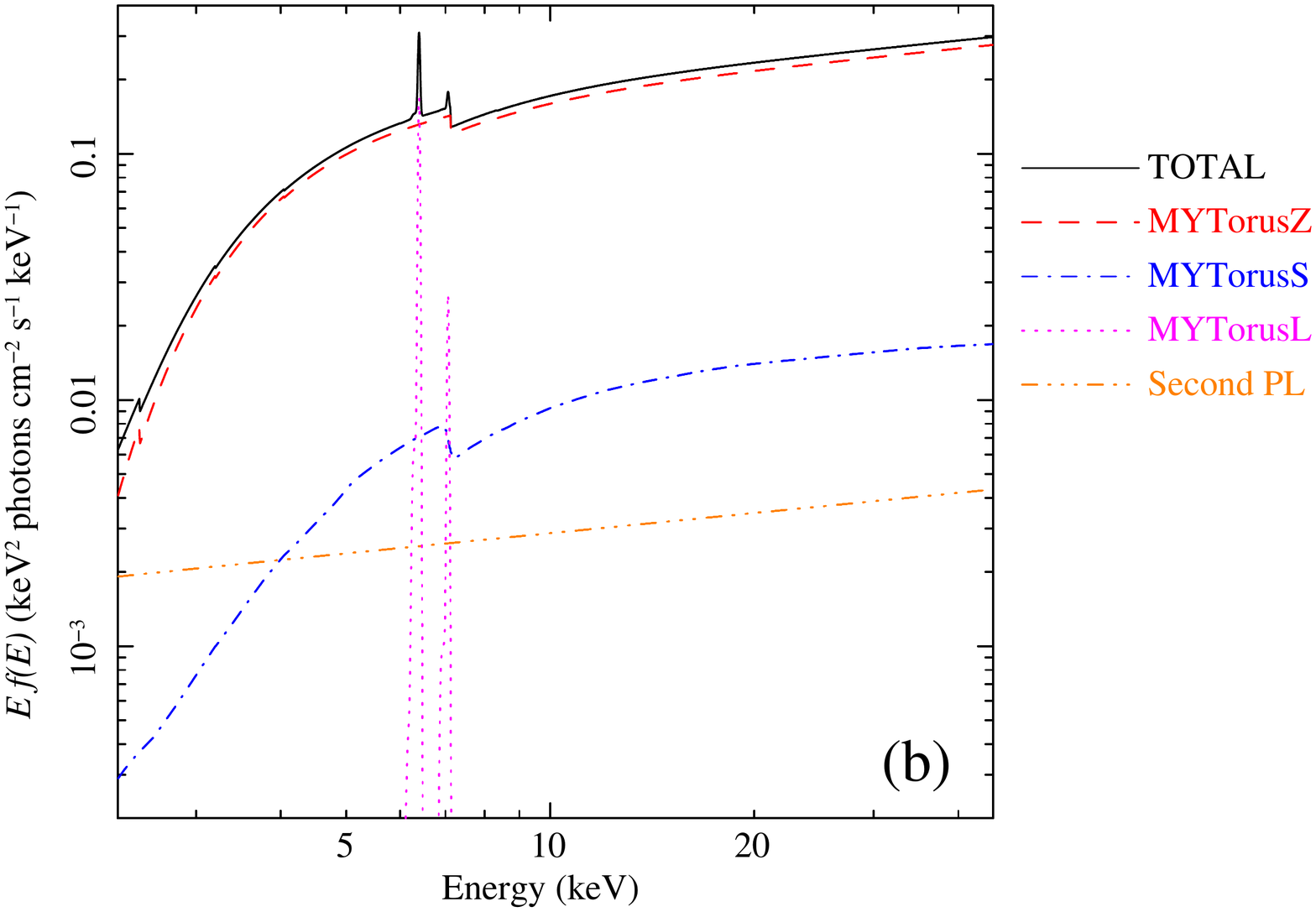}
      \vspace{-0pt}
    \end{minipage}\\
    
    \begin{minipage}[c]{0.5\textwidth}
        \includegraphics[trim=0 -300 0 360,clip,width=1.\textwidth,angle=0]{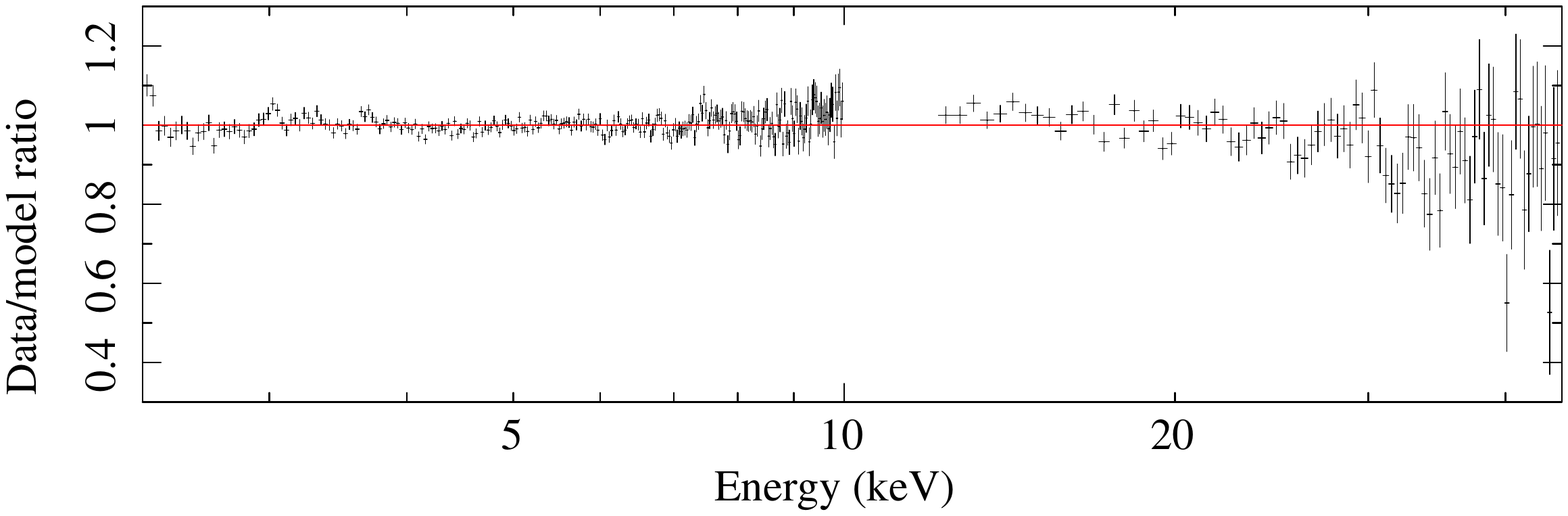}
    \end{minipage}
    \begin{minipage}[c]{0.5\textwidth}
        \includegraphics[trim=0 193 0 80,clip,width=\textwidth,angle=0]{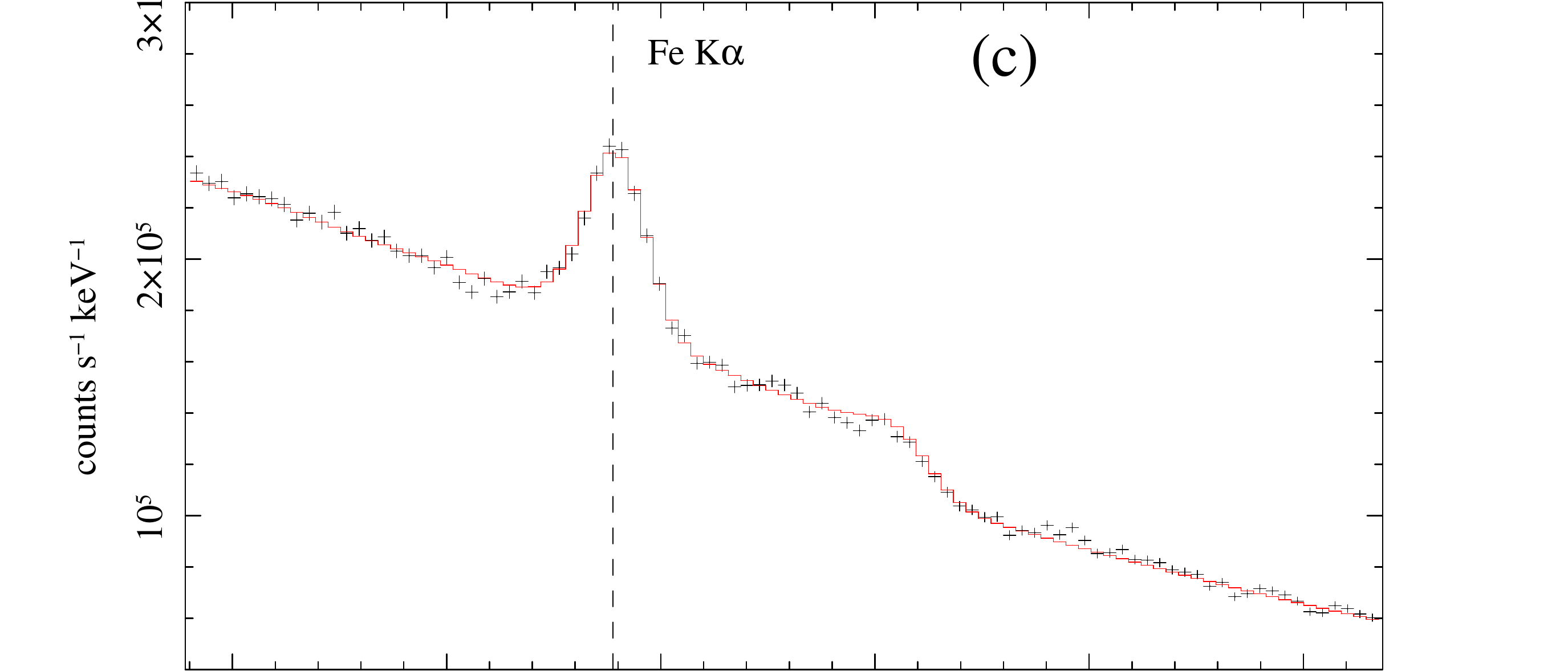}
      \begin{minipage}[c]{\textwidth}
        \includegraphics[trim=0 30 0 360,clip,width=1.\textwidth,angle=0]{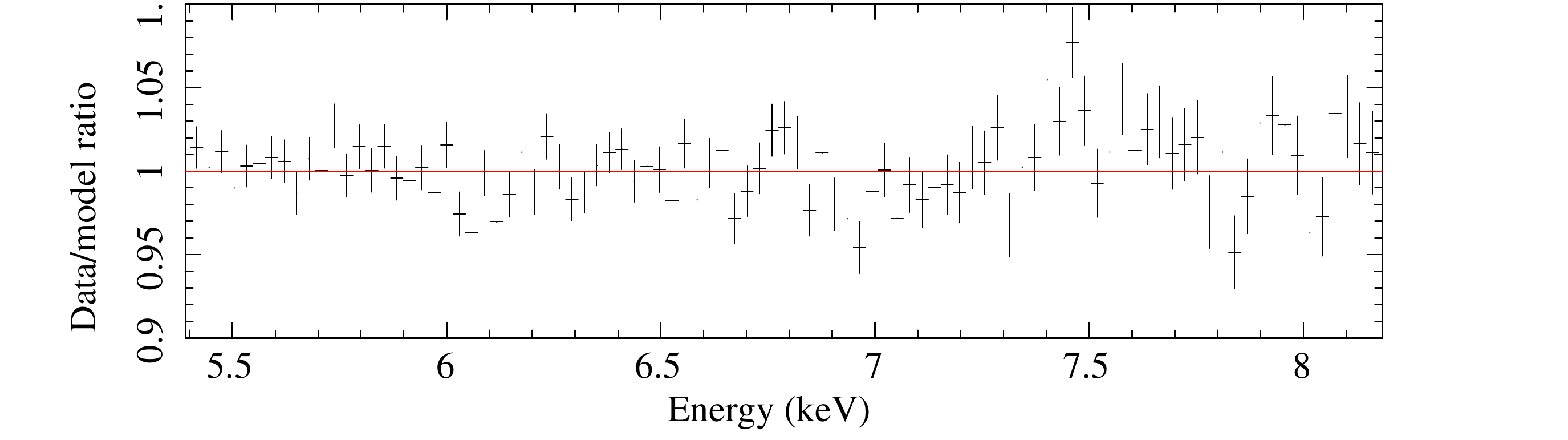}
      \end{minipage}
    \end{minipage}

    \caption{\footnotesize CenA 100005010 \label{fig-cena}}
\end{figure*}

\begin{figure*}[t!]
    \begin{minipage}[c]{0.5\textwidth}
      \includegraphics[trim=0 50 0 -200,clip,width=1.\textwidth,angle=0]{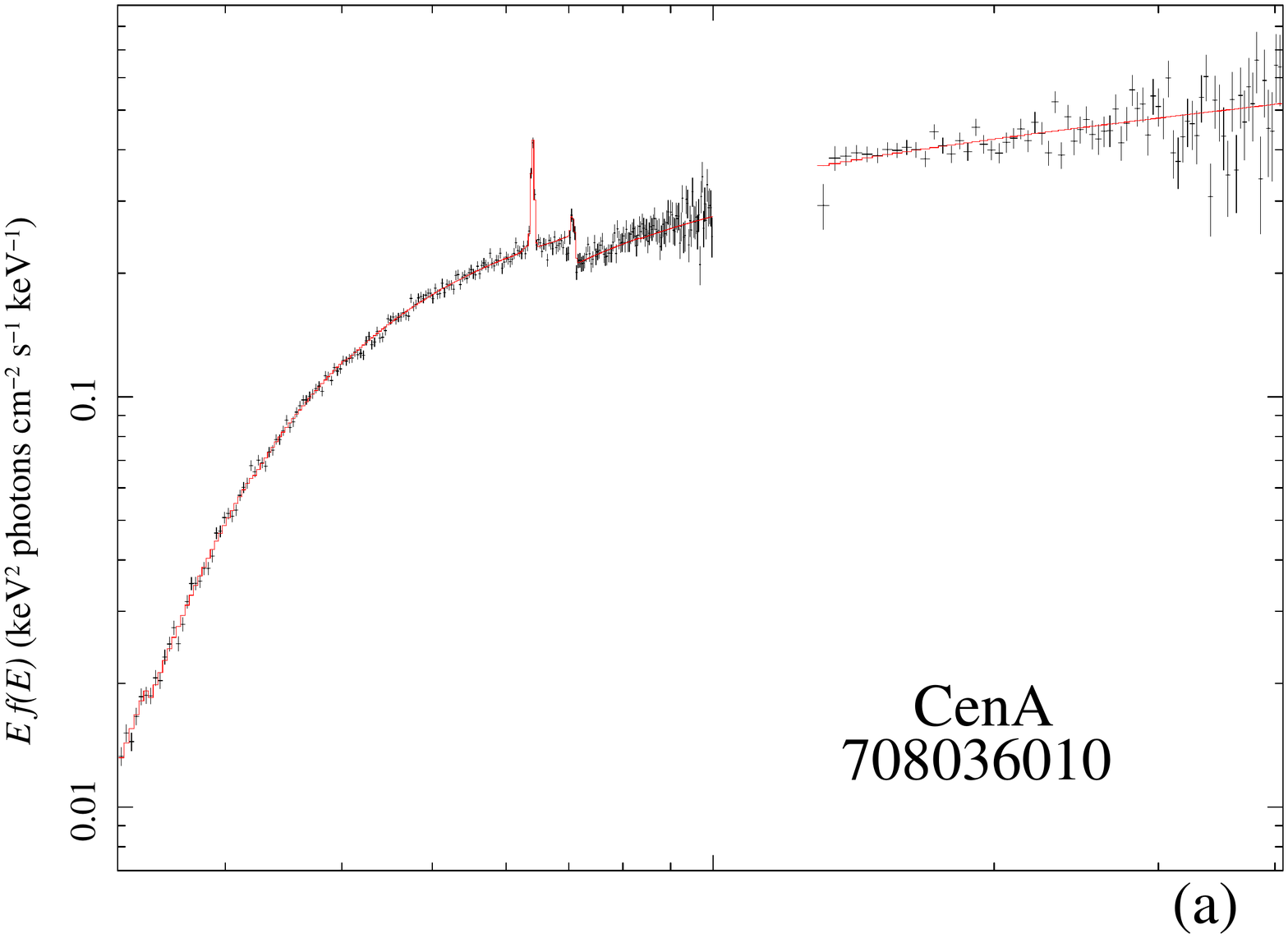}
    \end{minipage}
    \begin{minipage}[c]{0.5\textwidth}\vspace{-0pt}
      \includegraphics[trim=0 30 0 50,clip,width=1.\textwidth,angle=0]{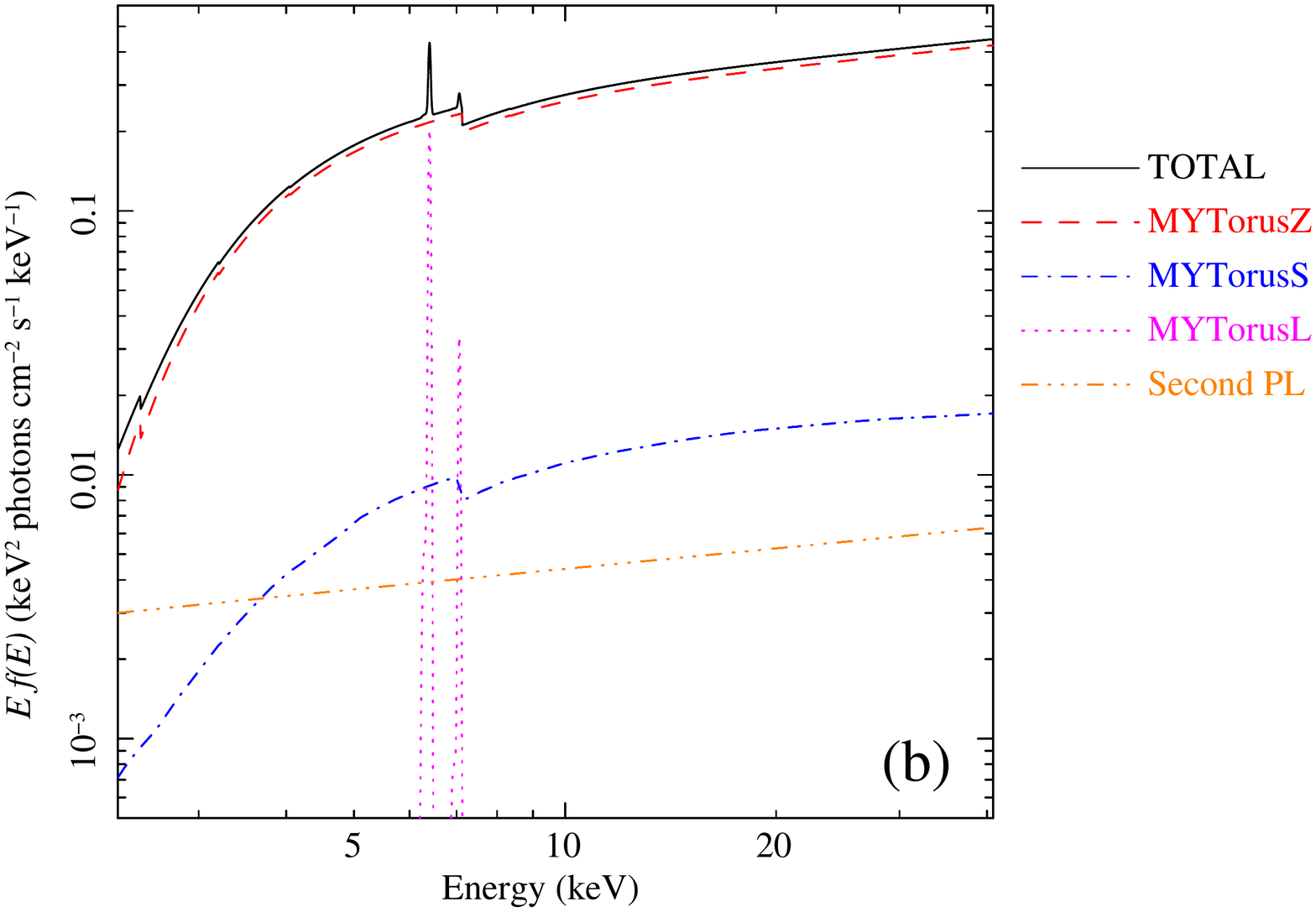}
      \vspace{-0pt}
    \end{minipage}\\
    
    \begin{minipage}[c]{0.5\textwidth}
        \includegraphics[trim=0 -300 0 360,clip,width=1.\textwidth,angle=0]{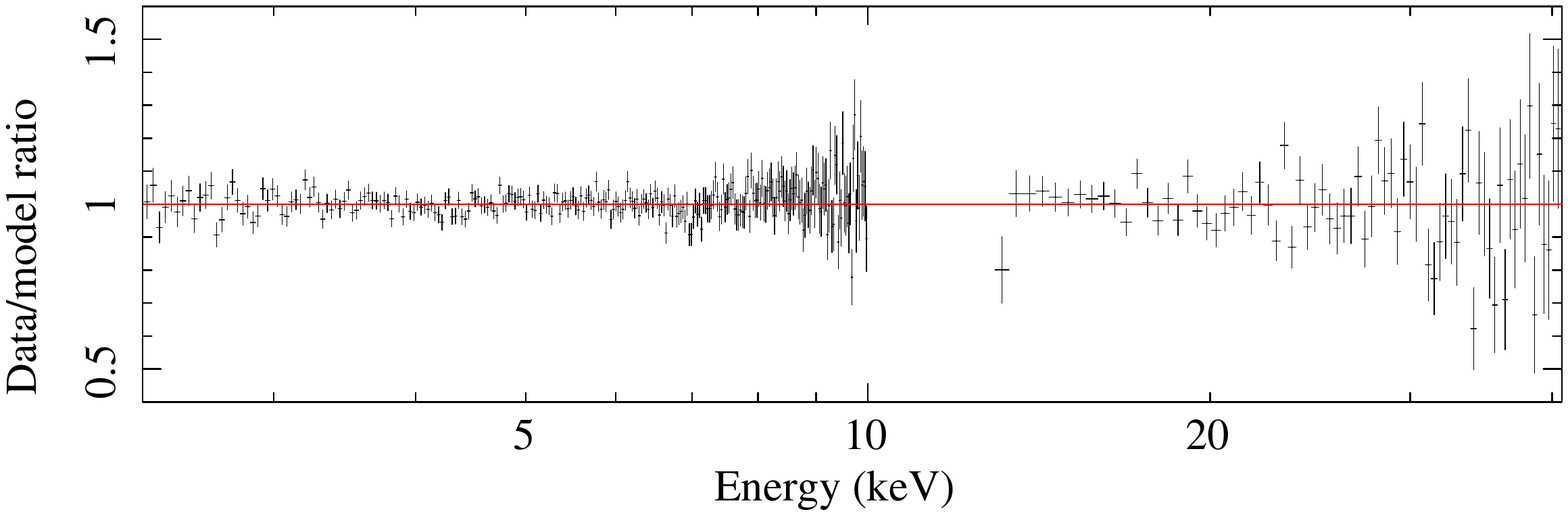}
    \end{minipage}
    \begin{minipage}[c]{0.5\textwidth}
        \includegraphics[trim=0 193 0 80,clip,width=\textwidth,angle=0]{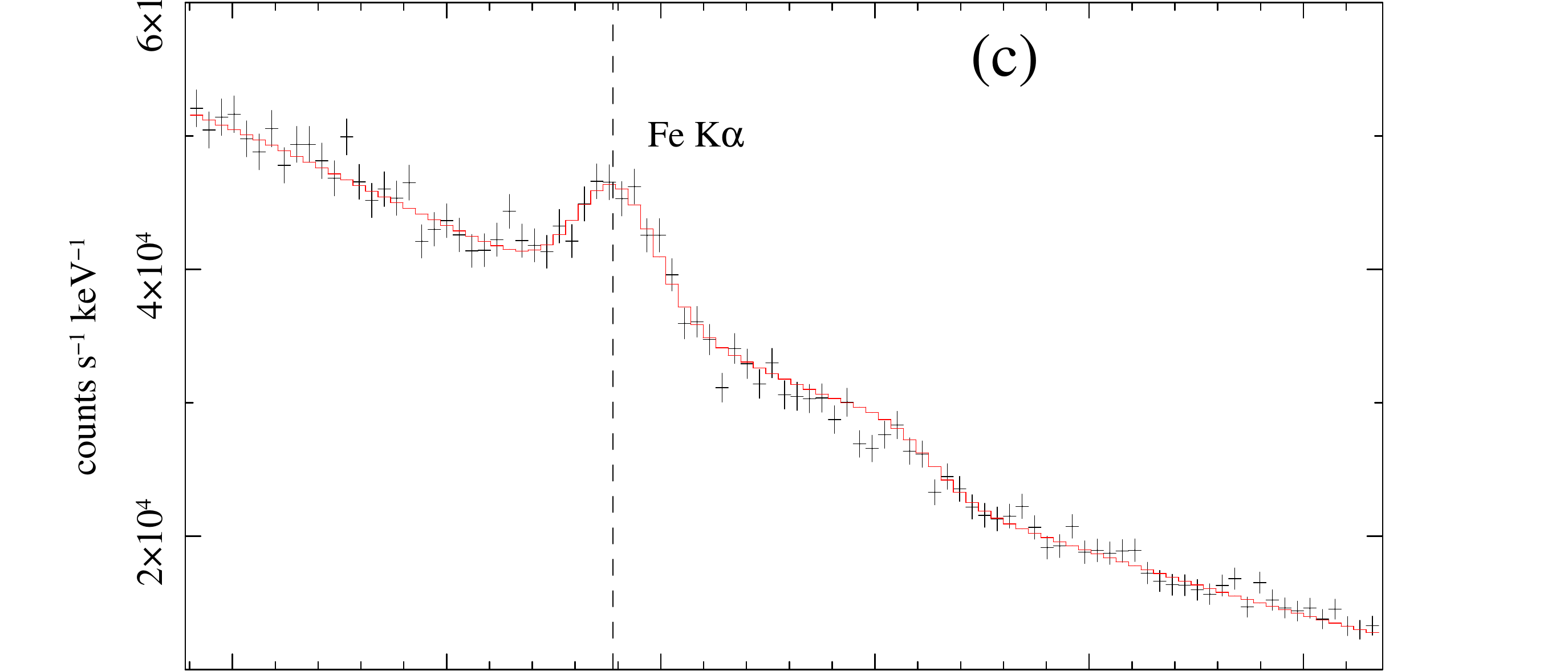}
      \begin{minipage}[c]{\textwidth}
        \includegraphics[trim=0 30 0 360,clip,width=1.\textwidth,angle=0]{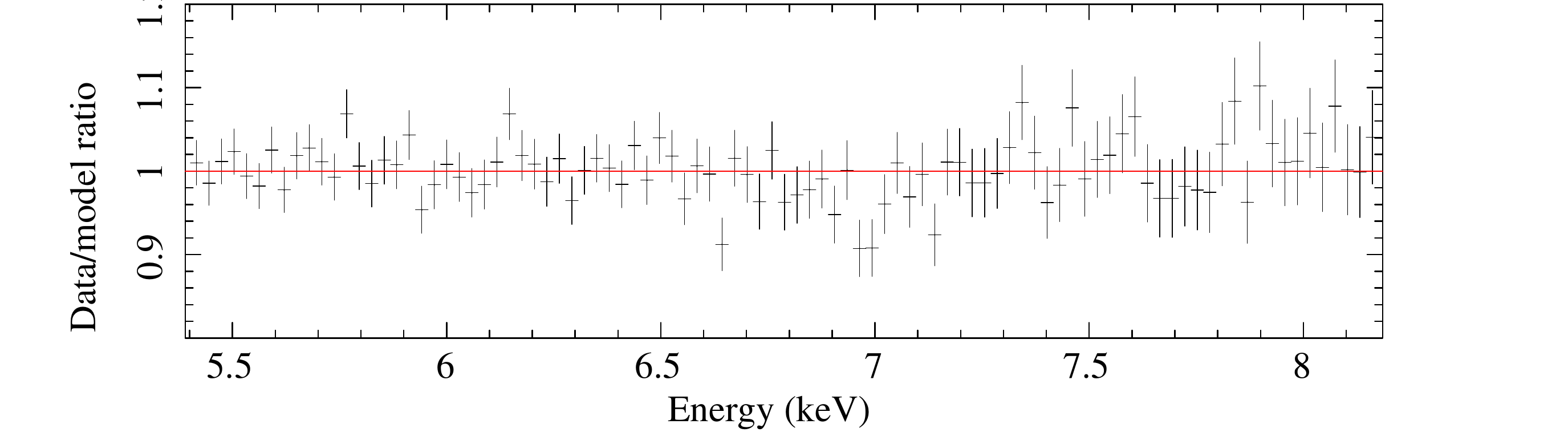}
      \end{minipage}
    \end{minipage}

    \caption{\footnotesize CenA 708036010 \label{fig-cen708-10}}
\end{figure*}

\begin{figure*}[t!]
    \begin{minipage}[c]{0.5\textwidth}
      \includegraphics[trim=0 50 0 -200,clip,width=1.\textwidth,angle=0]{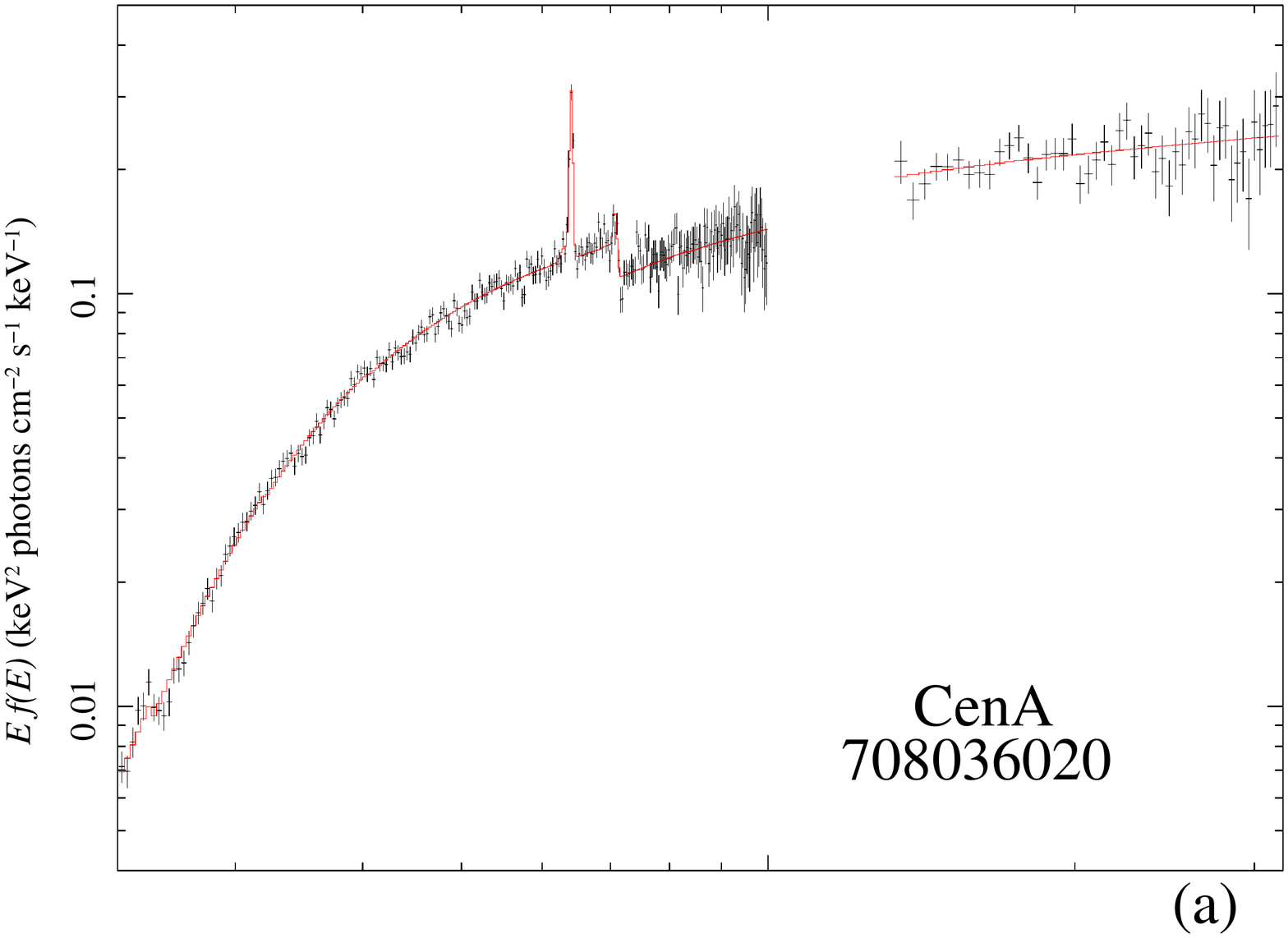}
    \end{minipage}
    \begin{minipage}[c]{0.5\textwidth}\vspace{-0pt}
      \includegraphics[trim=0 30 0 50,clip,width=1.\textwidth,angle=0]{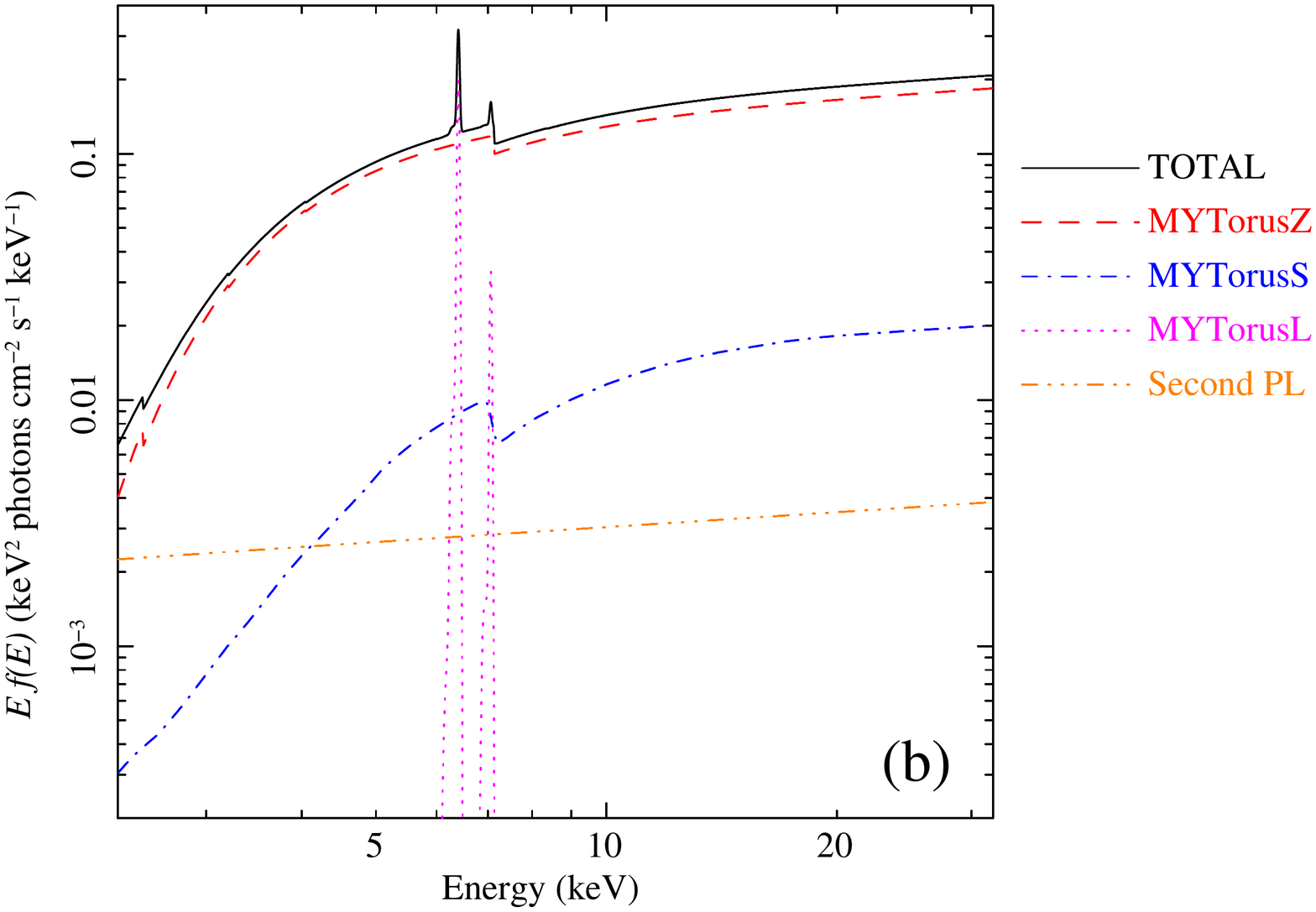}
      \vspace{-0pt}
    \end{minipage}\\
    
    \begin{minipage}[c]{0.5\textwidth}
        \includegraphics[trim=0 -300 0 360,clip,width=1.\textwidth,angle=0]{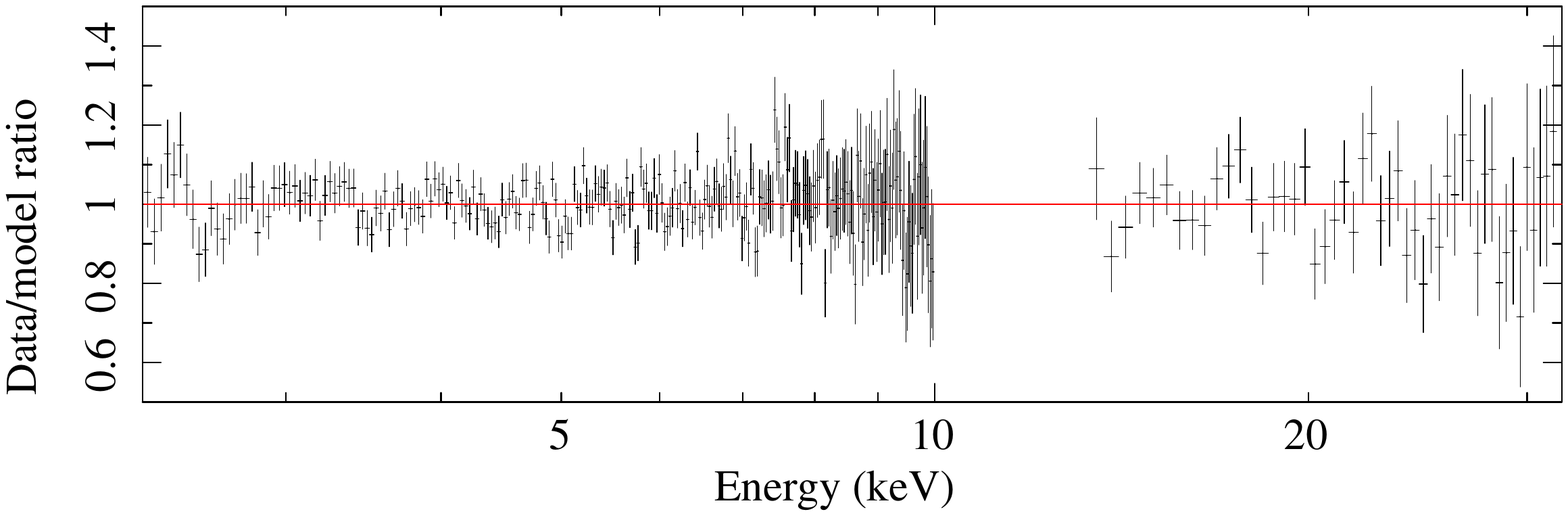}
    \end{minipage}
    \begin{minipage}[c]{0.5\textwidth}
        \includegraphics[trim=0 193 0 80,clip,width=\textwidth,angle=0]{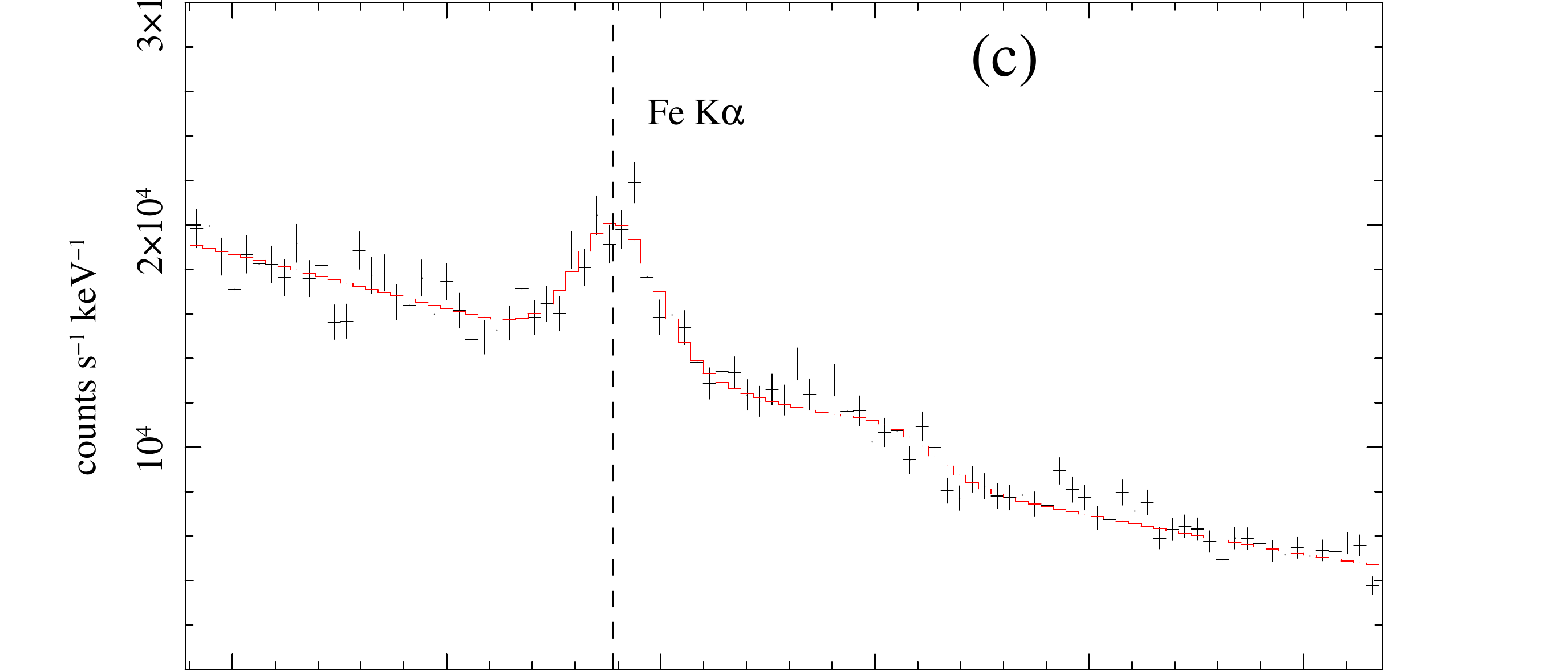}
      \begin{minipage}[c]{\textwidth}
        \includegraphics[trim=0 30 0 360,clip,width=1.\textwidth,angle=0]{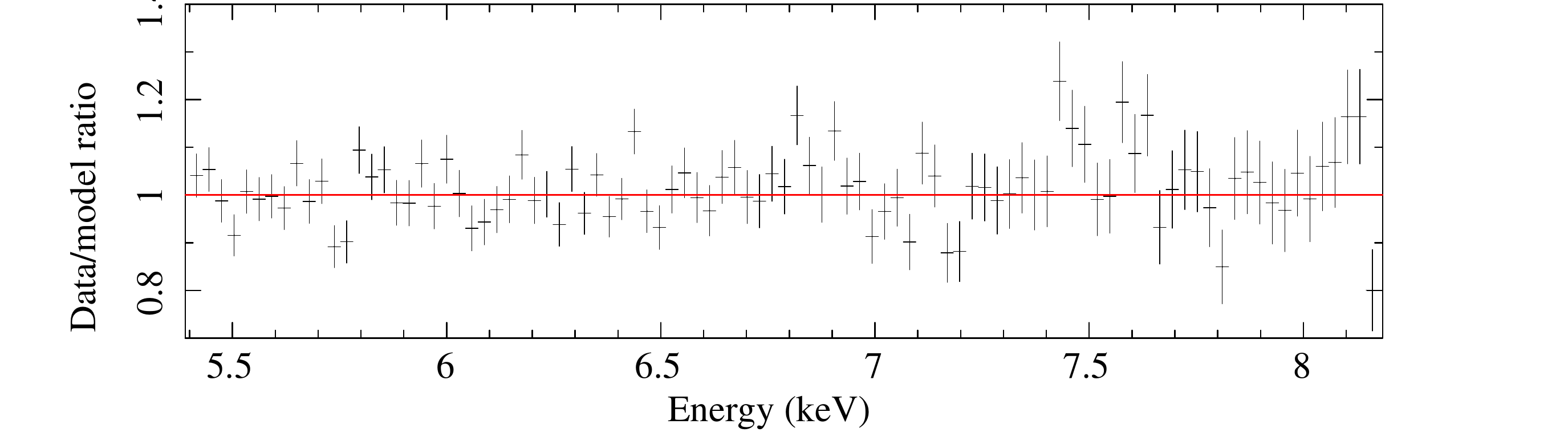}
      \end{minipage}
    \end{minipage}

    \caption{\footnotesize CenA 708036020 \label{fig-cen708-20}}
\end{figure*}

\begin{figure*}[t!]
    \begin{minipage}[c]{0.5\textwidth}
      \includegraphics[trim=0 50 0 -200,clip,width=1.\textwidth,angle=0]{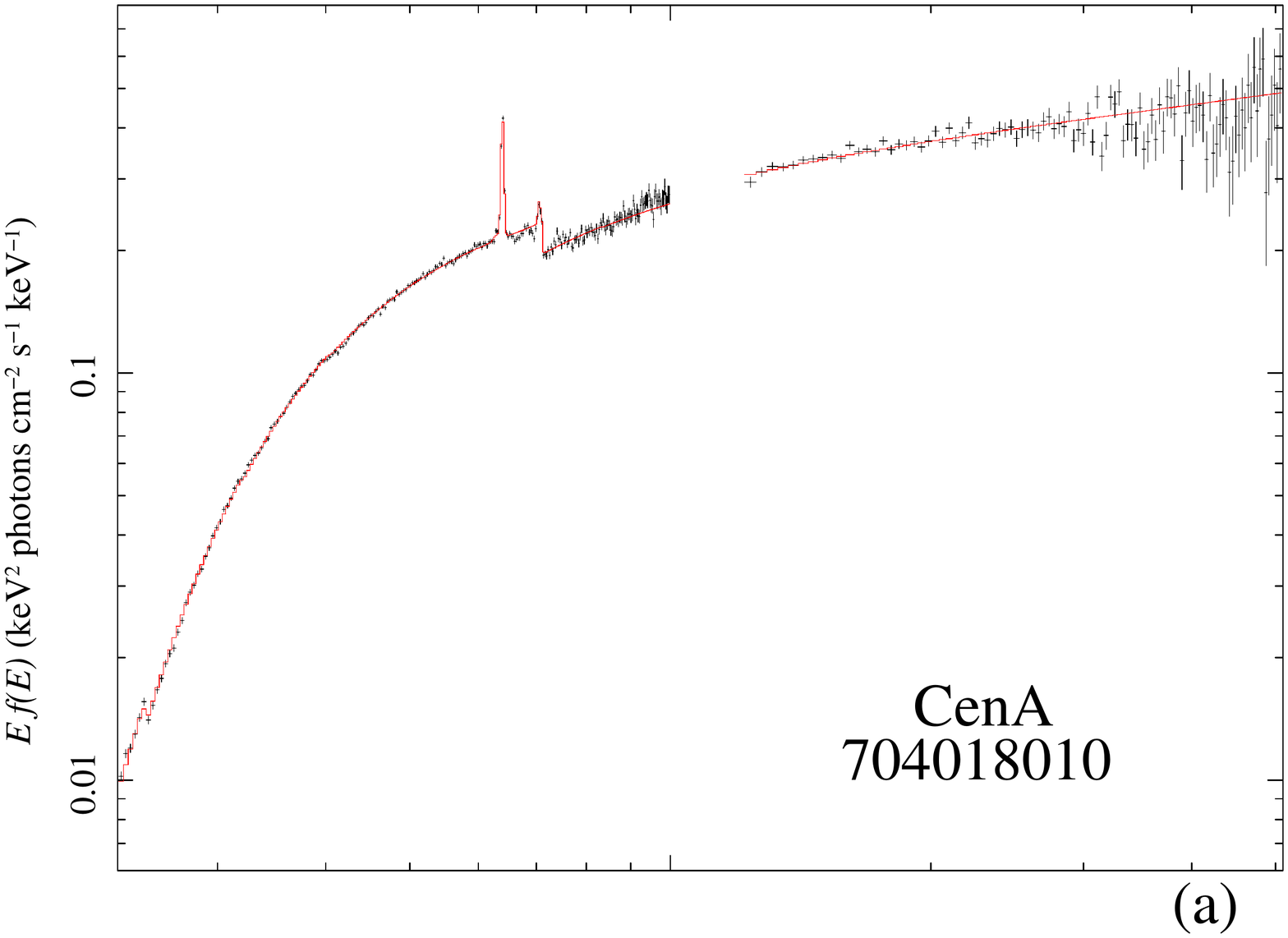}
    \end{minipage}
    \begin{minipage}[c]{0.5\textwidth}\vspace{-0pt}
      \includegraphics[trim=0 30 0 50,clip,width=1.\textwidth,angle=0]{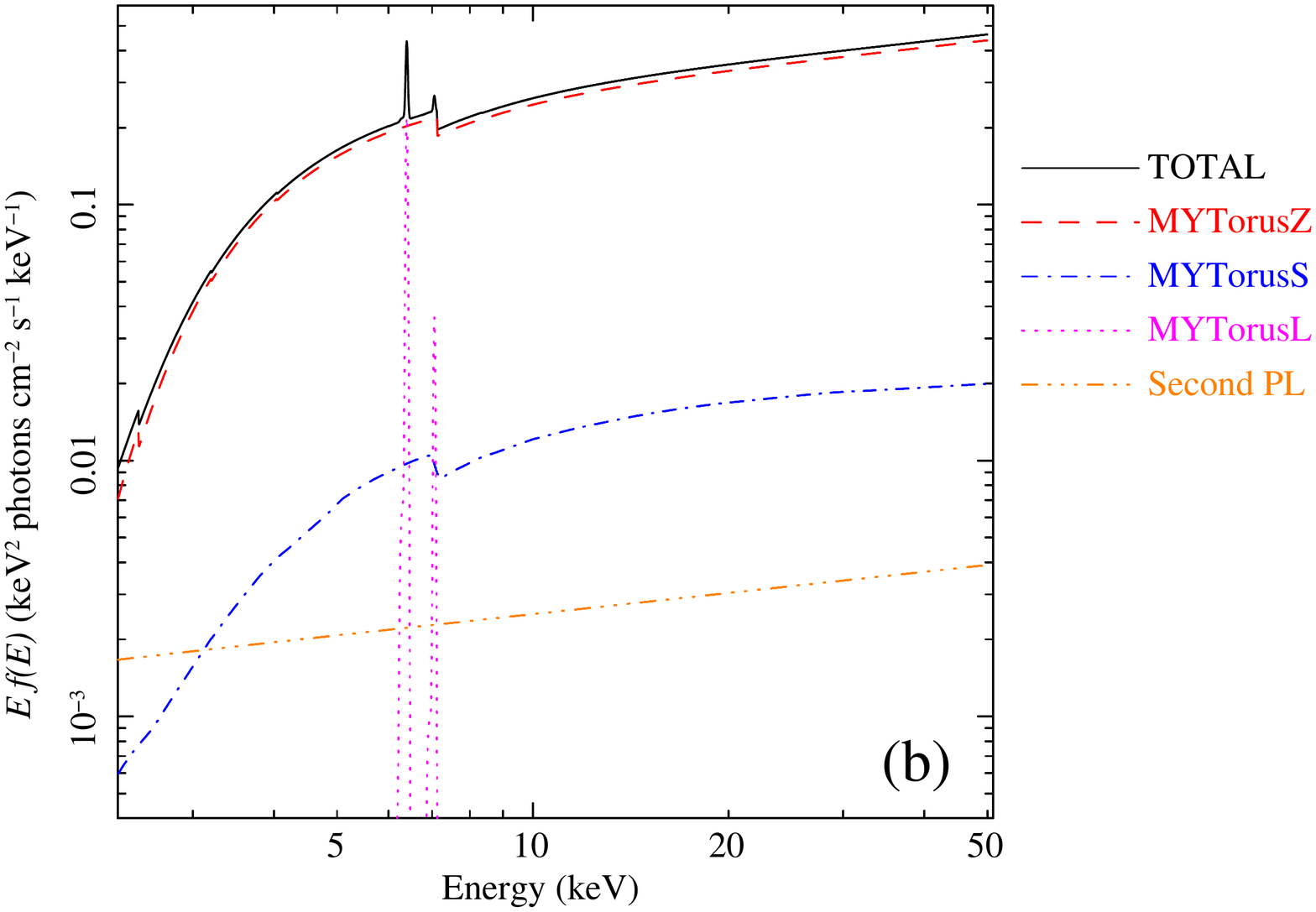}
      \vspace{-0pt}
    \end{minipage}\\
    
    \begin{minipage}[c]{0.5\textwidth}
        \includegraphics[trim=0 -300 0 360,clip,width=1.\textwidth,angle=0]{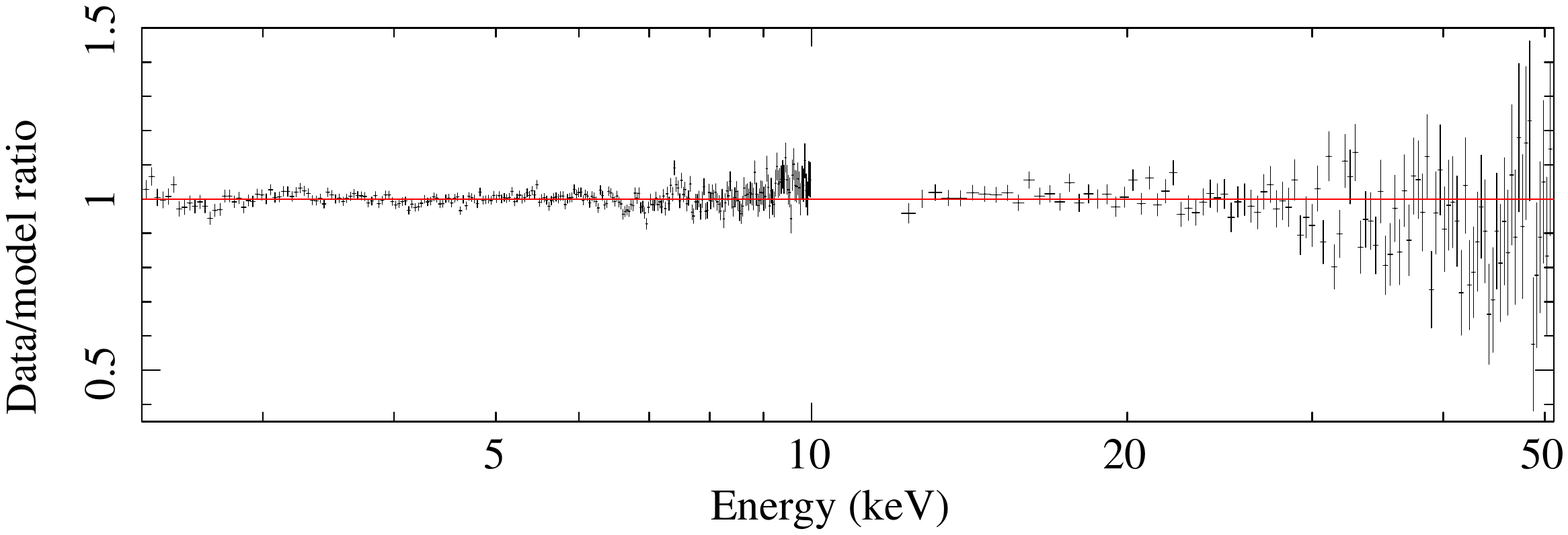}
    \end{minipage}
    \begin{minipage}[c]{0.5\textwidth}
        \includegraphics[trim=0 193 0 80,clip,width=\textwidth,angle=0]{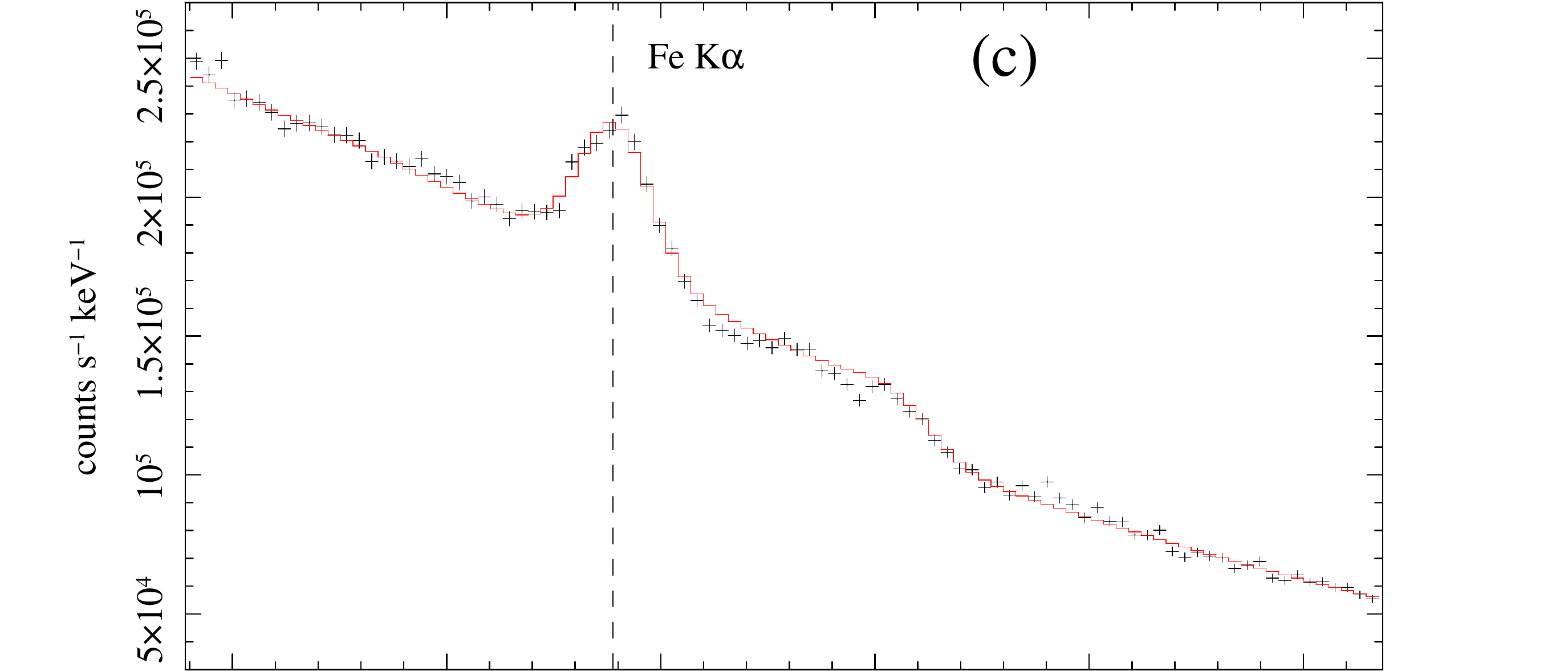}
      \begin{minipage}[c]{\textwidth}
        \includegraphics[trim=0 30 0 360,clip,width=1.\textwidth,angle=0]{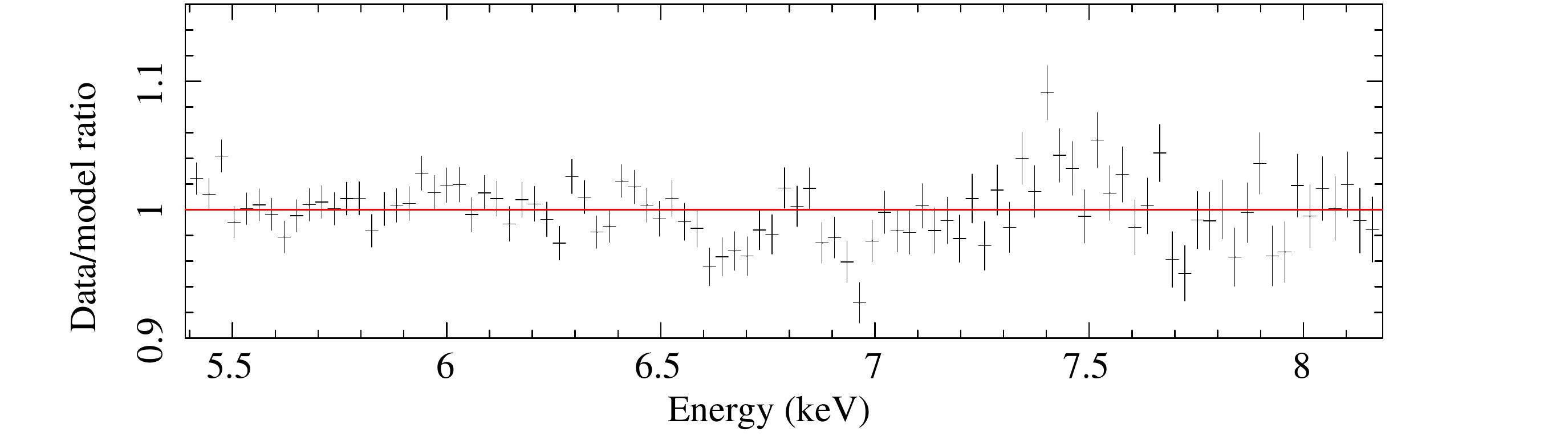}
      \end{minipage}
    \end{minipage}

    \caption{\footnotesize CenA 704018010 \label{fig-cen704-10}}
\end{figure*}

\begin{figure*}[t!]
    \begin{minipage}[c]{0.5\textwidth}
      \includegraphics[trim=0 50 0 -200,clip,width=1.\textwidth,angle=0]{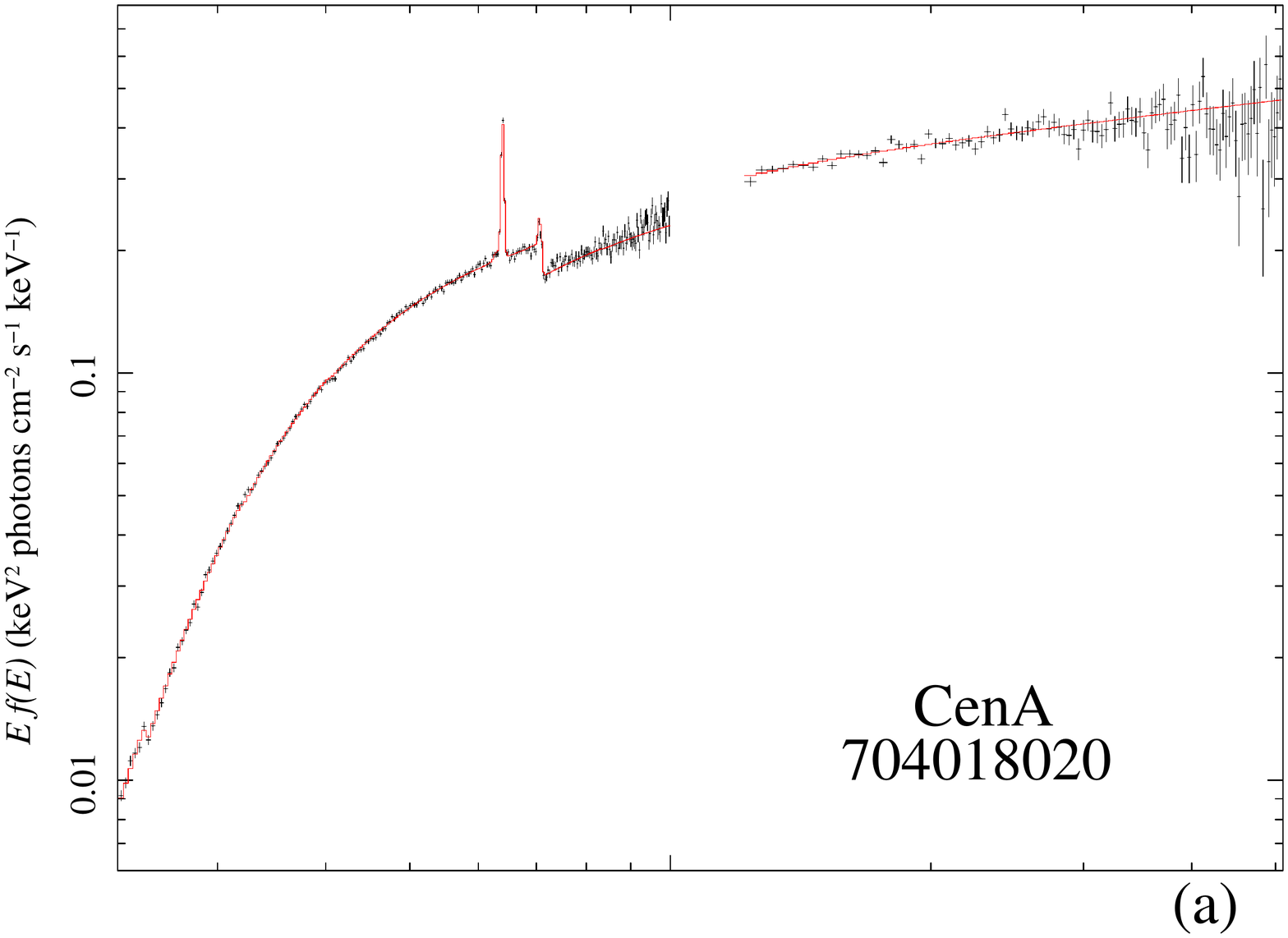}
    \end{minipage}
    \begin{minipage}[c]{0.5\textwidth}\vspace{-0pt}
      \includegraphics[trim=0 30 0 50,clip,width=1.\textwidth,angle=0]{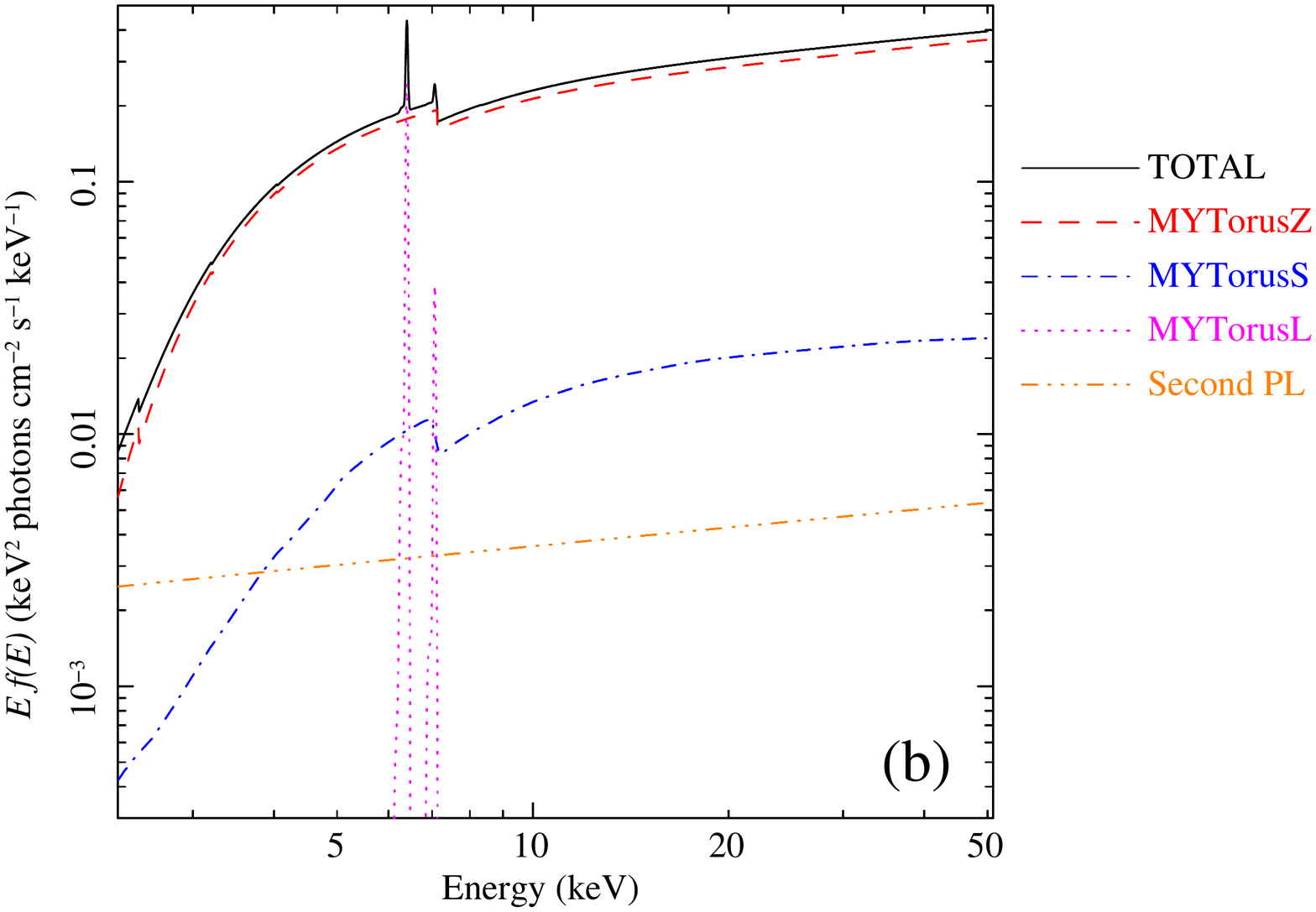}
      \vspace{-0pt}
    \end{minipage}\\
    
    \begin{minipage}[c]{0.5\textwidth}
        \includegraphics[trim=0 -300 0 360,clip,width=1.\textwidth,angle=0]{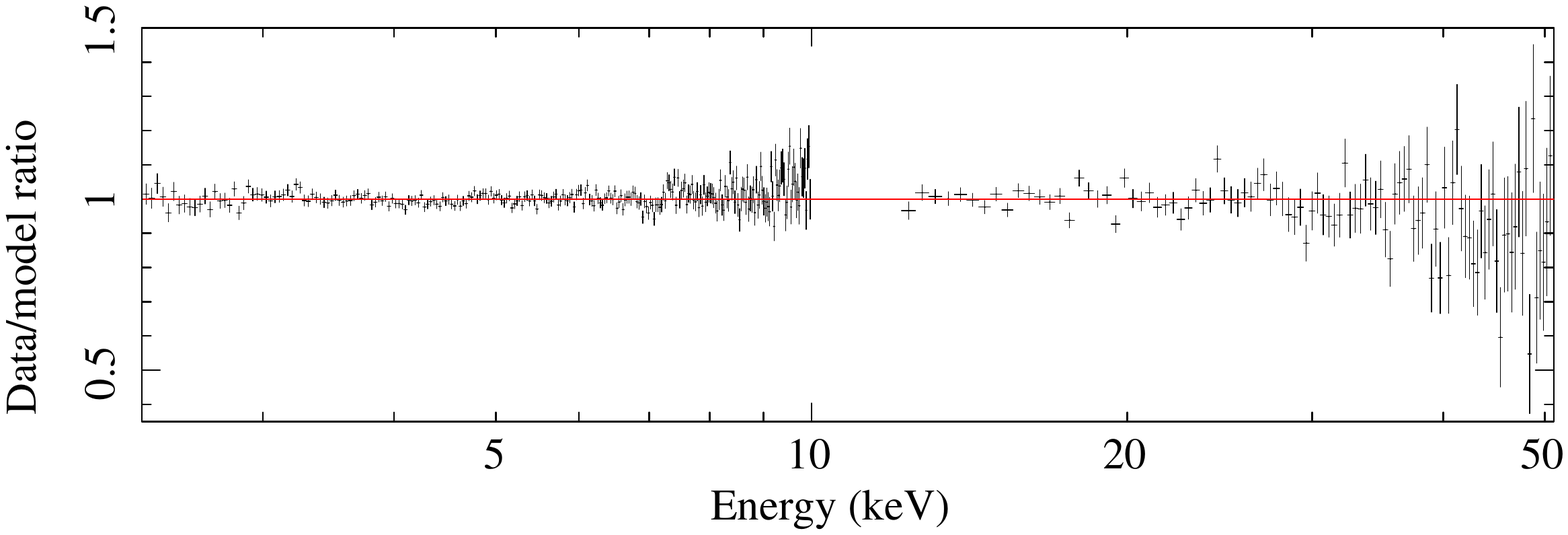}
    \end{minipage}
    \begin{minipage}[c]{0.5\textwidth}
        \includegraphics[trim=0 193 0 80,clip,width=\textwidth,angle=0]{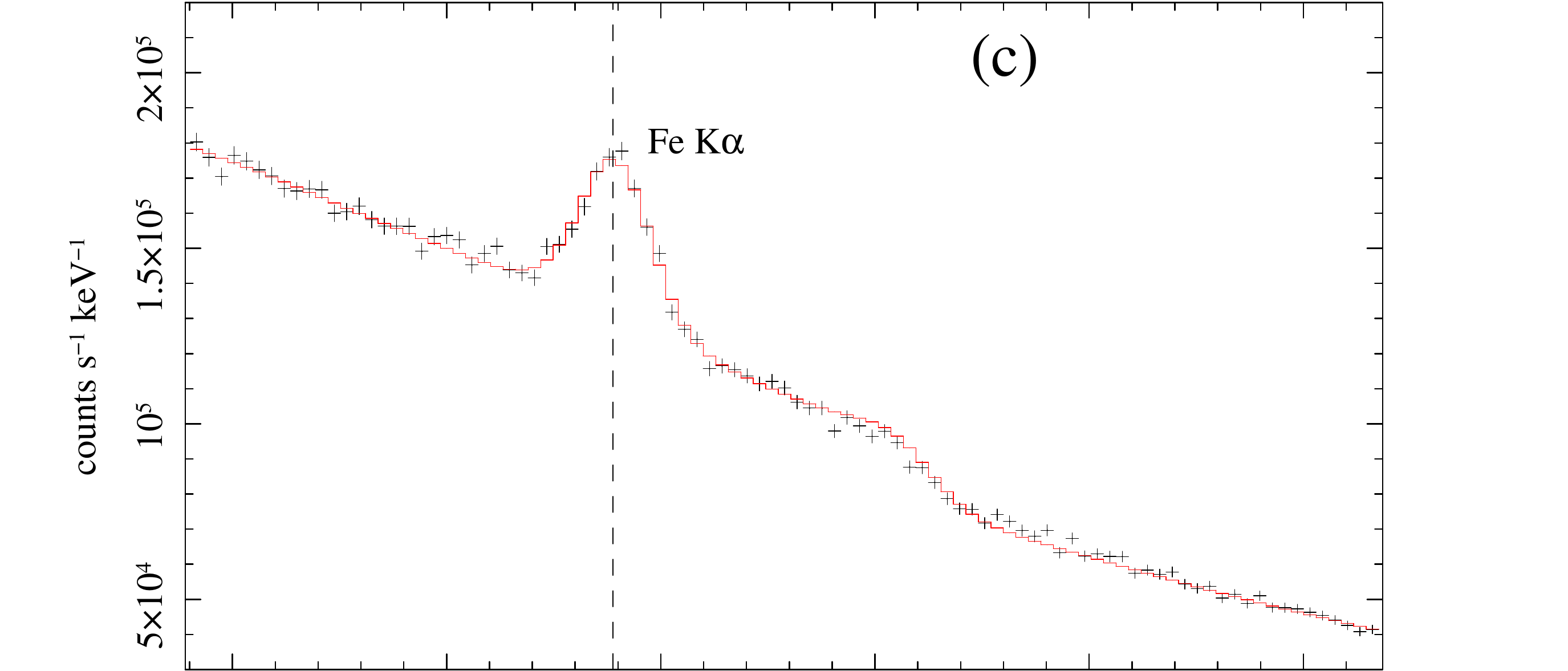}
      \begin{minipage}[c]{\textwidth}
        \includegraphics[trim=0 30 0 360,clip,width=1.\textwidth,angle=0]{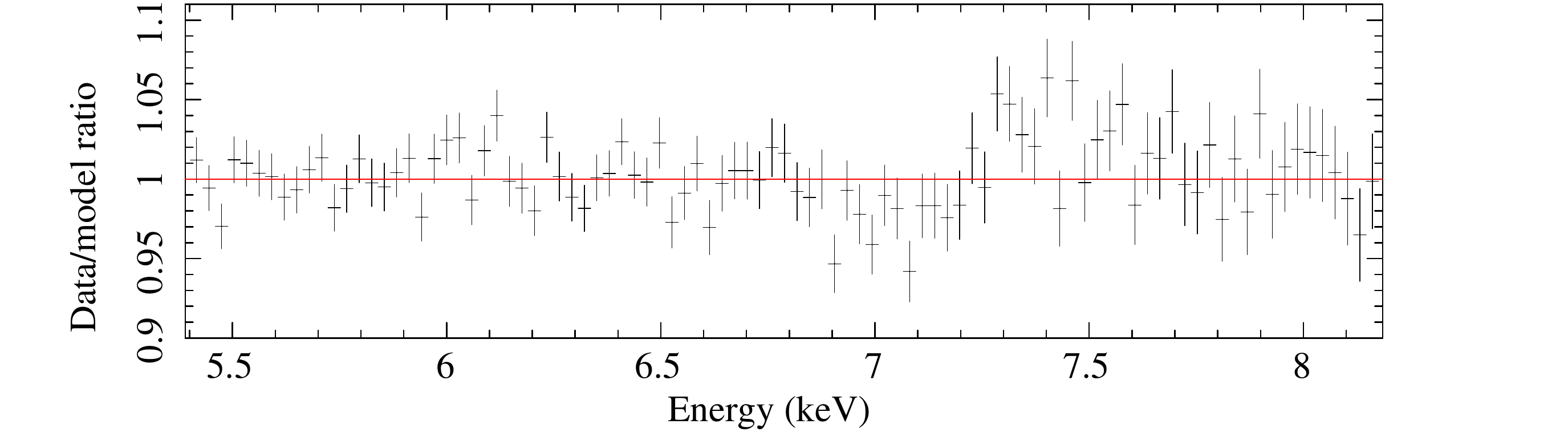}
      \end{minipage}
    \end{minipage}

    \caption{\footnotesize CenA 704018020 \label{fig-cen704-20}}
\end{figure*}

\begin{figure*}[t!]
    \begin{minipage}[c]{0.5\textwidth}
      \includegraphics[trim=0 50 0 -200,clip,width=1.\textwidth,angle=0]{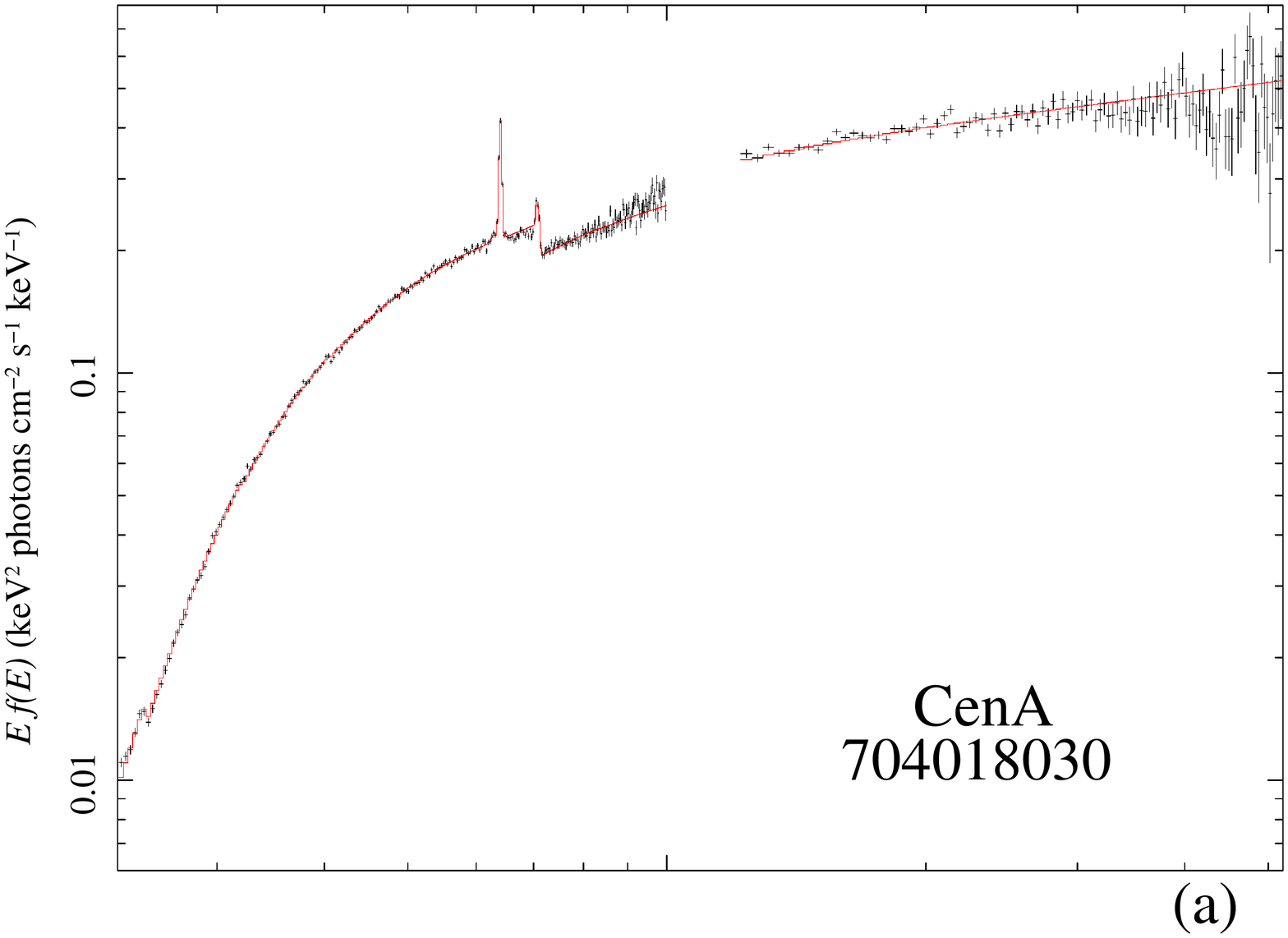}
    \end{minipage}
    \begin{minipage}[c]{0.5\textwidth}\vspace{-0pt}
      \includegraphics[trim=0 30 0 50,clip,width=1.\textwidth,angle=0]{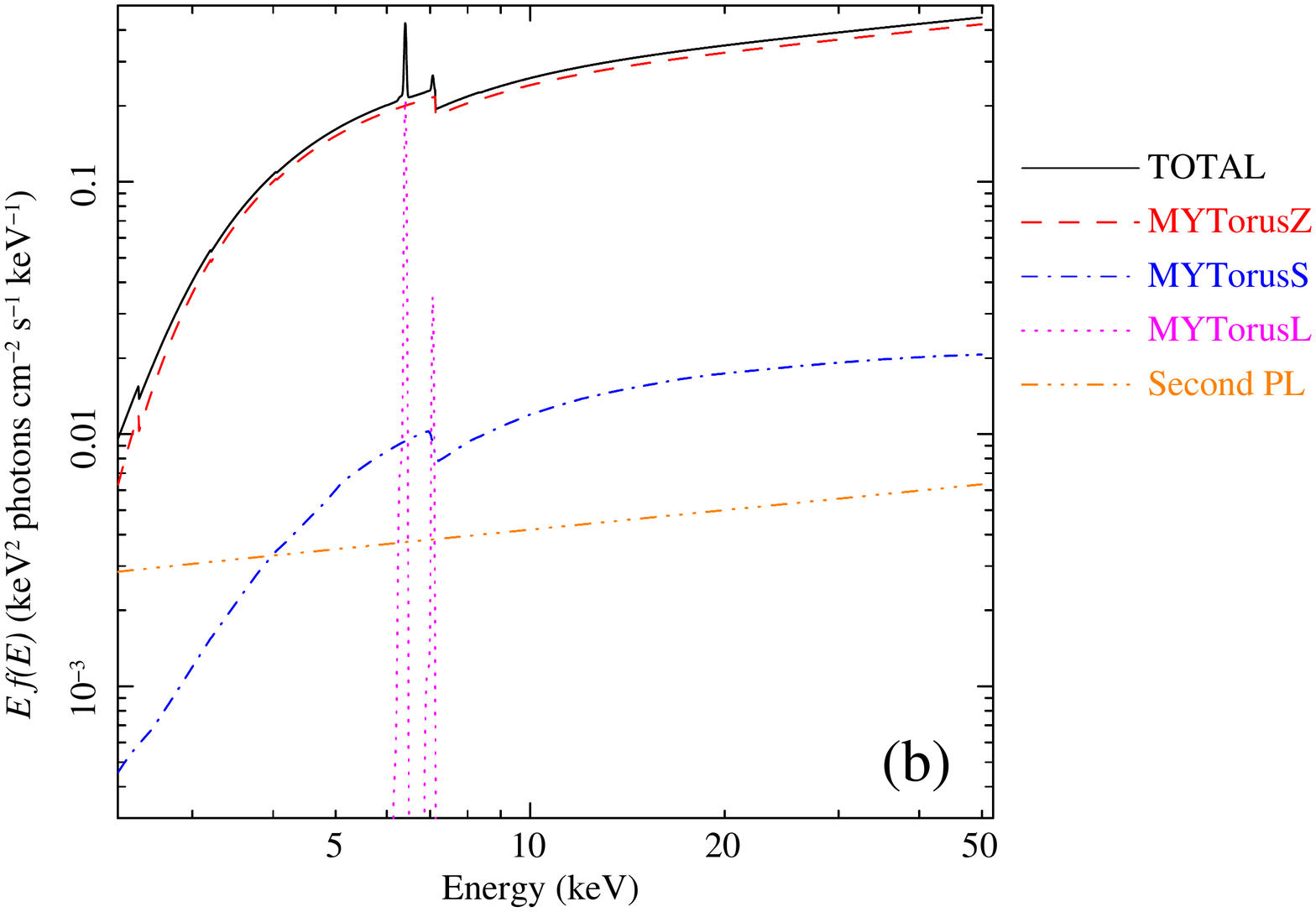}
      \vspace{-0pt}
    \end{minipage}\\
    
    \begin{minipage}[c]{0.5\textwidth}
        \includegraphics[trim=0 -300 0 360,clip,width=1.\textwidth,angle=0]{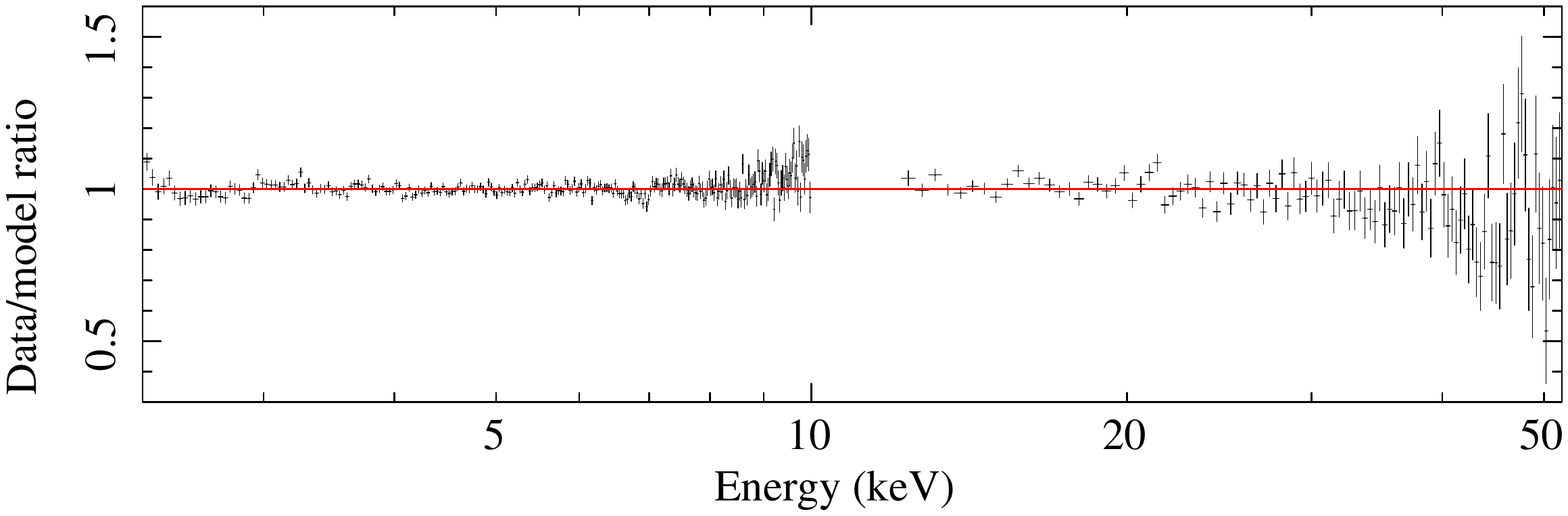}
    \end{minipage}
    \begin{minipage}[c]{0.5\textwidth}
        \includegraphics[trim=0 193 0 80,clip,width=\textwidth,angle=0]{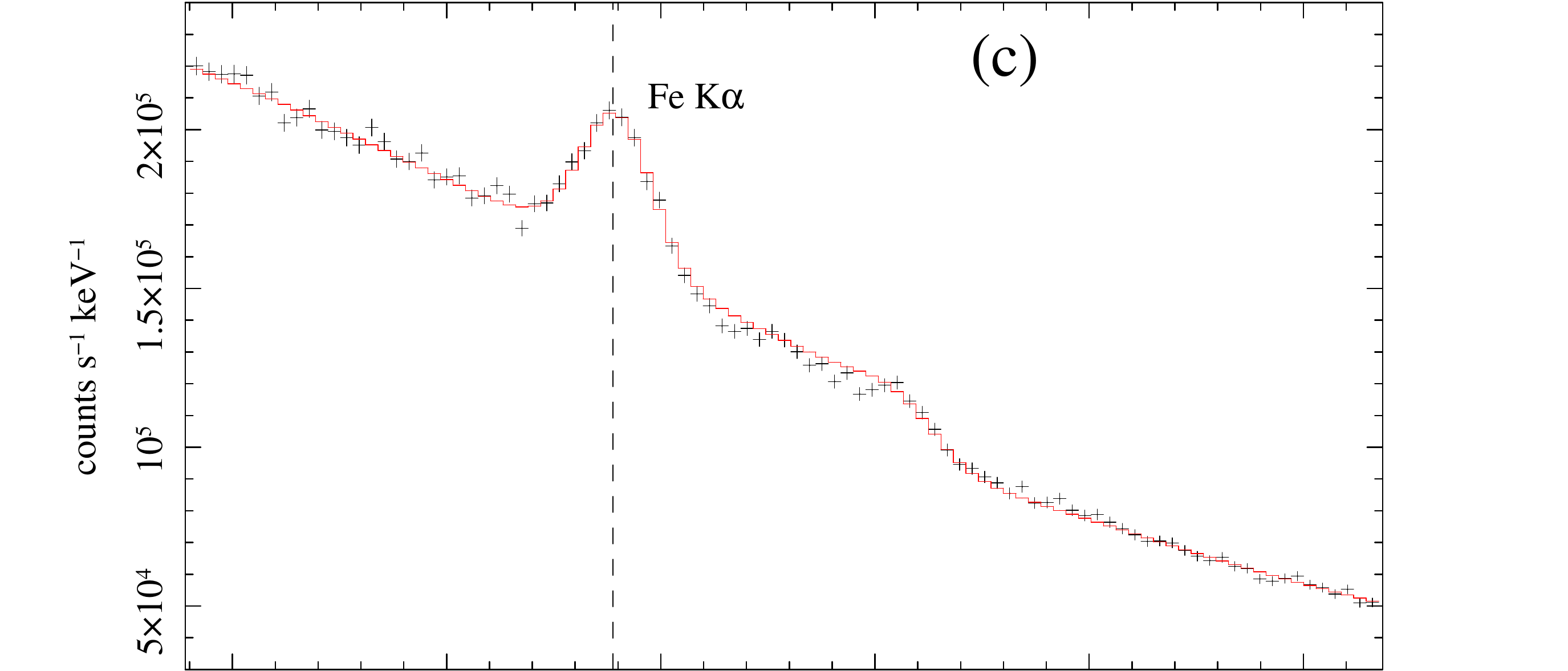}
      \begin{minipage}[c]{\textwidth}
        \includegraphics[trim=0 30 0 360,clip,width=1.\textwidth,angle=0]{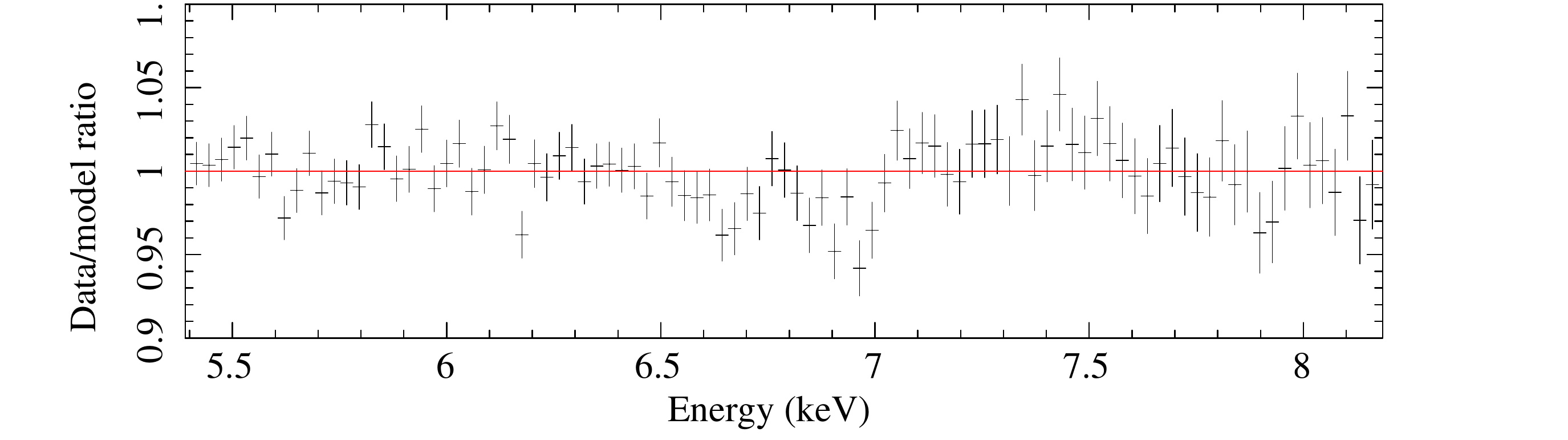}
      \end{minipage}
    \end{minipage}

    \caption{\footnotesize CenA 704018030 \label{fig-cen704-30}}
\end{figure*}

\begin{figure*}[t!]
    \begin{minipage}[c]{0.5\textwidth}
      \includegraphics[trim=0 50 0 -200,clip,width=1.\textwidth,angle=0]{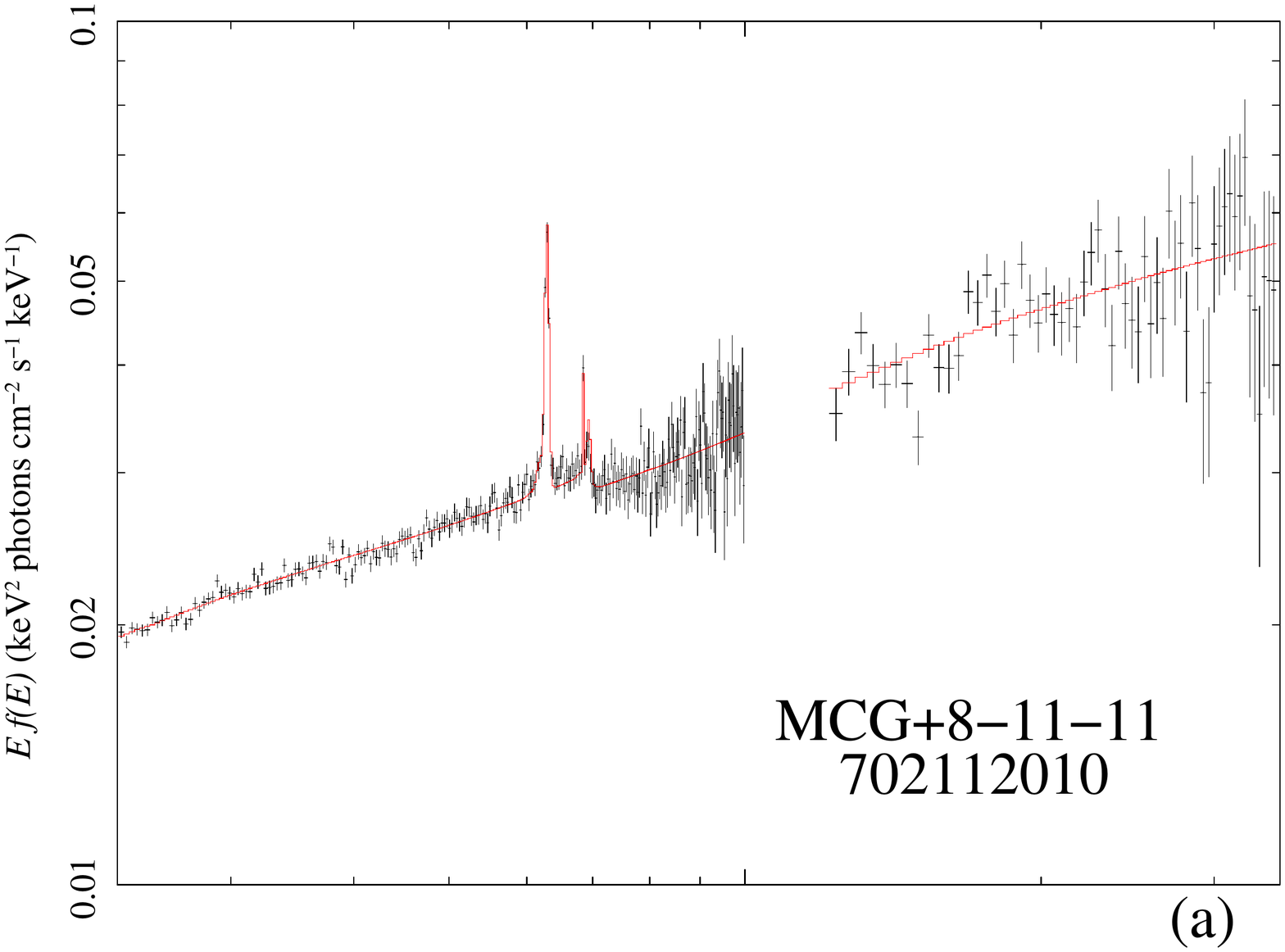}
    \end{minipage}
    \begin{minipage}[c]{0.5\textwidth}\vspace{-0pt}
      \includegraphics[trim=0 30 0 50,clip,width=1.\textwidth,angle=0]{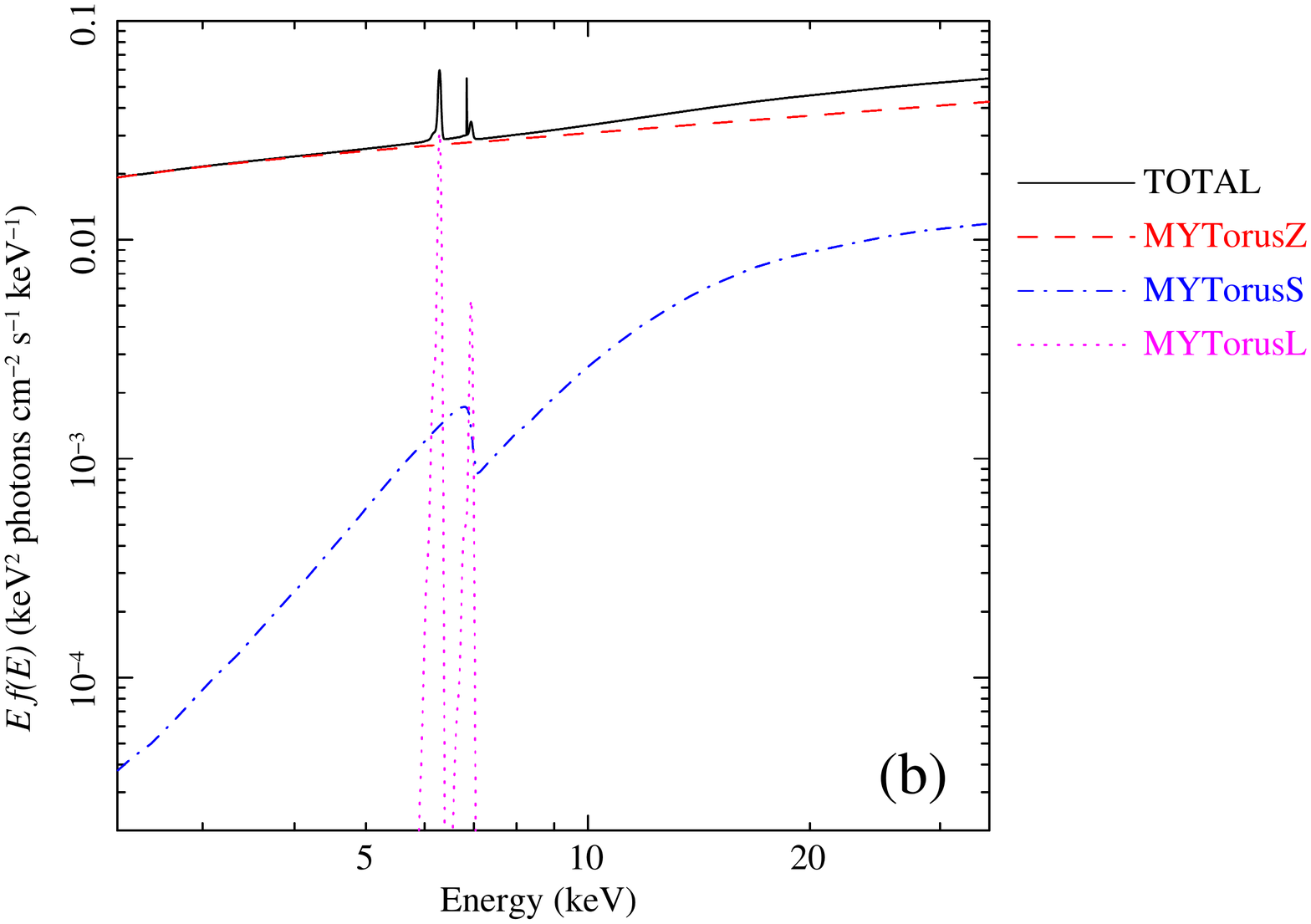}
      \vspace{-0pt}
    \end{minipage}\\
    
    \begin{minipage}[c]{0.5\textwidth}
        \includegraphics[trim=0 -300 0 360,clip,width=1.\textwidth,angle=0]{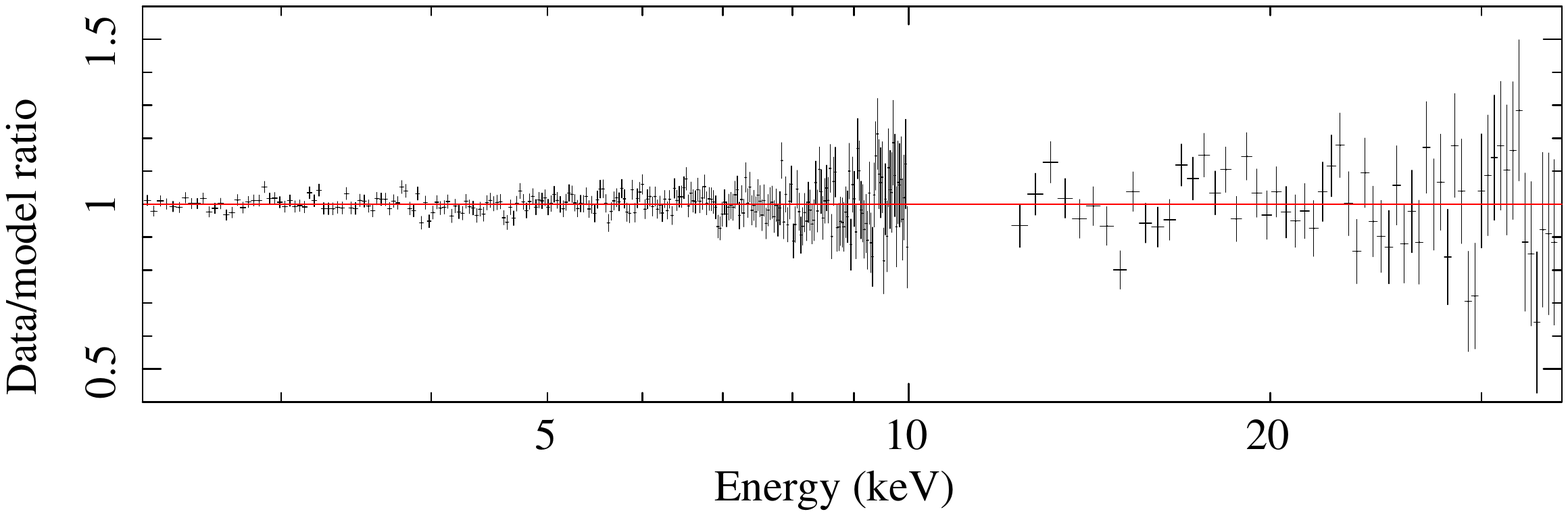}
    \end{minipage}
    \begin{minipage}[c]{0.5\textwidth}
        \includegraphics[trim=0 193 0 80,clip,width=\textwidth,angle=0]{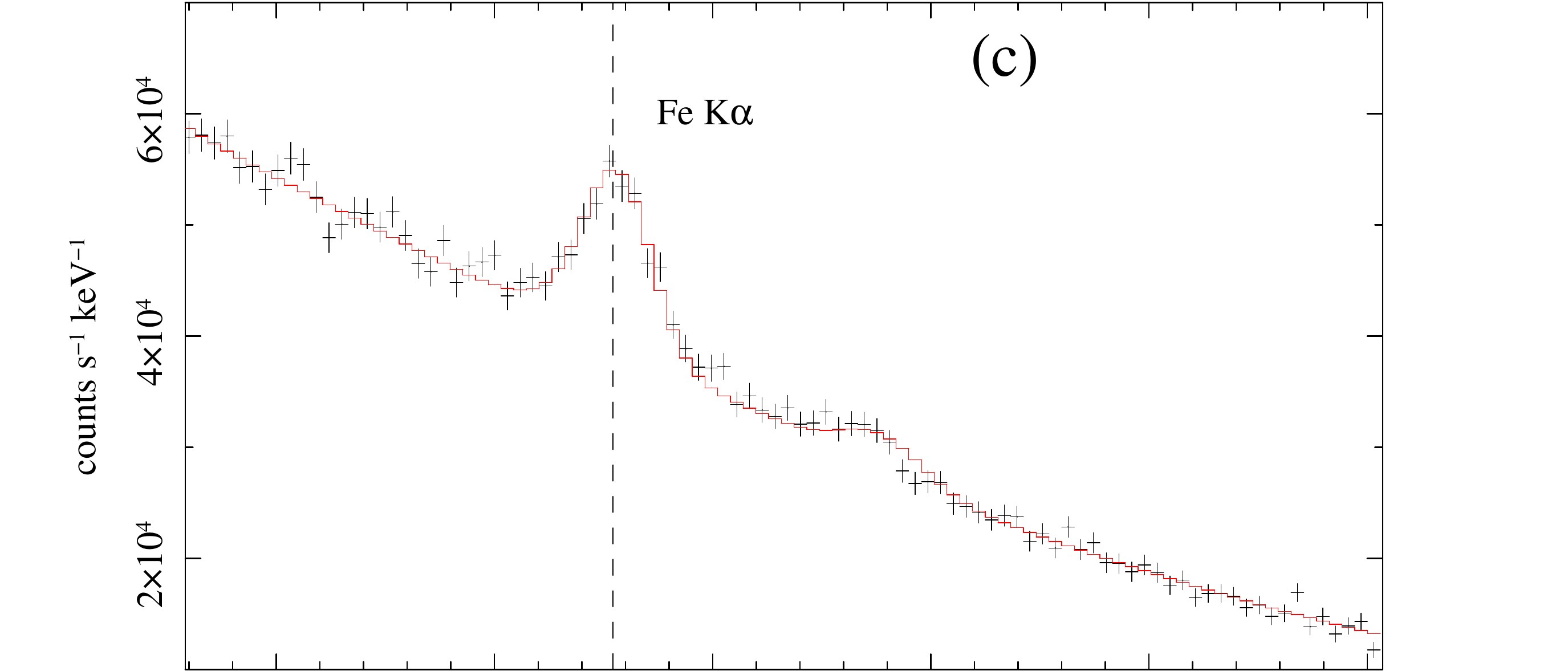}
      \begin{minipage}[c]{\textwidth}
        \includegraphics[trim=0 30 0 360,clip,width=1.\textwidth,angle=0]{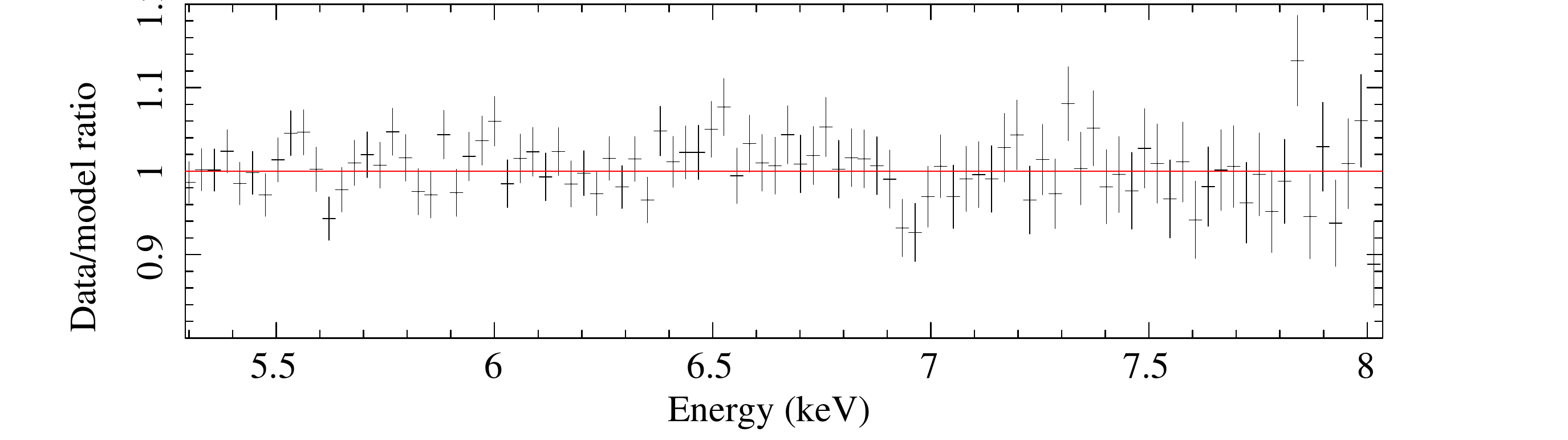}
      \end{minipage}
    \end{minipage}

    \caption{\footnotesize MCG+8-11-11 702112010 \label{fig-mcg8}}
\end{figure*}

\begin{figure*}[t!]
    \begin{minipage}[c]{0.5\textwidth}
      \includegraphics[trim=0 50 0 -200,clip,width=1.\textwidth,angle=0]{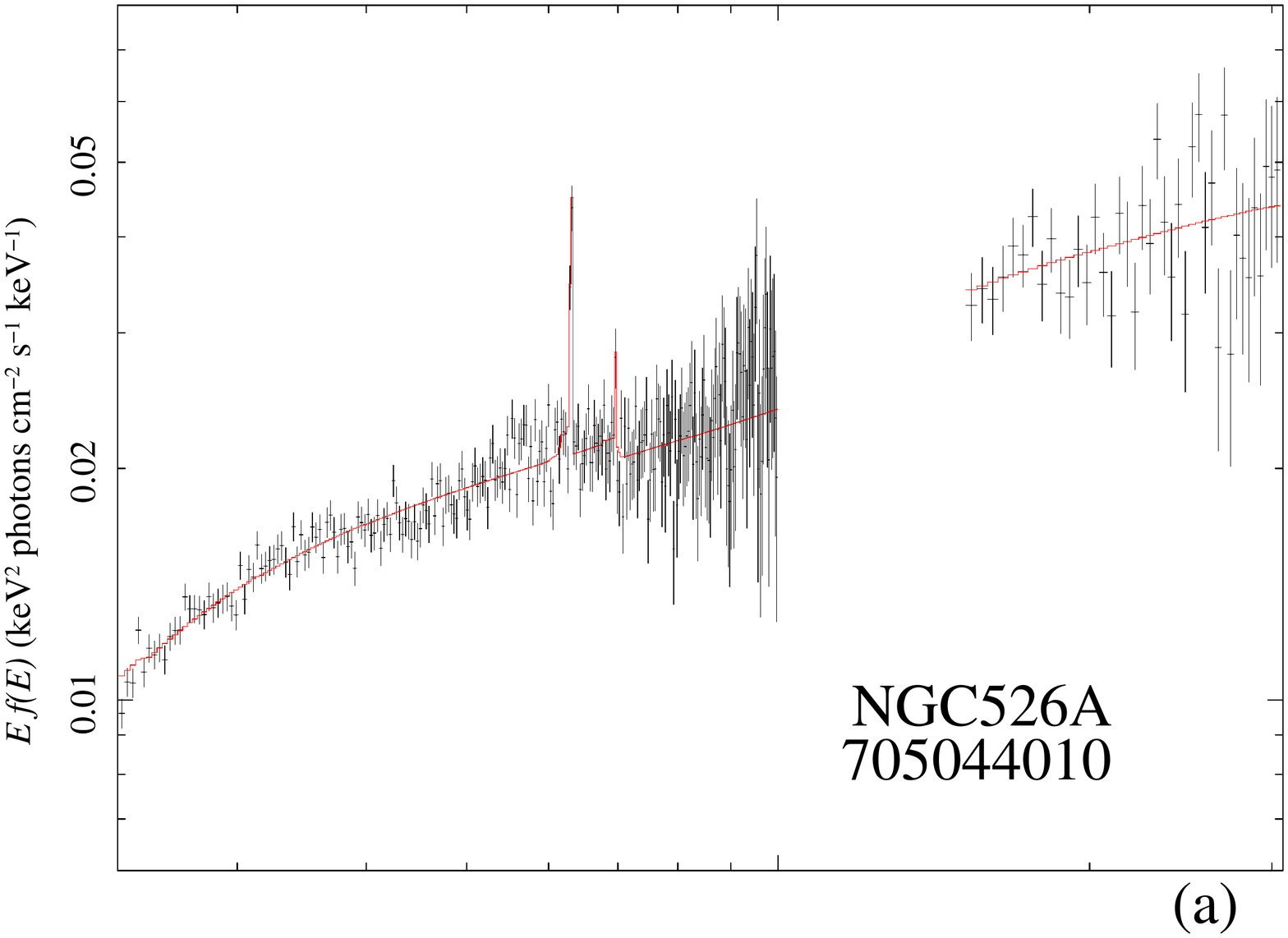}
    \end{minipage}
    \begin{minipage}[c]{0.5\textwidth}\vspace{-0pt}
      \includegraphics[trim=0 30 0 50,clip,width=1.\textwidth,angle=0]{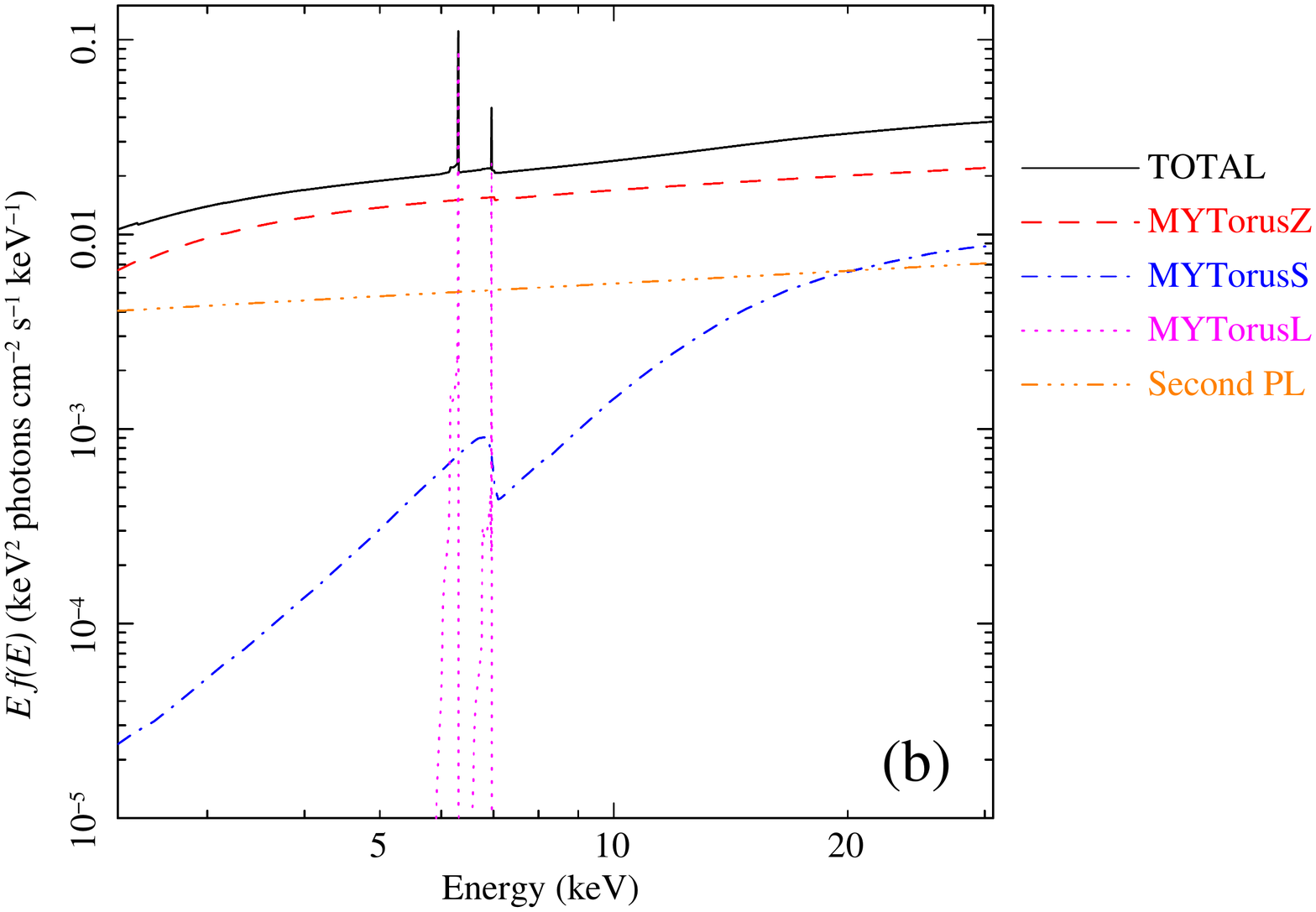}
      \vspace{-0pt}
    \end{minipage}\\
    
    \begin{minipage}[c]{0.5\textwidth}
        \includegraphics[trim=0 -300 0 360,clip,width=1.\textwidth,angle=0]{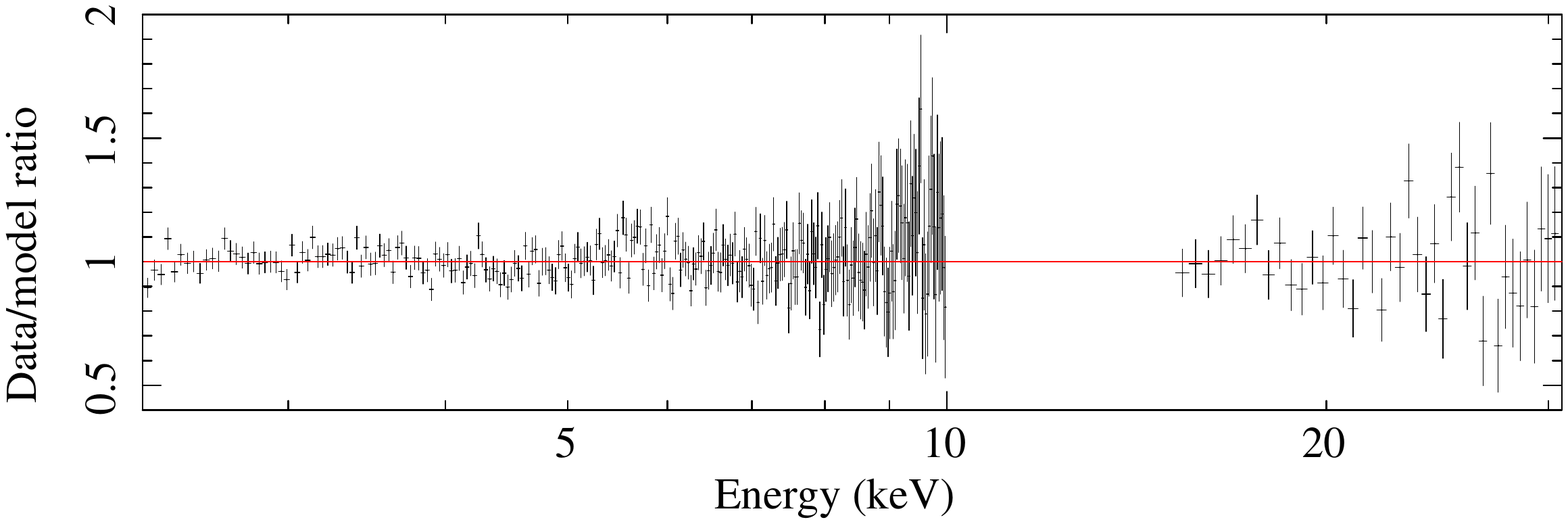}
    \end{minipage}
    \begin{minipage}[c]{0.5\textwidth}
        \includegraphics[trim=0 193 0 80,clip,width=\textwidth,angle=0]{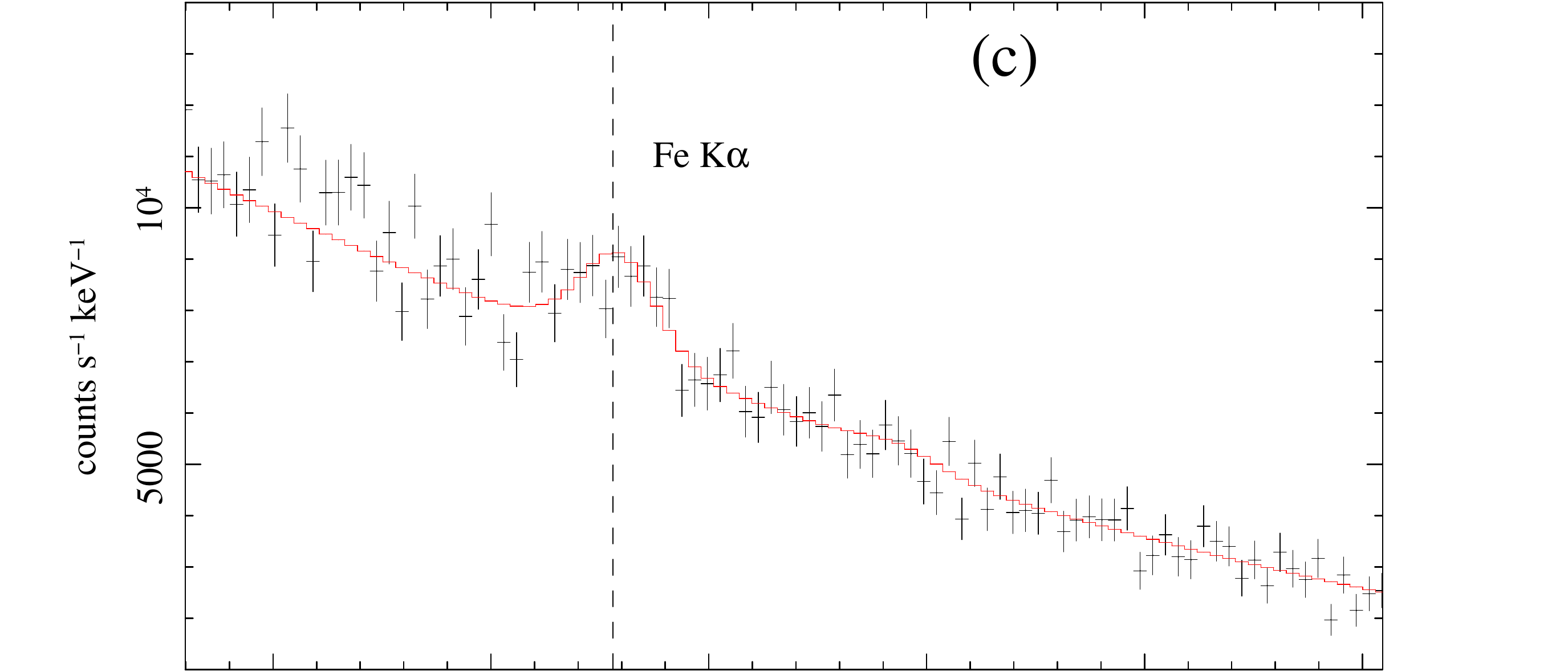}
      \begin{minipage}[c]{\textwidth}
        \includegraphics[trim=0 30 0 360,clip,width=1.\textwidth,angle=0]{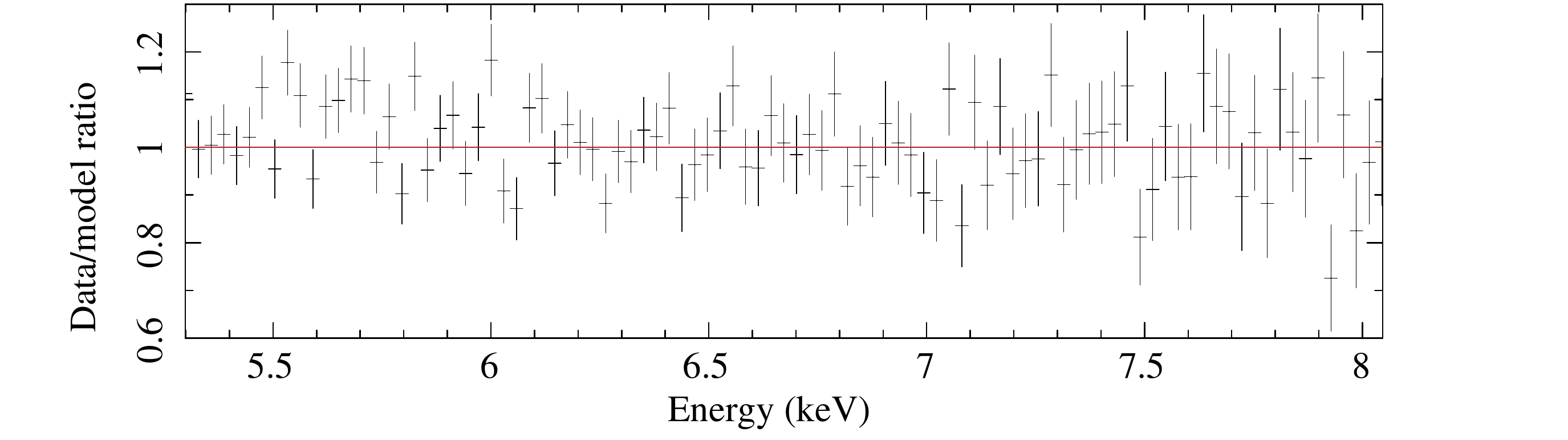}
      \end{minipage}
    \end{minipage}

    \caption{\footnotesize NGC526A 705044010 \label{fig-n526}}
\end{figure*}

\begin{figure*}[t!]
    \begin{minipage}[c]{0.5\textwidth}
      \includegraphics[trim=0 50 0 -200,clip,width=1.\textwidth,angle=0]{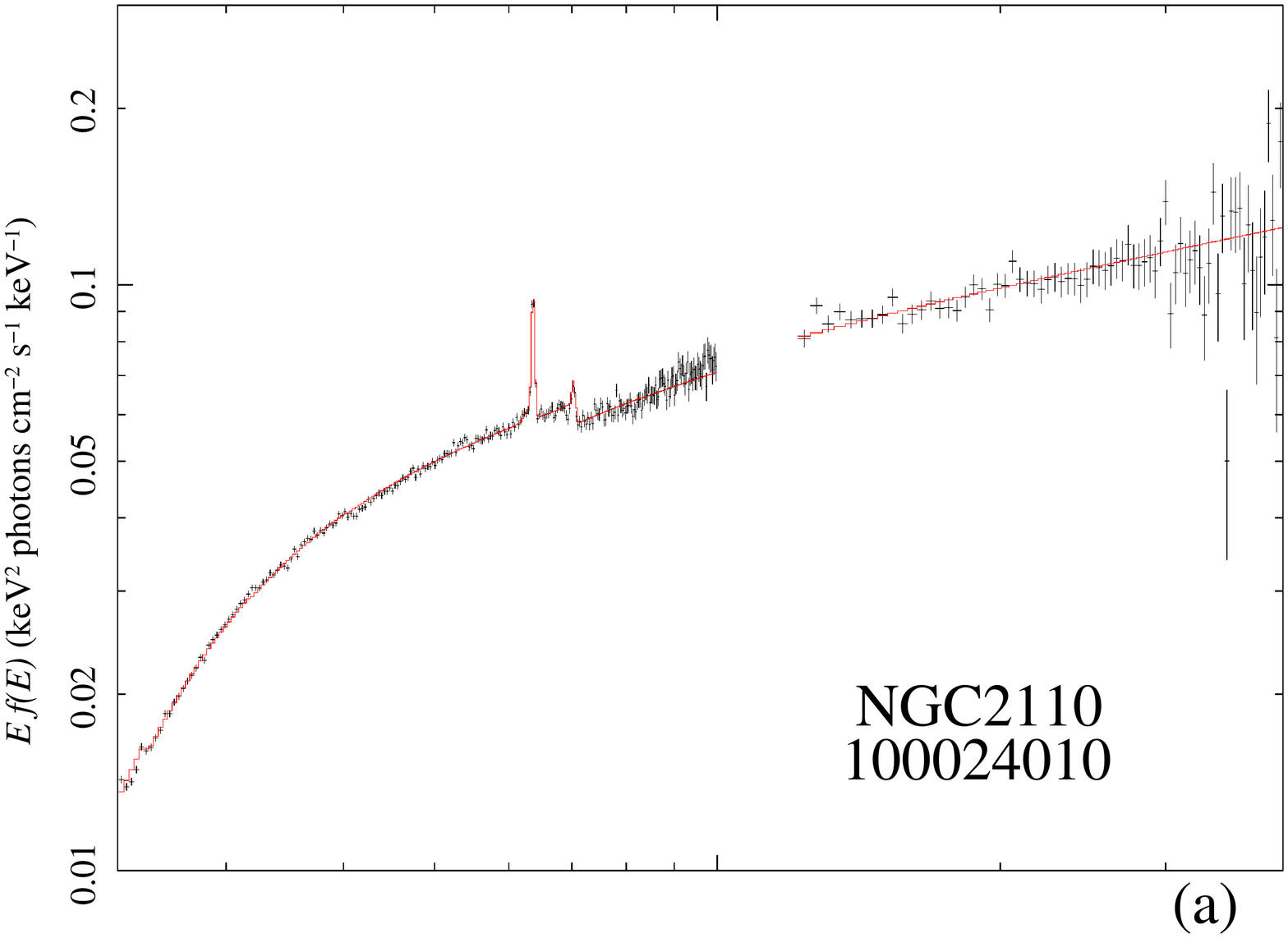}
    \end{minipage}
    \begin{minipage}[c]{0.5\textwidth}\vspace{-0pt}
      \includegraphics[trim=0 30 0 50,clip,width=1.\textwidth,angle=0]{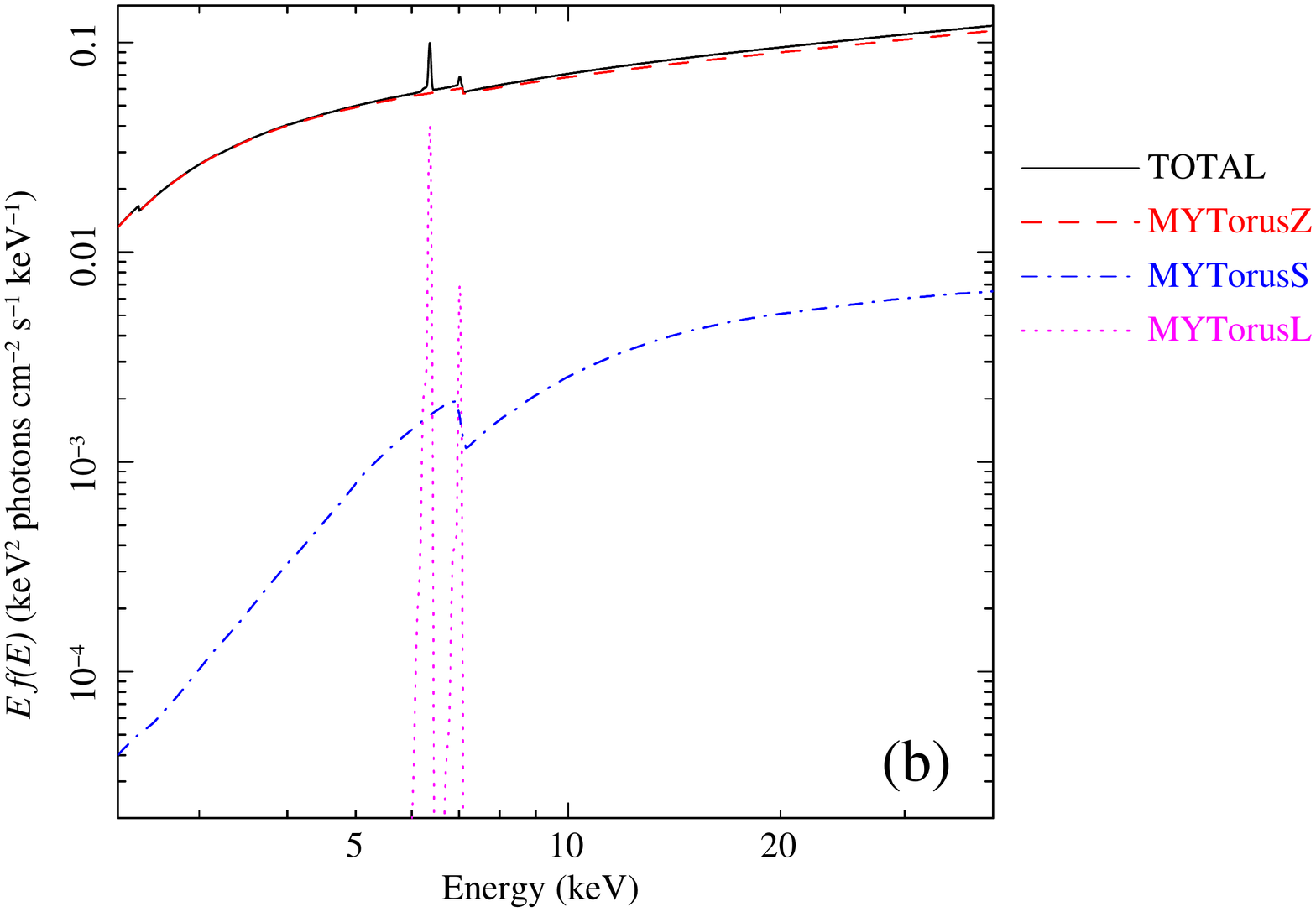}
      \vspace{-0pt}
    \end{minipage}\\
    
    \begin{minipage}[c]{0.5\textwidth}
        \includegraphics[trim=0 -300 0 360,clip,width=1.\textwidth,angle=0]{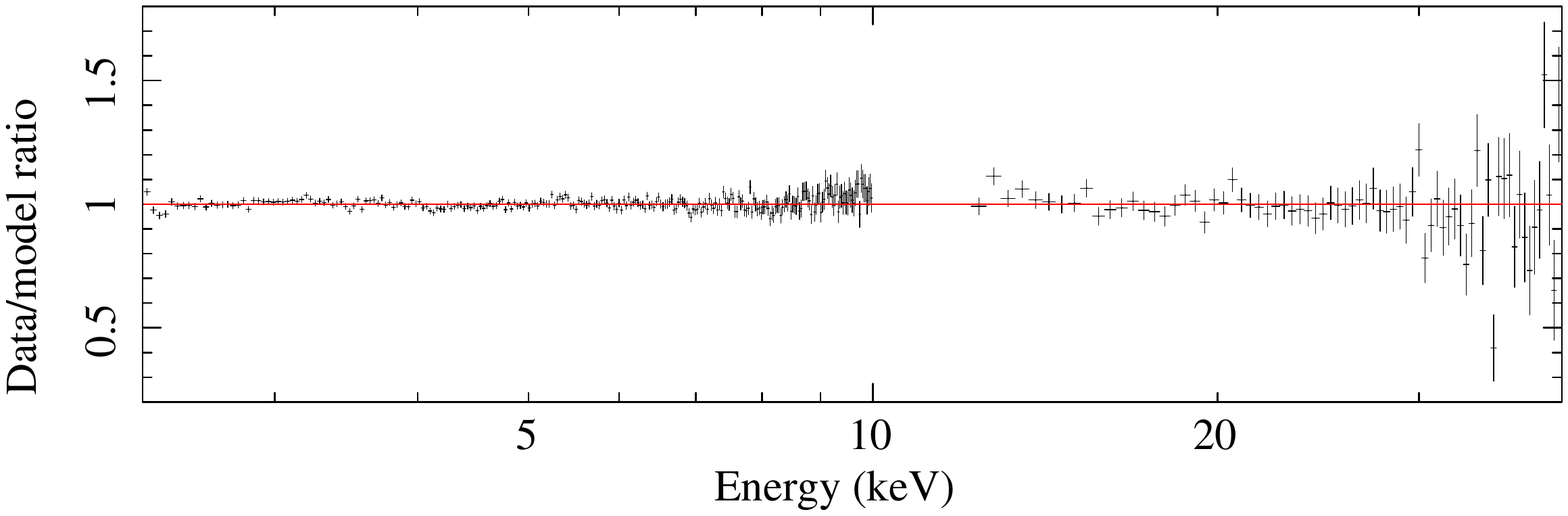}
    \end{minipage}
    \begin{minipage}[c]{0.5\textwidth}
        \includegraphics[trim=0 193 0 80,clip,width=\textwidth,angle=0]{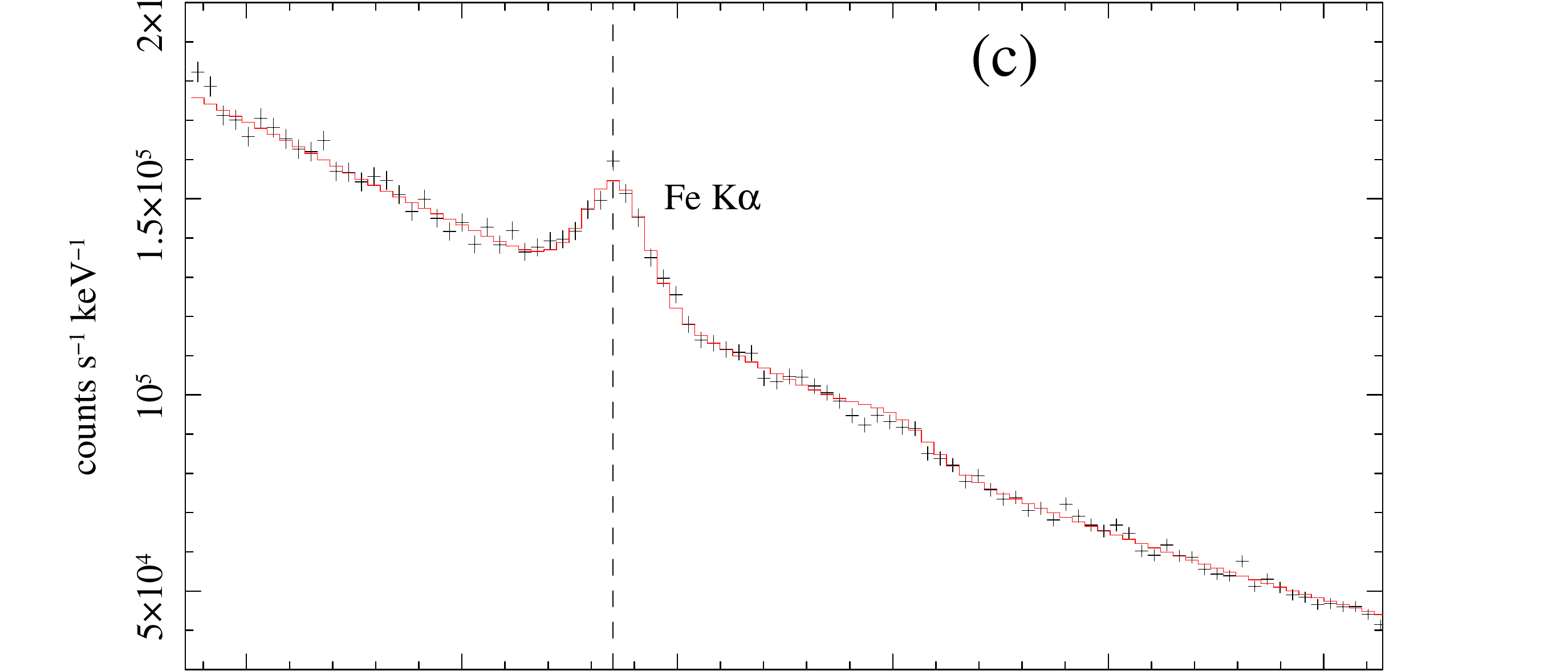}
      \begin{minipage}[c]{\textwidth}
        \includegraphics[trim=0 30 0 360,clip,width=1.\textwidth,angle=0]{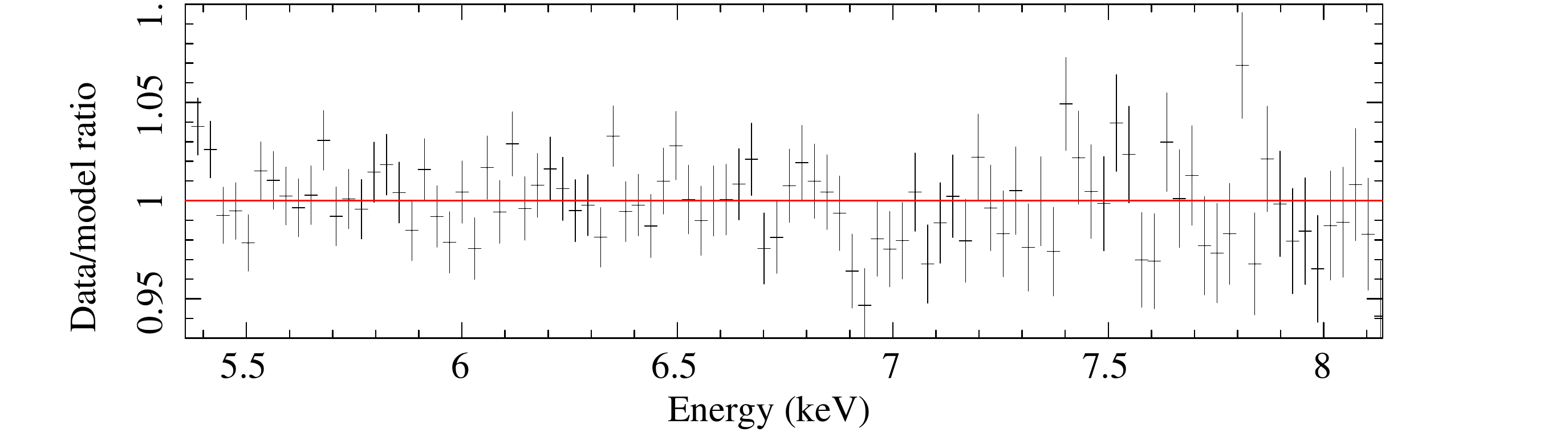}
      \end{minipage}
    \end{minipage}

    \caption{\footnotesize NGC2110 100024010 \label{fig-n2110-100}}
\end{figure*}

\begin{figure*}[t!]
    \begin{minipage}[c]{0.5\textwidth}
      \includegraphics[trim=0 50 0 -200,clip,width=1.\textwidth,angle=0]{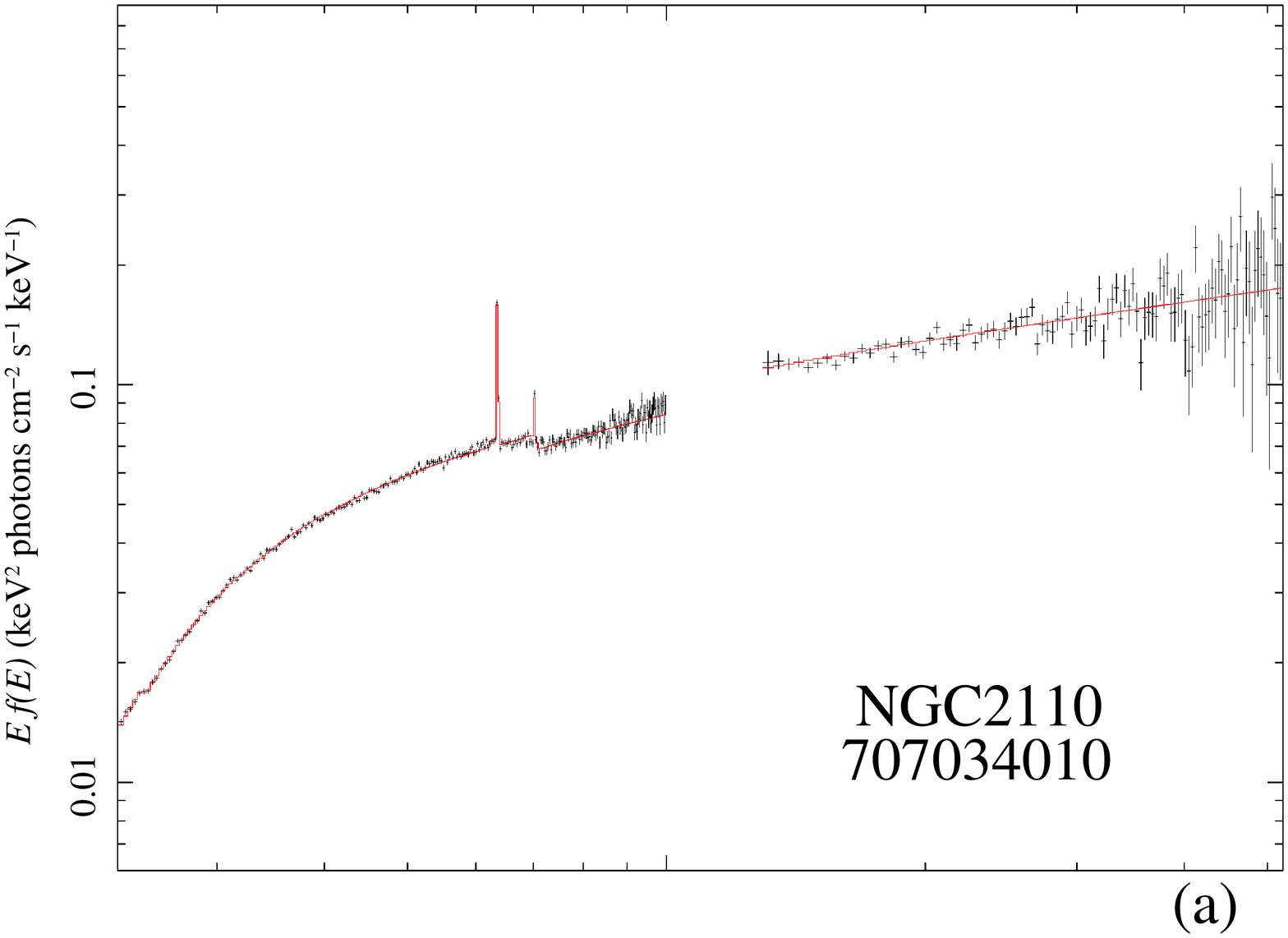}
    \end{minipage}
    \begin{minipage}[c]{0.5\textwidth}\vspace{-0pt}
      \includegraphics[trim=0 30 0 50,clip,width=1.\textwidth,angle=0]{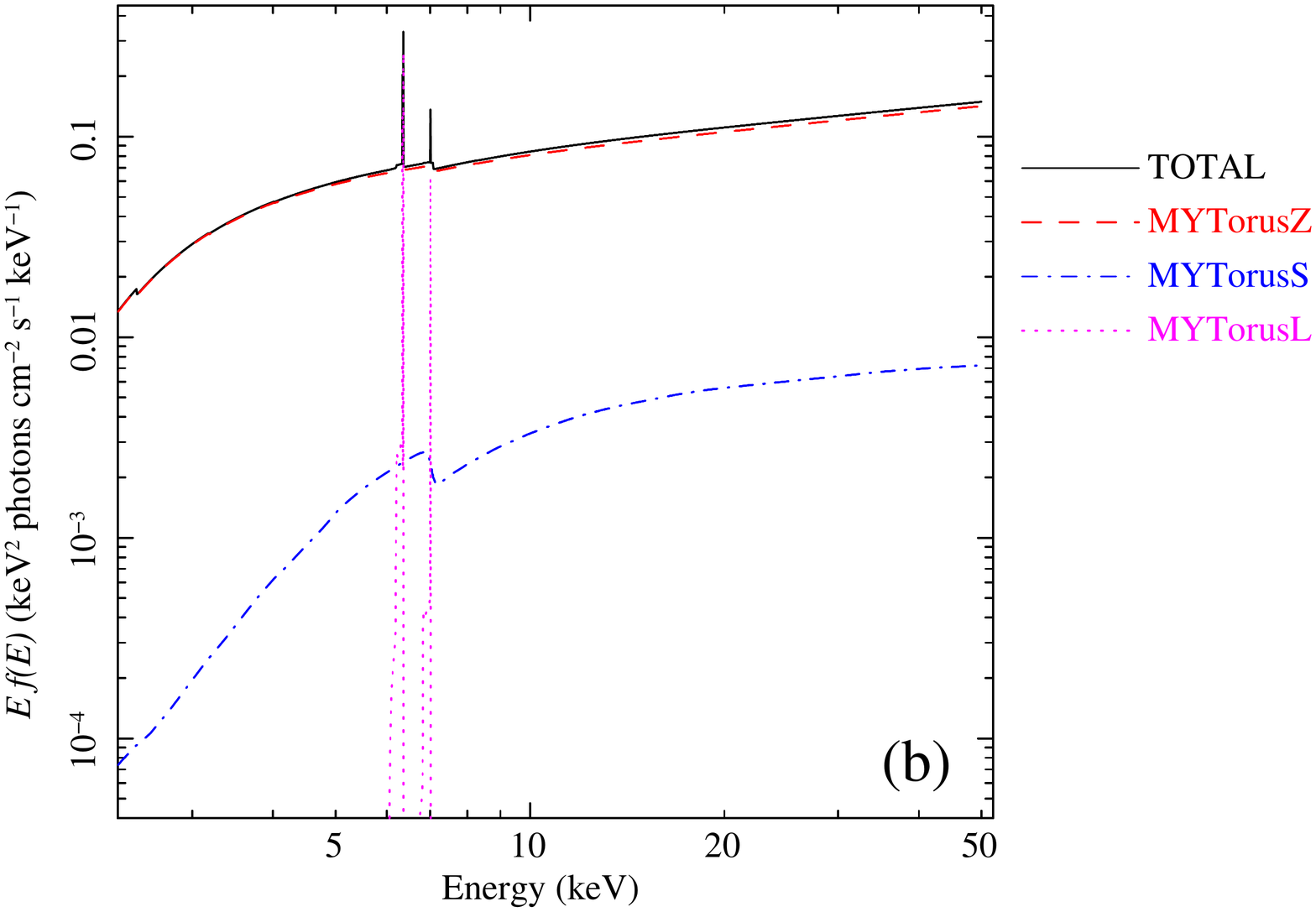}
      \vspace{-0pt}
    \end{minipage}\\
    
    \begin{minipage}[c]{0.5\textwidth}
        \includegraphics[trim=0 -300 0 360,clip,width=1.\textwidth,angle=0]{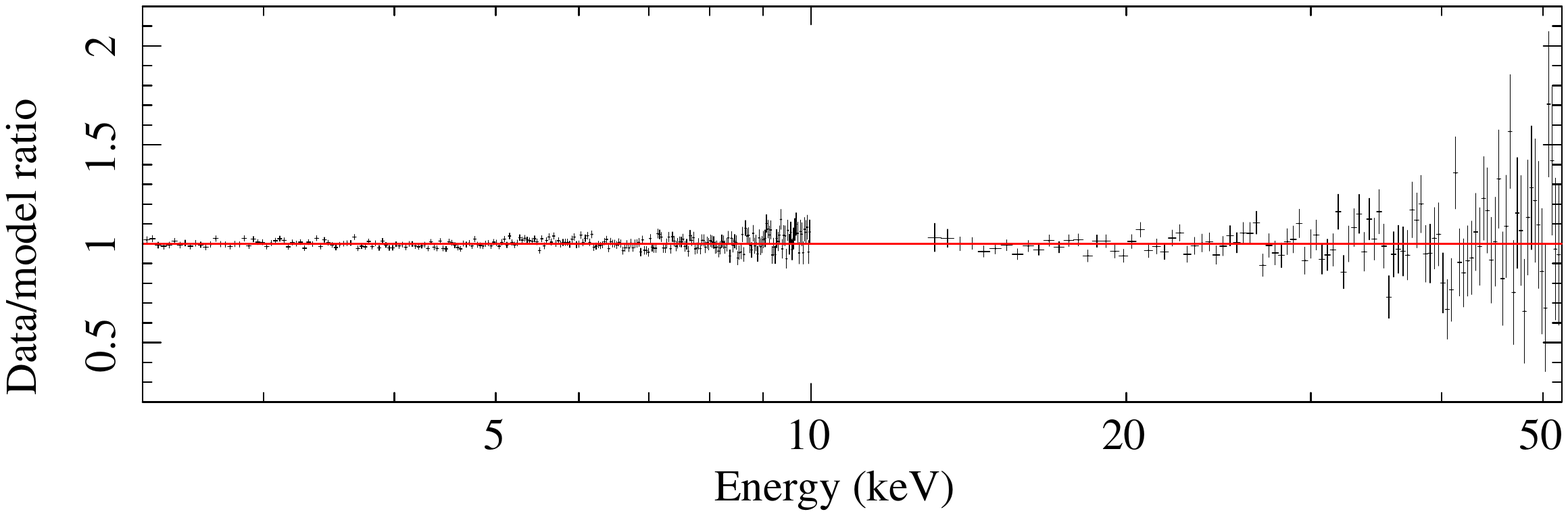}
    \end{minipage}
    \begin{minipage}[c]{0.5\textwidth}
        \includegraphics[trim=0 193 0 80,clip,width=\textwidth,angle=0]{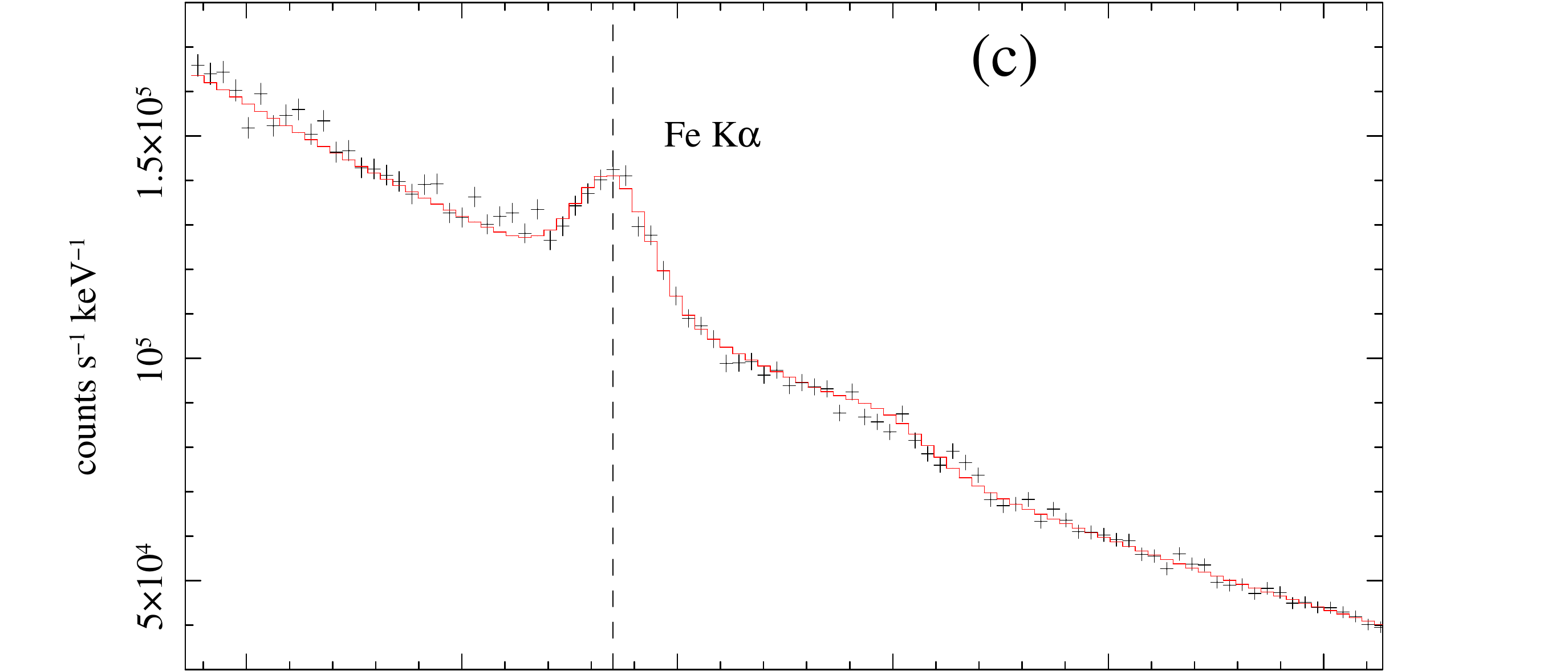}
      \begin{minipage}[c]{\textwidth}
        \includegraphics[trim=0 30 0 360,clip,width=1.\textwidth,angle=0]{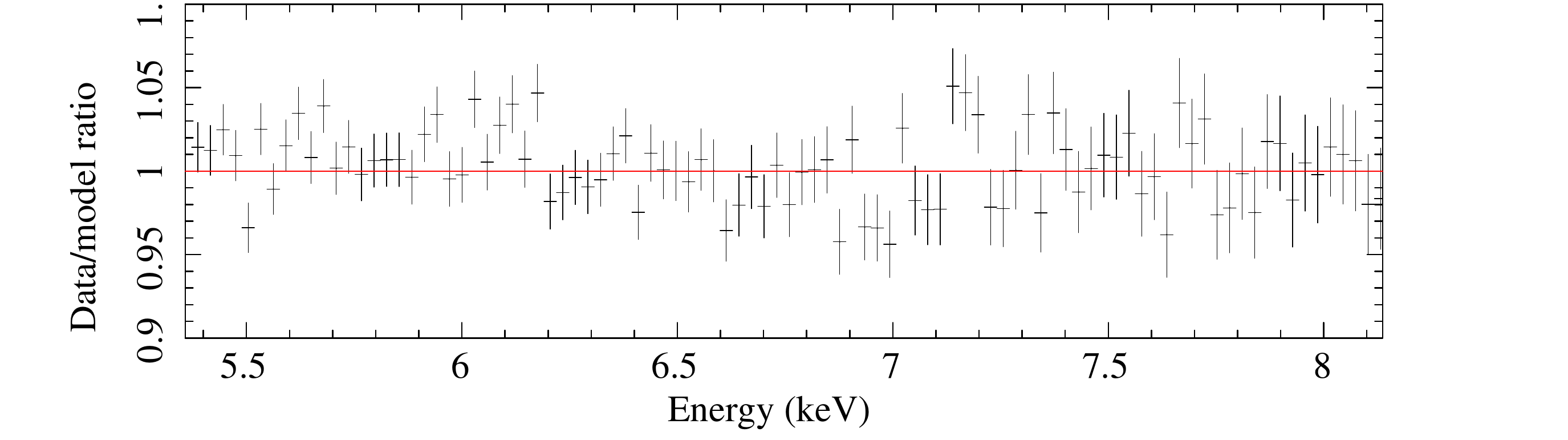}
      \end{minipage}
    \end{minipage}

    \caption{\footnotesize NGC2110 707034010 \label{fig-n2110-707}}
\end{figure*}

\begin{figure*}[t!]
    \begin{minipage}[c]{0.5\textwidth}
      \includegraphics[trim=0 50 0 -200,clip,width=1.\textwidth,angle=0]{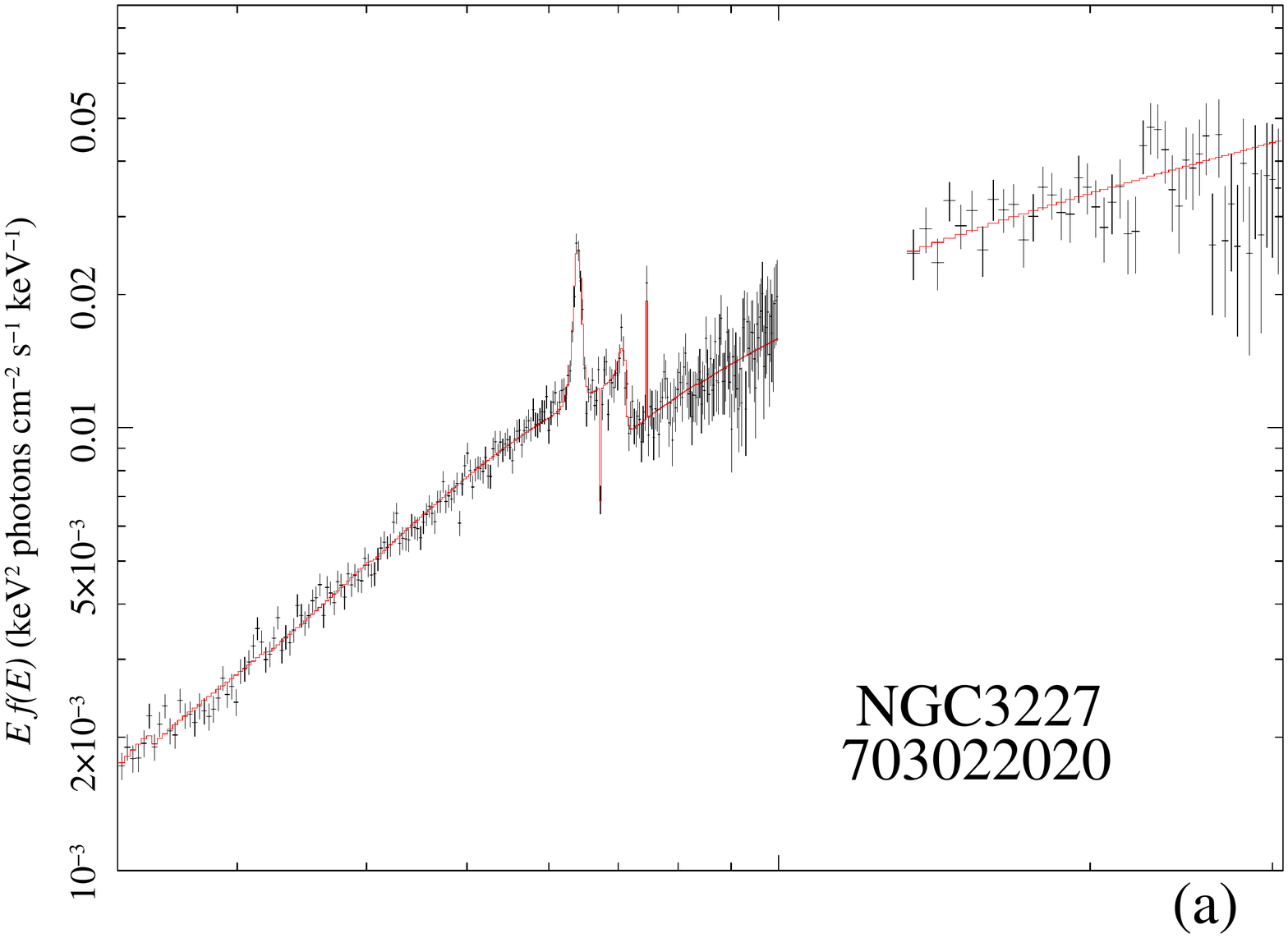}
    \end{minipage}
    \begin{minipage}[c]{0.5\textwidth}\vspace{-0pt}
      \includegraphics[trim=0 30 0 50,clip,width=1.\textwidth,angle=0]{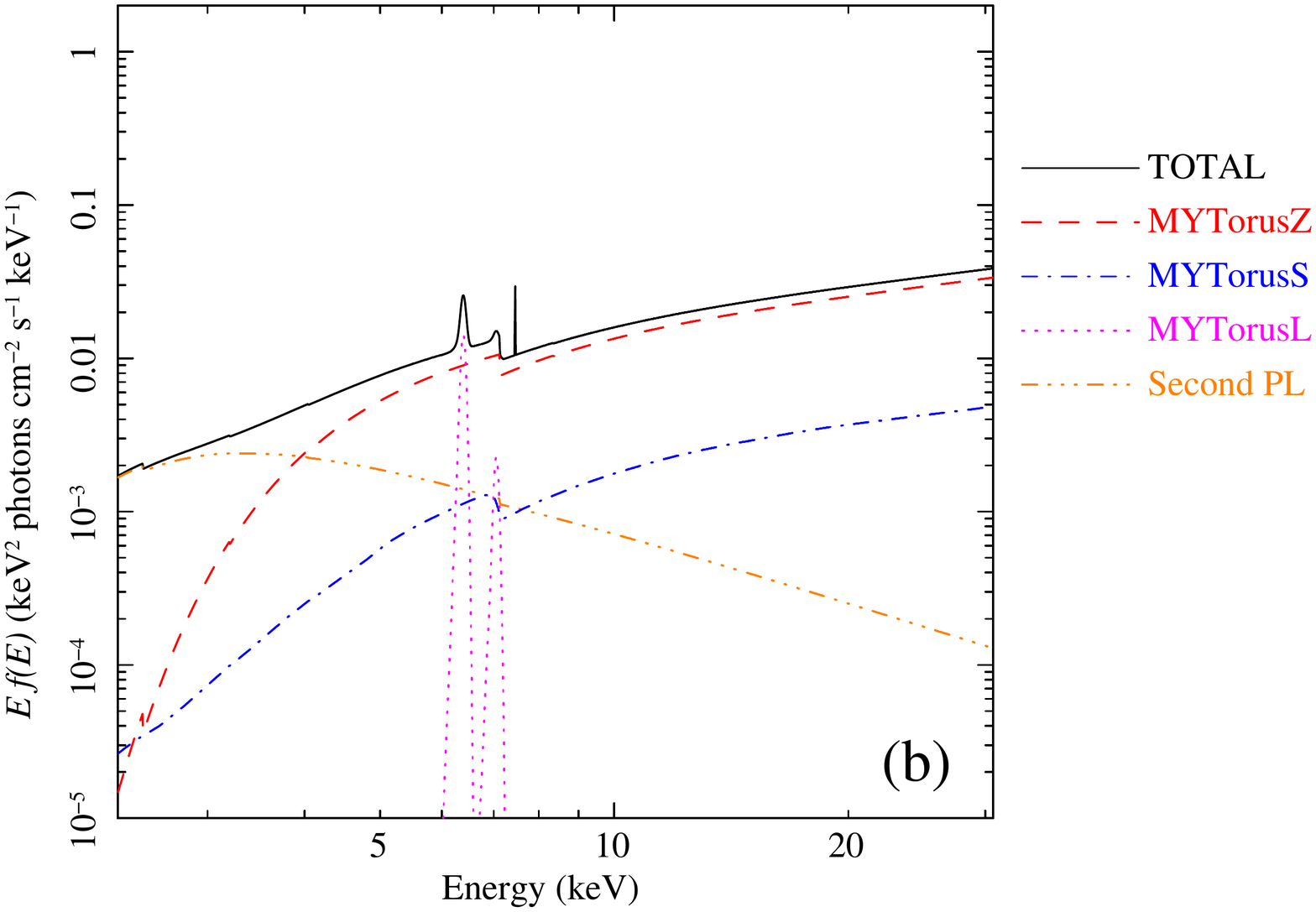}
      \vspace{-0pt}
    \end{minipage}\\
    
    \begin{minipage}[c]{0.5\textwidth}
        \includegraphics[trim=0 -300 0 360,clip,width=1.\textwidth,angle=0]{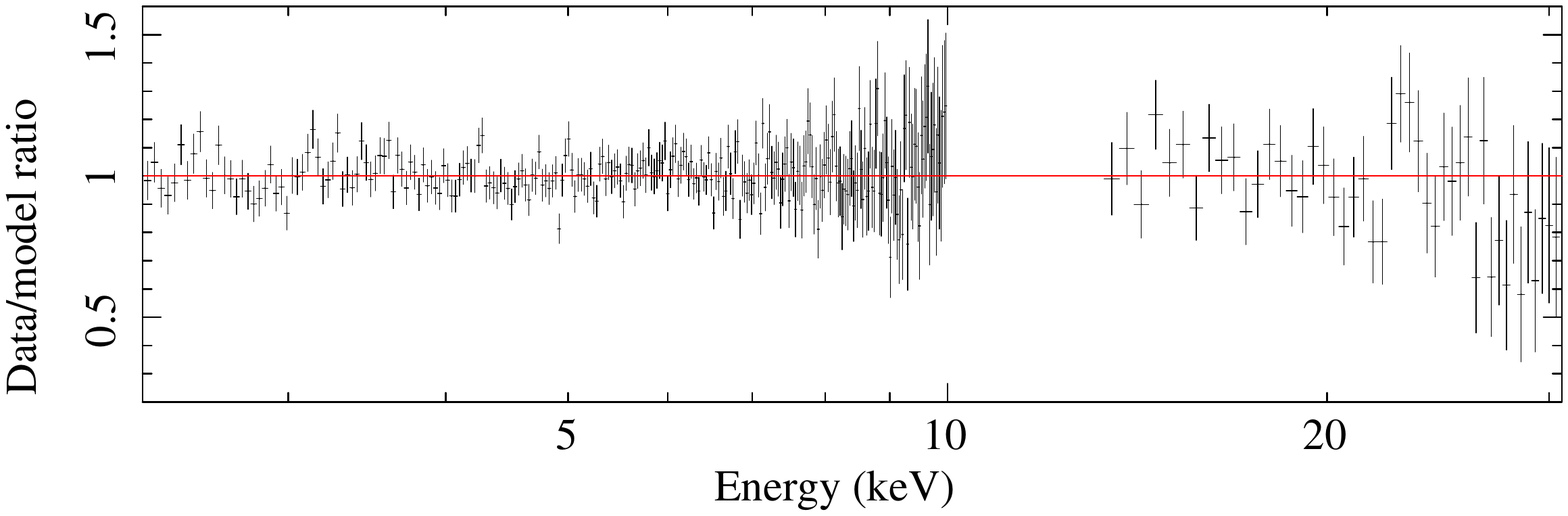}
    \end{minipage}
    \begin{minipage}[c]{0.5\textwidth}
        \includegraphics[trim=0 193 0 80,clip,width=\textwidth,angle=0]{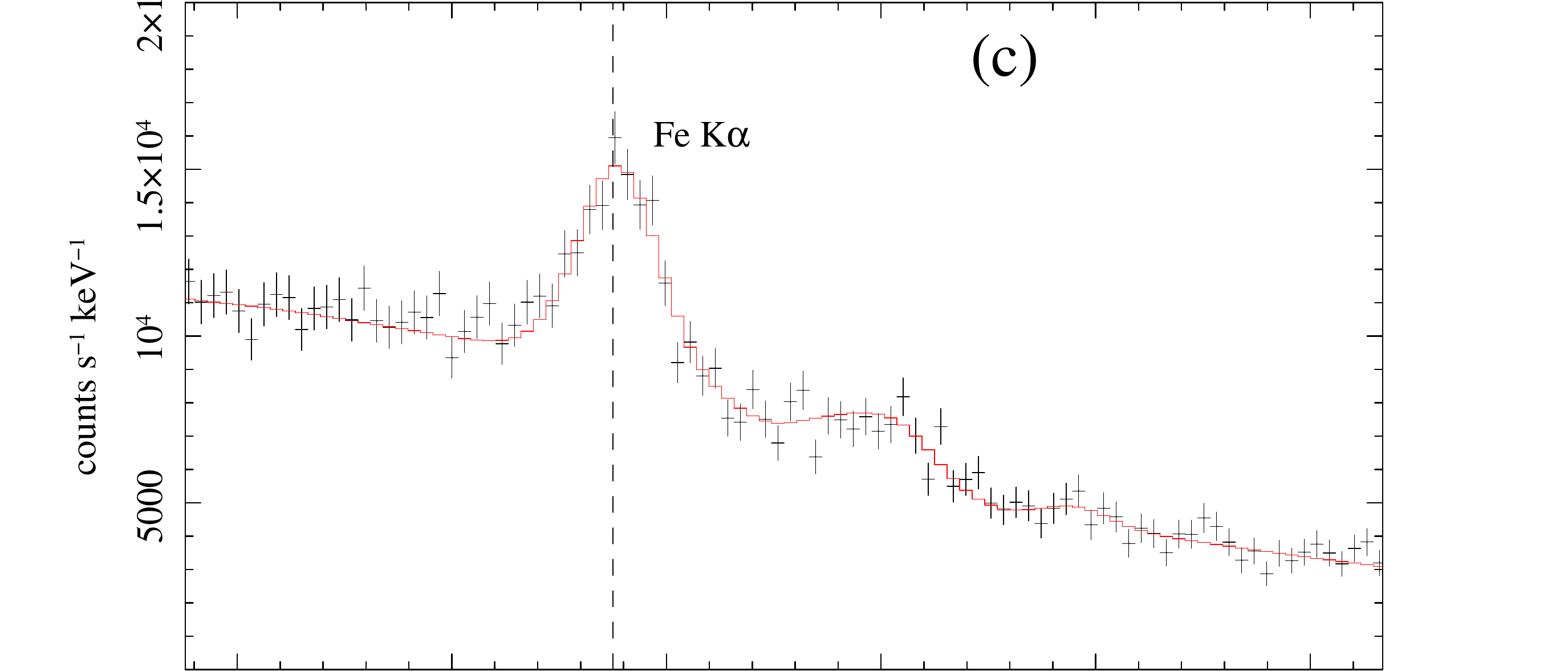}
      \begin{minipage}[c]{\textwidth}
        \includegraphics[trim=0 30 0 360,clip,width=1.\textwidth,angle=0]{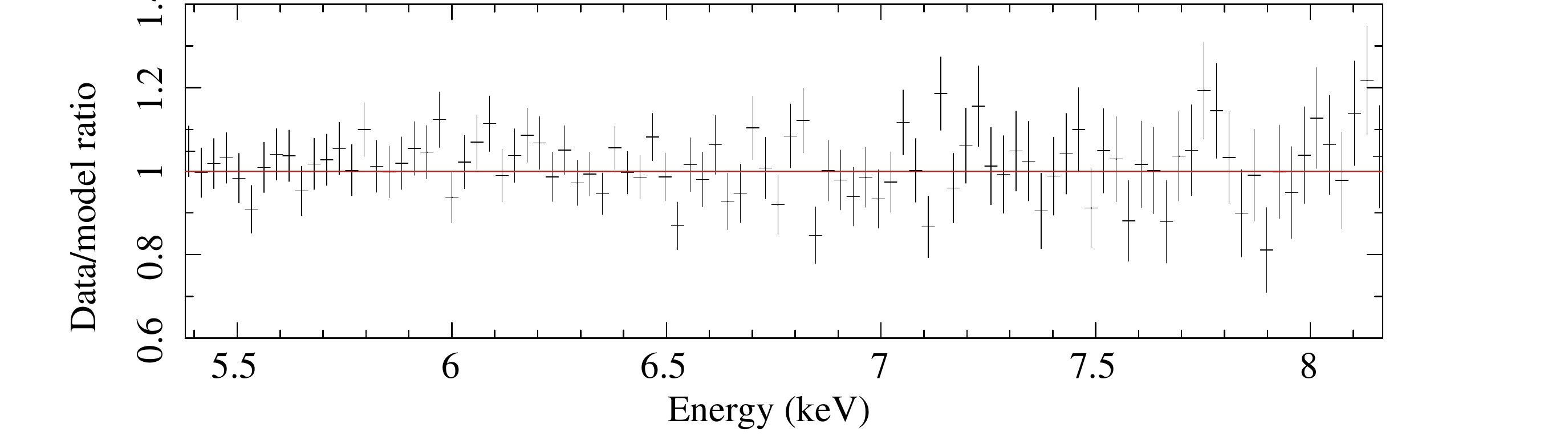}
      \end{minipage}
    \end{minipage}

    \caption{\footnotesize NGC3227 703022020 \label{fig-n3227-20}}
\end{figure*}

\begin{figure*}[t!]
    \begin{minipage}[c]{0.5\textwidth}
      \includegraphics[trim=0 50 0 -200,clip,width=1.\textwidth,angle=0]{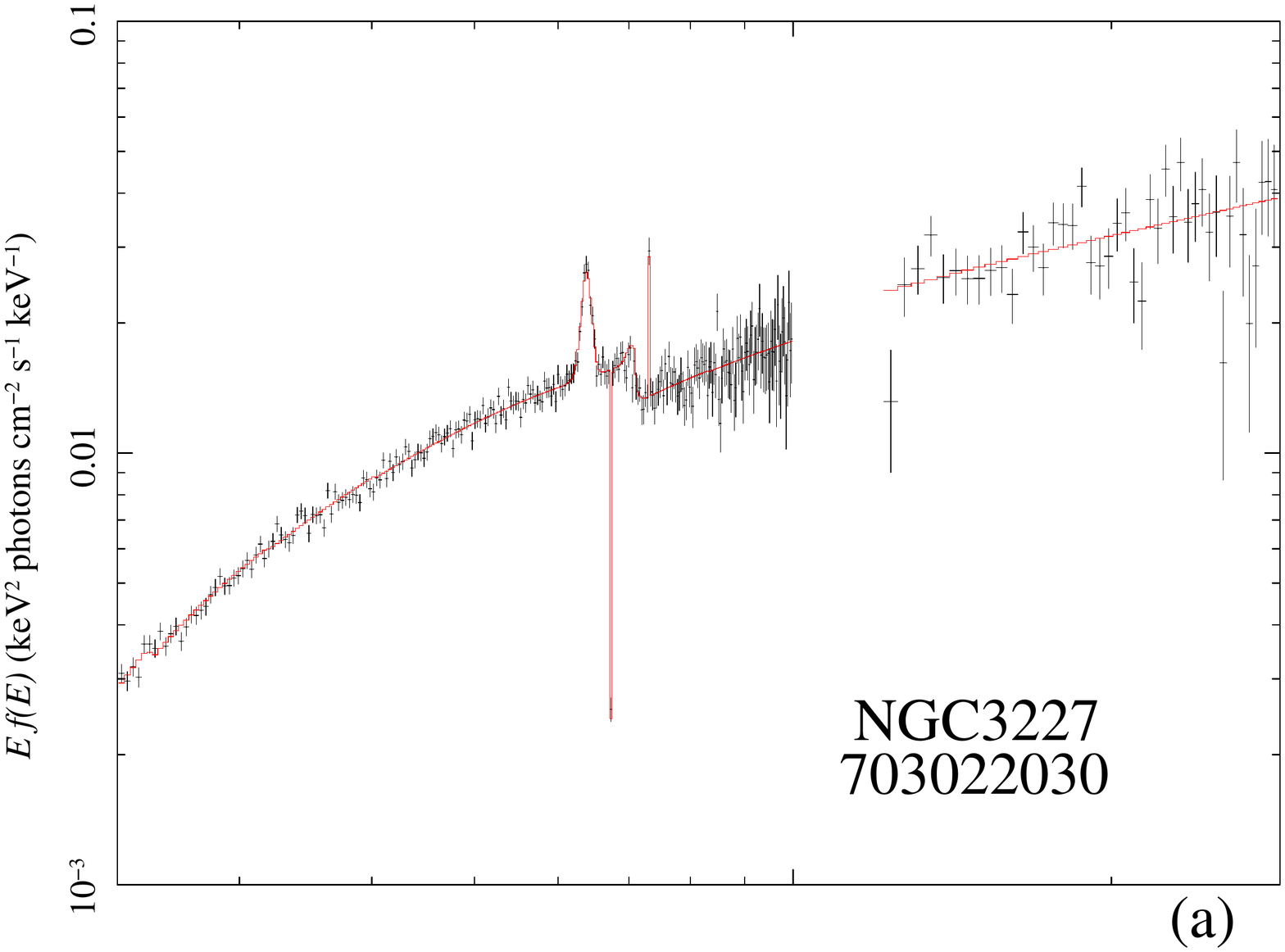}
    \end{minipage}
    \begin{minipage}[c]{0.5\textwidth}\vspace{-0pt}
      \includegraphics[trim=0 30 0 50,clip,width=1.\textwidth,angle=0]{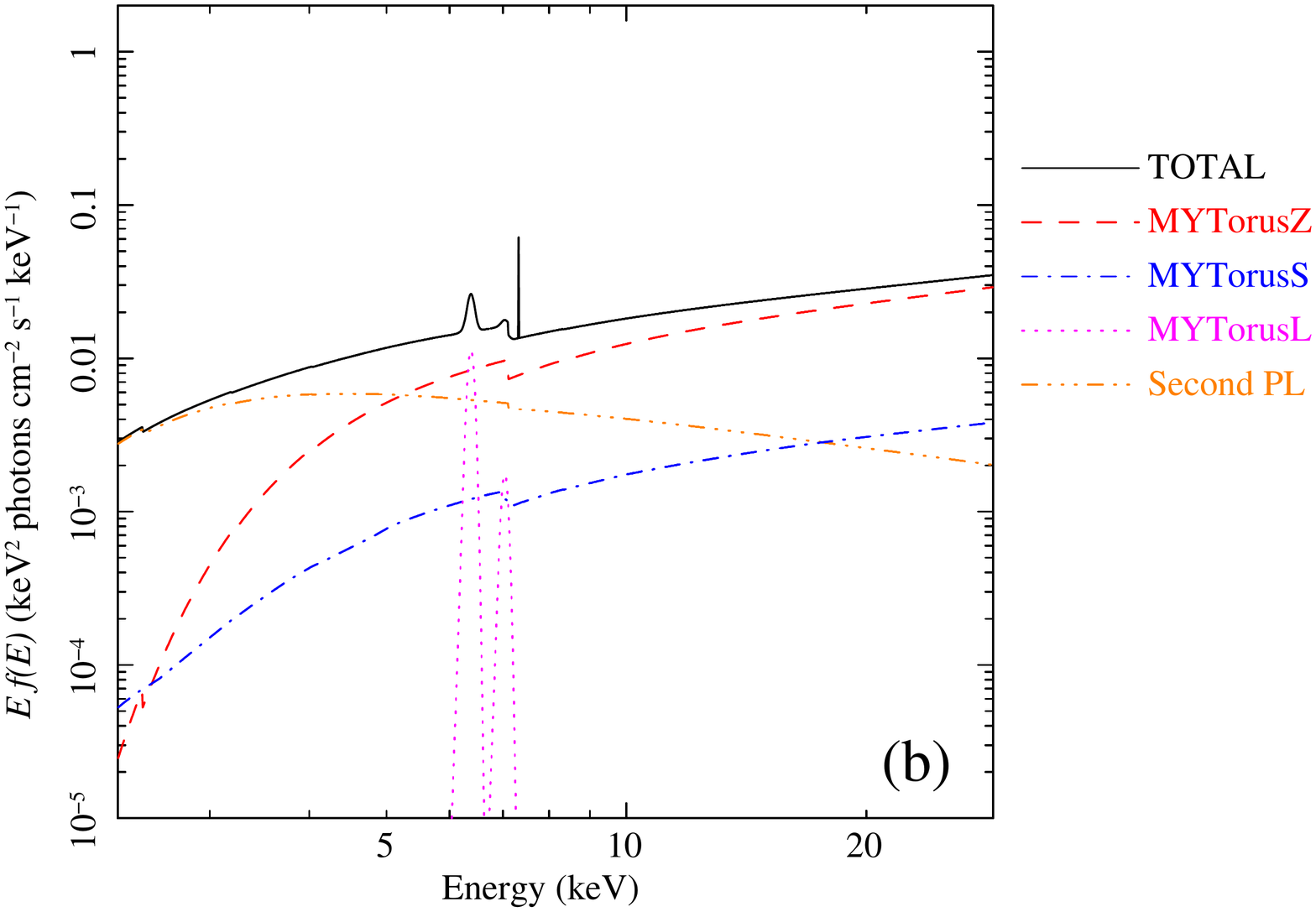}
      \vspace{-0pt}
    \end{minipage}\\
    
    \begin{minipage}[c]{0.5\textwidth}
        \includegraphics[trim=0 -300 0 360,clip,width=1.\textwidth,angle=0]{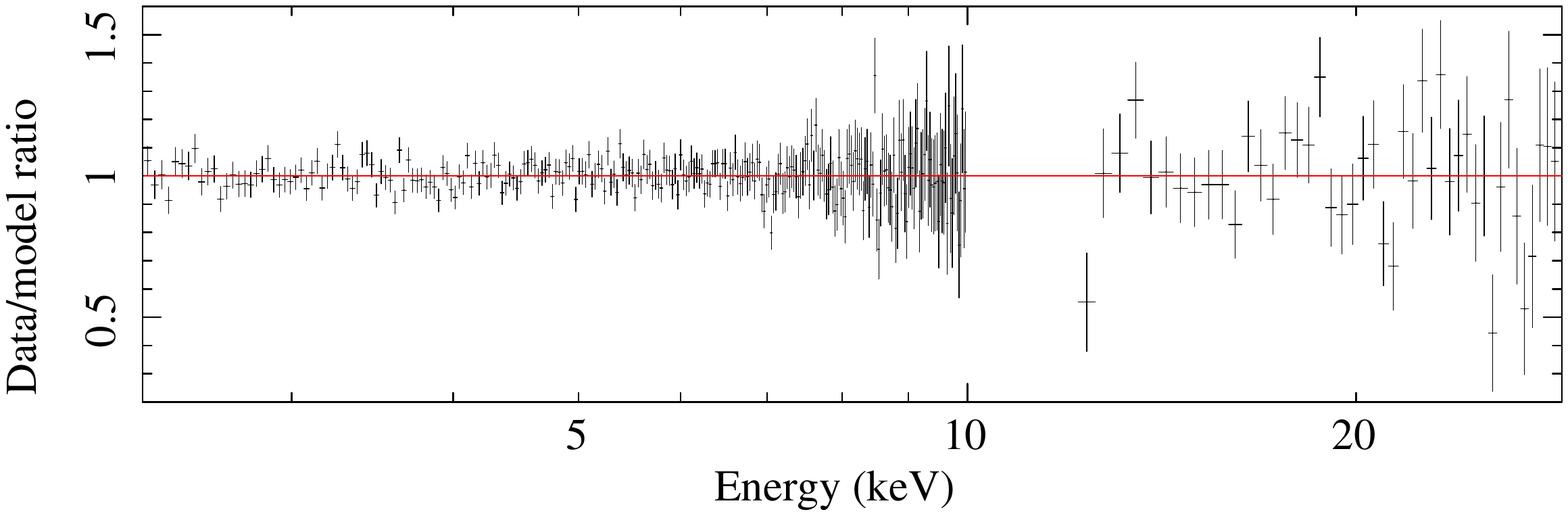}
    \end{minipage}
    \begin{minipage}[c]{0.5\textwidth}
        \includegraphics[trim=0 193 0 80,clip,width=\textwidth,angle=0]{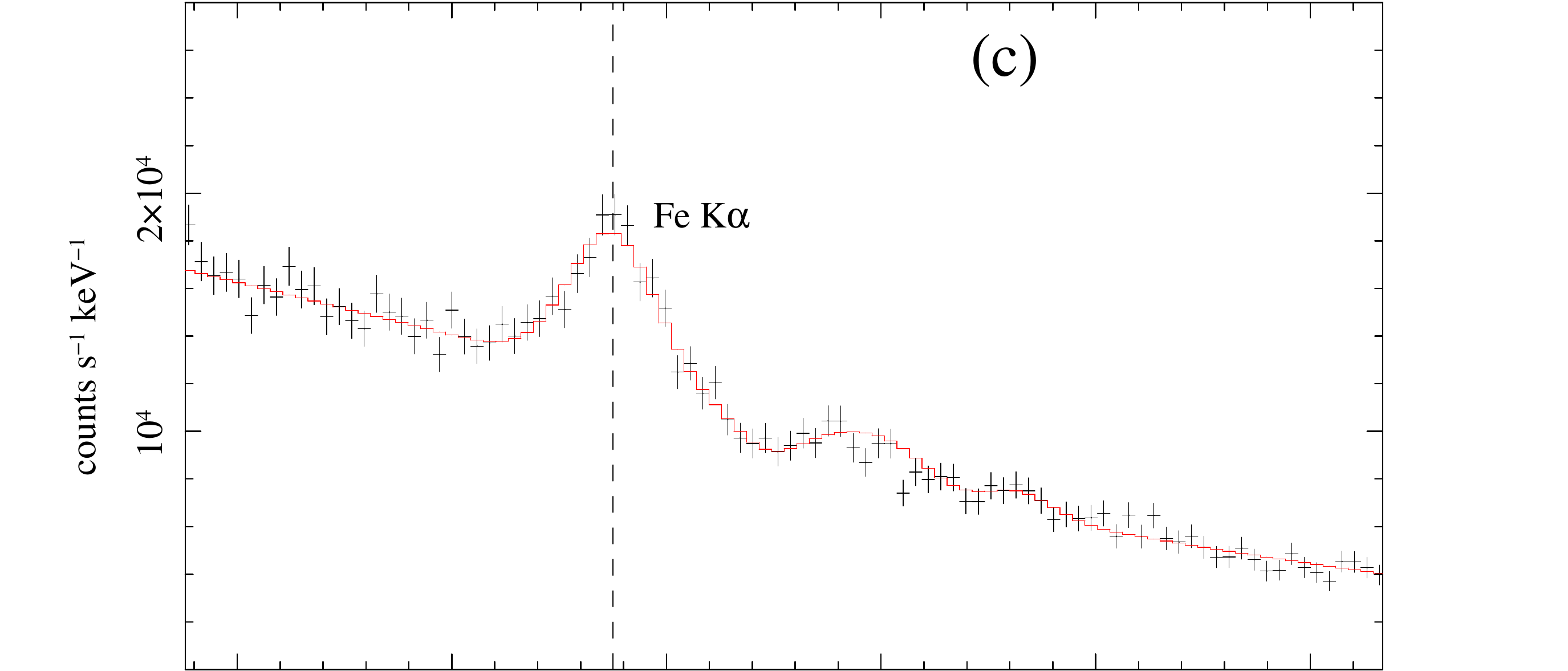}
      \begin{minipage}[c]{\textwidth}
        \includegraphics[trim=0 30 0 360,clip,width=1.\textwidth,angle=0]{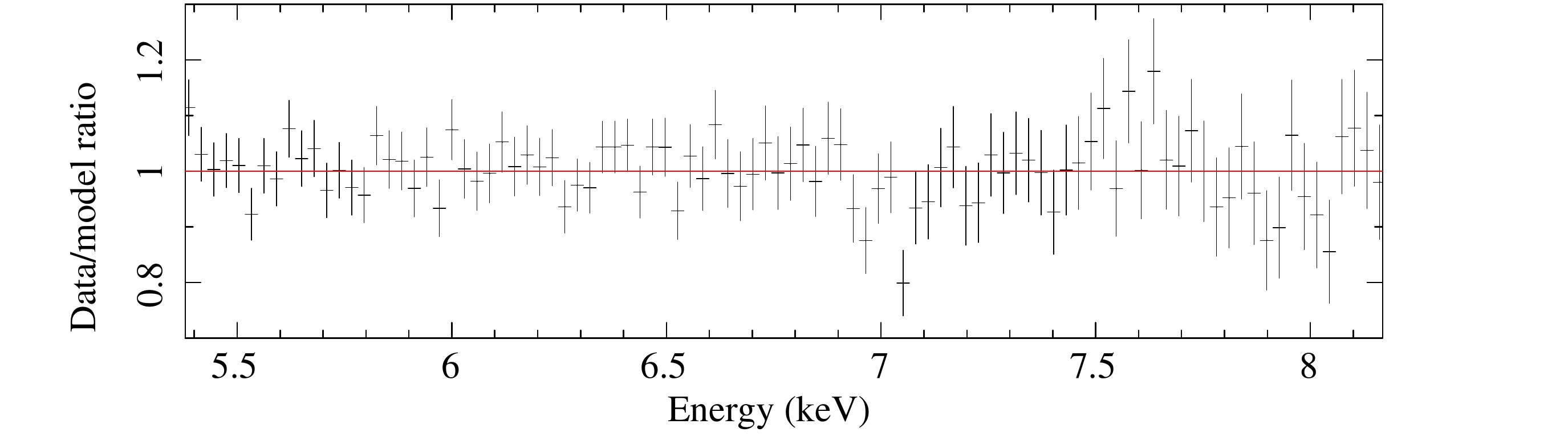}
      \end{minipage}
    \end{minipage}

    \caption{\footnotesize NGC3227 703022030 \label{fig-n3227-30}}
\end{figure*}

\begin{figure*}[t!]
    \begin{minipage}[c]{0.5\textwidth}
      \includegraphics[trim=0 50 0 -200,clip,width=1.\textwidth,angle=0]{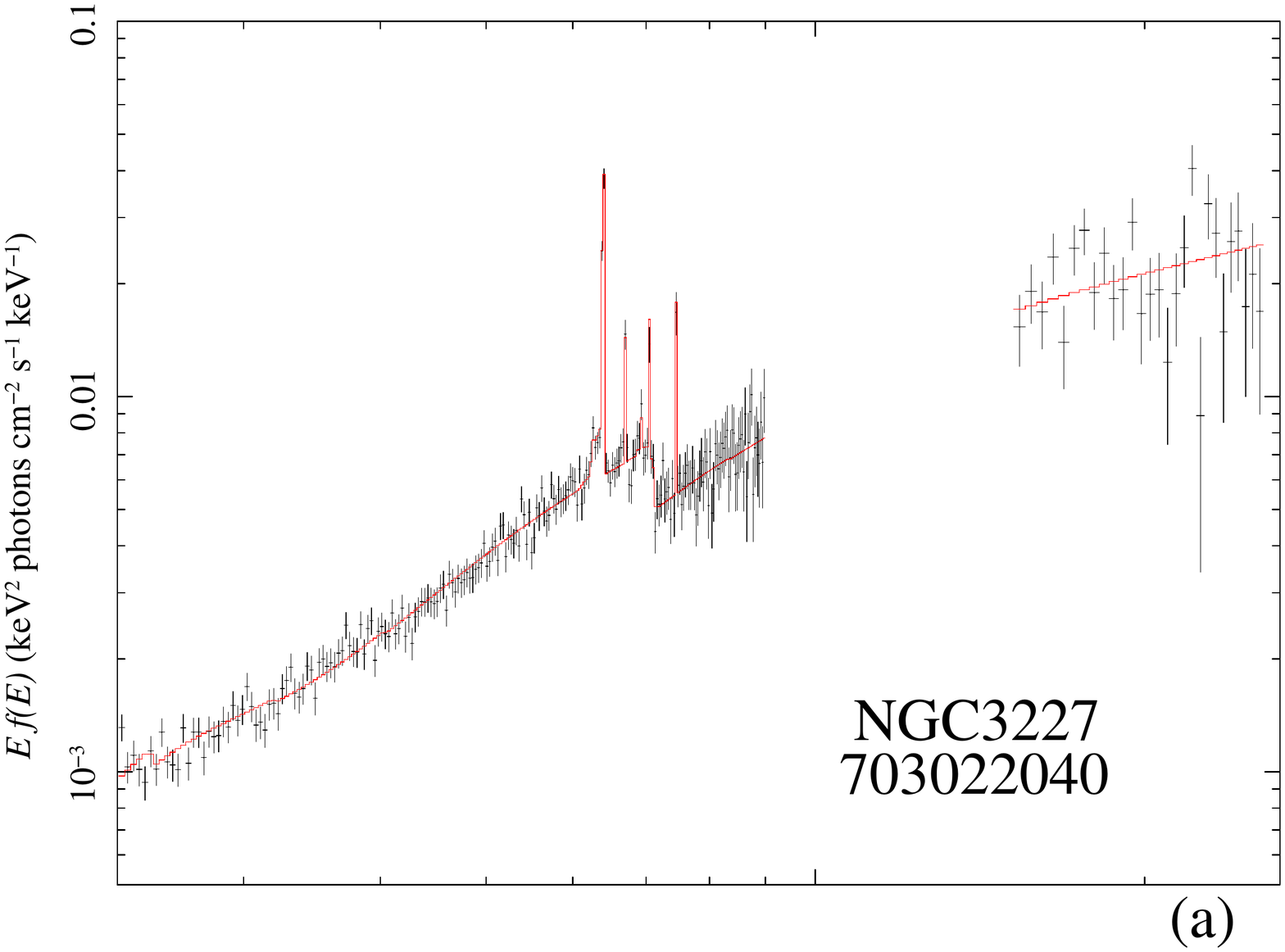}
    \end{minipage}
    \begin{minipage}[c]{0.5\textwidth}\vspace{-0pt}
      \includegraphics[trim=0 30 0 50,clip,width=1.\textwidth,angle=0]{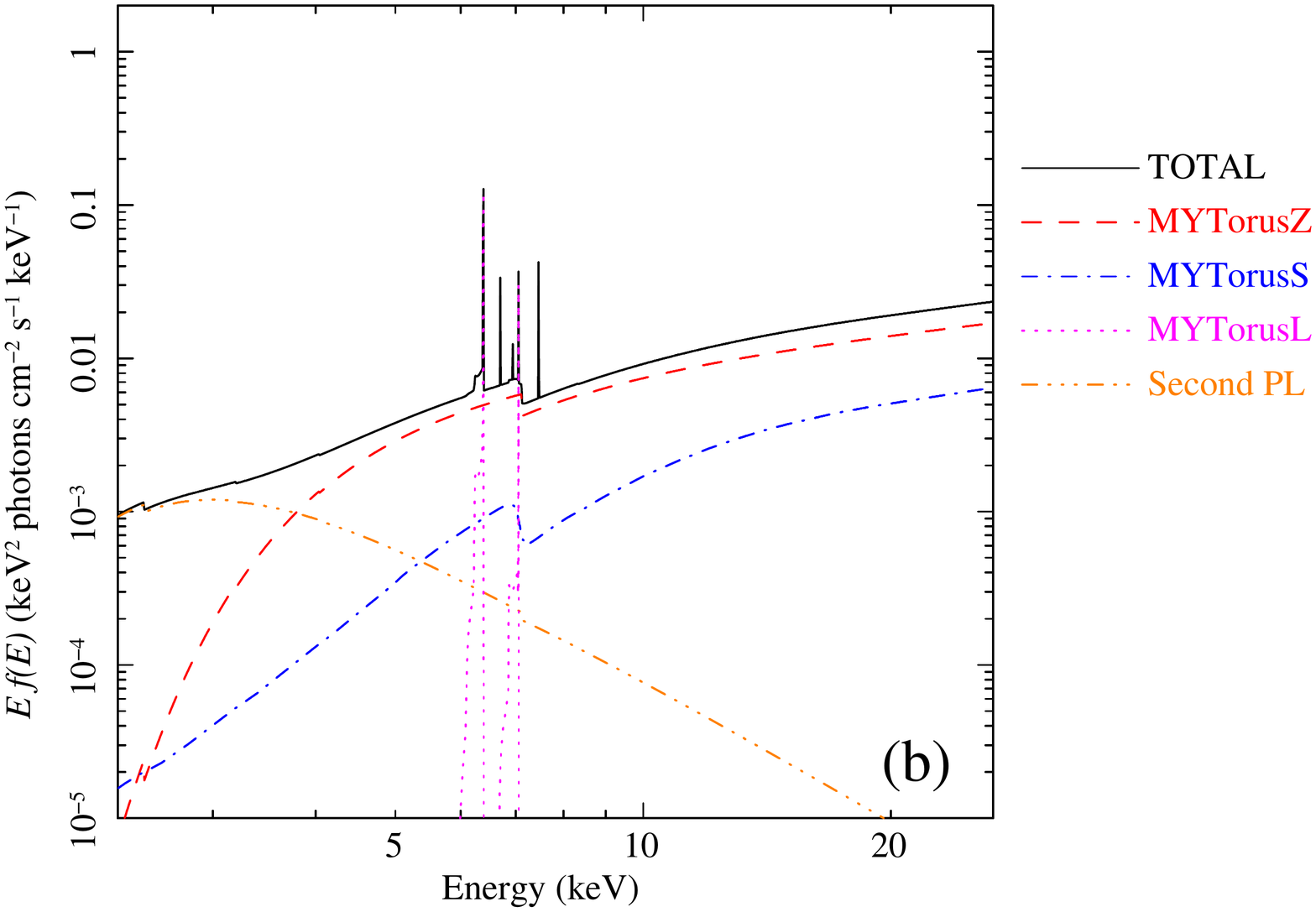}
      \vspace{-0pt}
    \end{minipage}\\
    
    \begin{minipage}[c]{0.5\textwidth}
        \includegraphics[trim=0 -300 0 360,clip,width=1.\textwidth,angle=0]{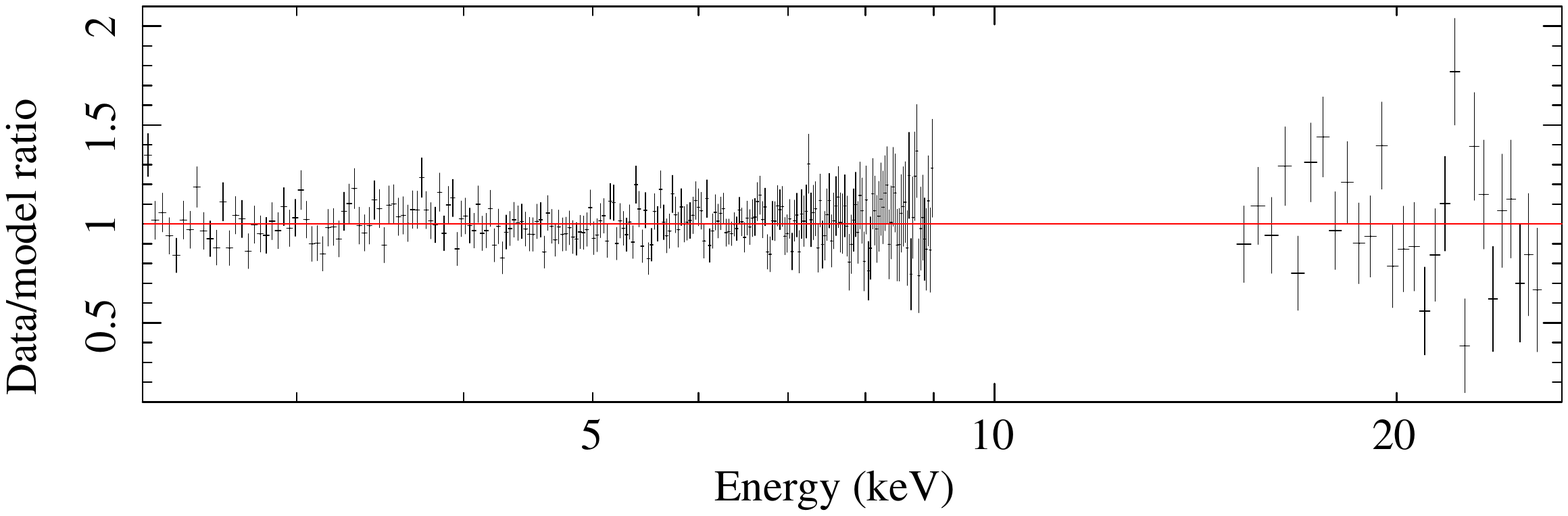}
    \end{minipage}
    \begin{minipage}[c]{0.5\textwidth}
        \includegraphics[trim=0 193 0 80,clip,width=\textwidth,angle=0]{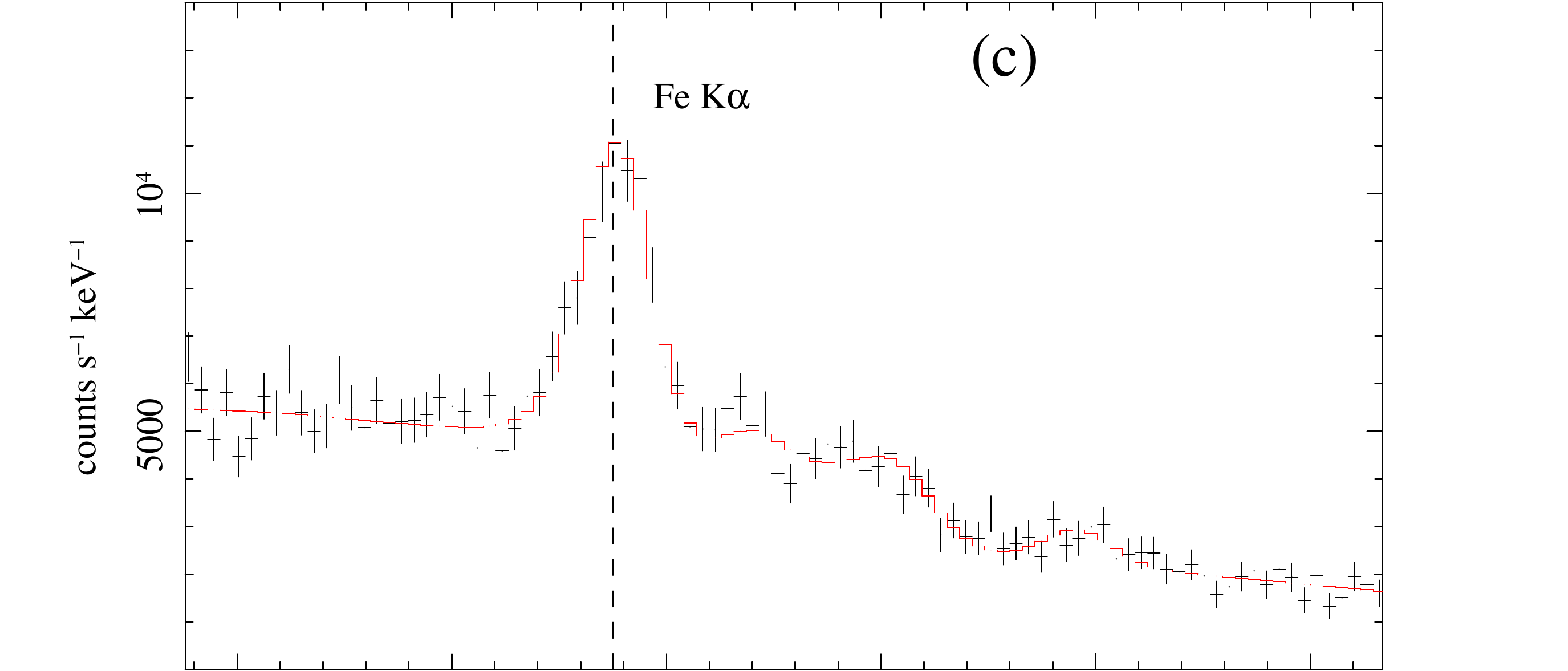}
      \begin{minipage}[c]{\textwidth}
        \includegraphics[trim=0 30 0 360,clip,width=1.\textwidth,angle=0]{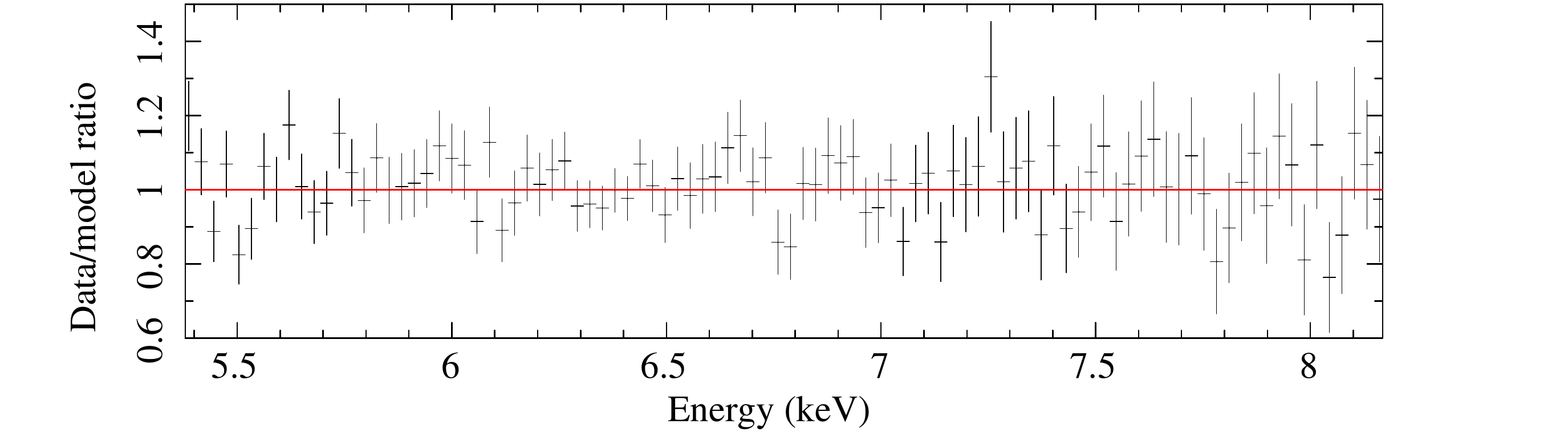}
      \end{minipage}
    \end{minipage}

    \caption{\footnotesize NGC3227 703022040 \label{fig-n3227-40}}
\end{figure*}

\begin{figure*}[t!]
    \begin{minipage}[c]{0.5\textwidth}
      \includegraphics[trim=0 50 0 -200,clip,width=1.\textwidth,angle=0]{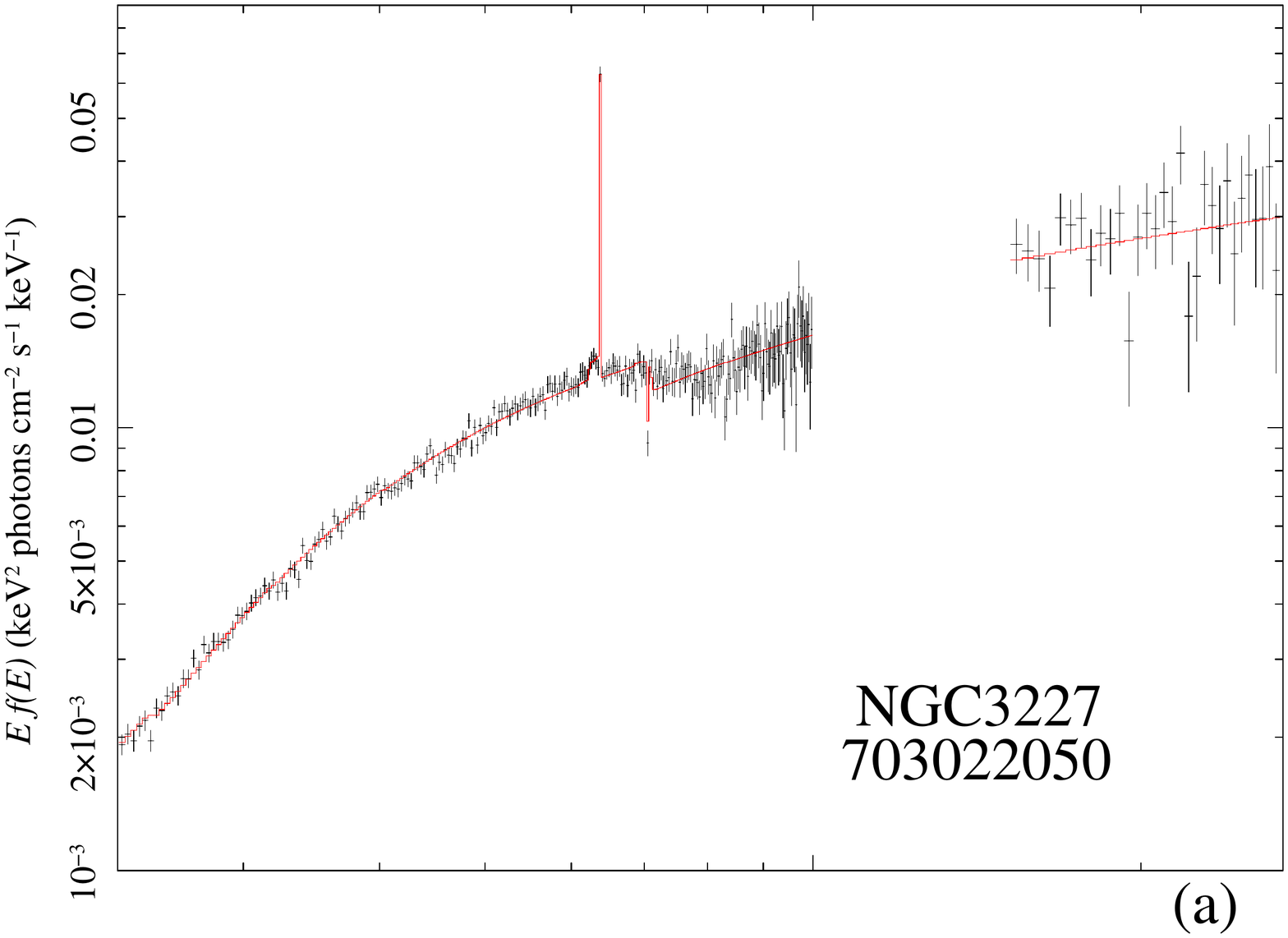}
    \end{minipage}
    \begin{minipage}[c]{0.5\textwidth}\vspace{-0pt}
      \includegraphics[trim=0 30 0 50,clip,width=1.\textwidth,angle=0]{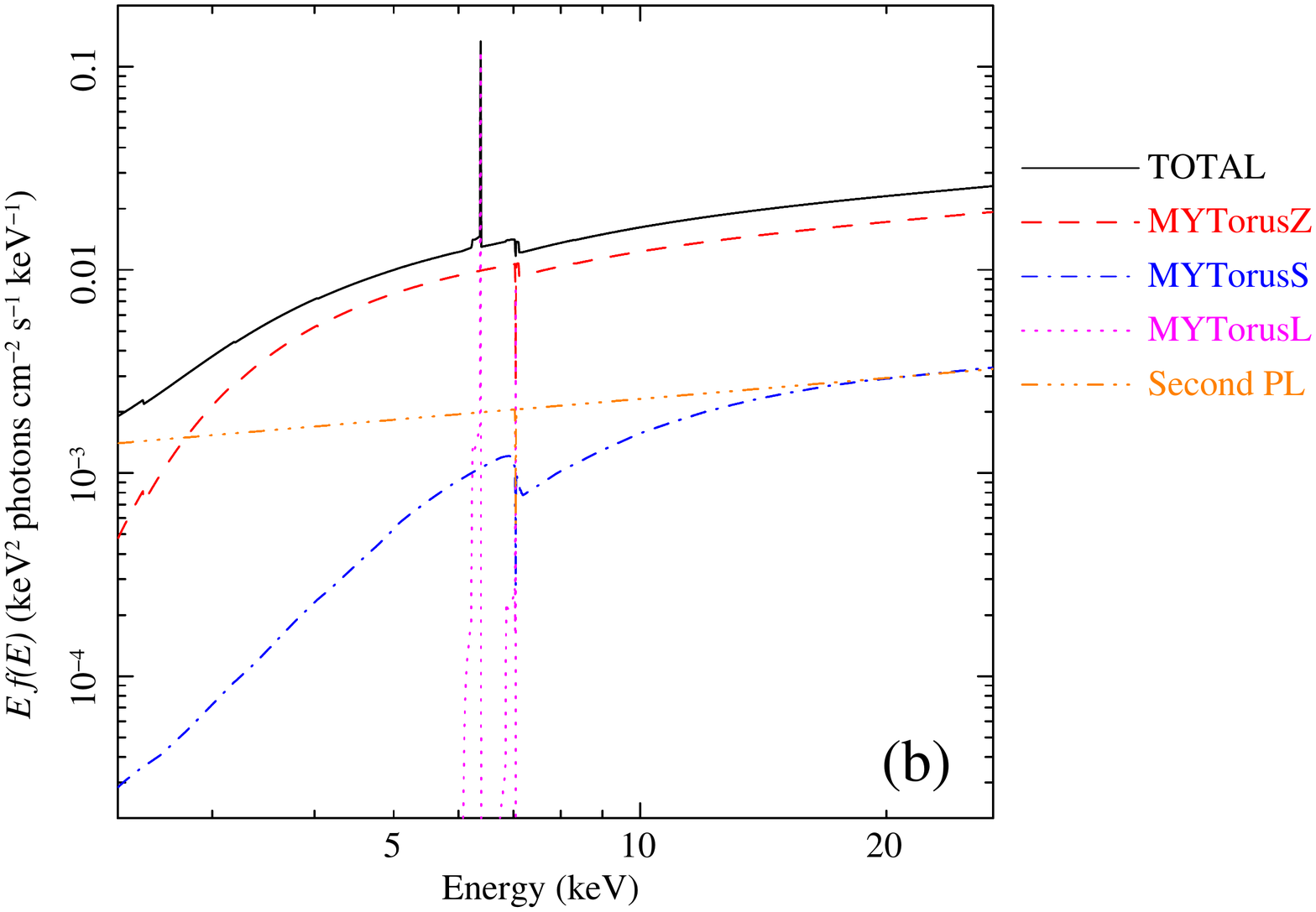}
      \vspace{-0pt}
    \end{minipage}\\
    
    \begin{minipage}[c]{0.5\textwidth}
        \includegraphics[trim=0 -300 0 360,clip,width=1.\textwidth,angle=0]{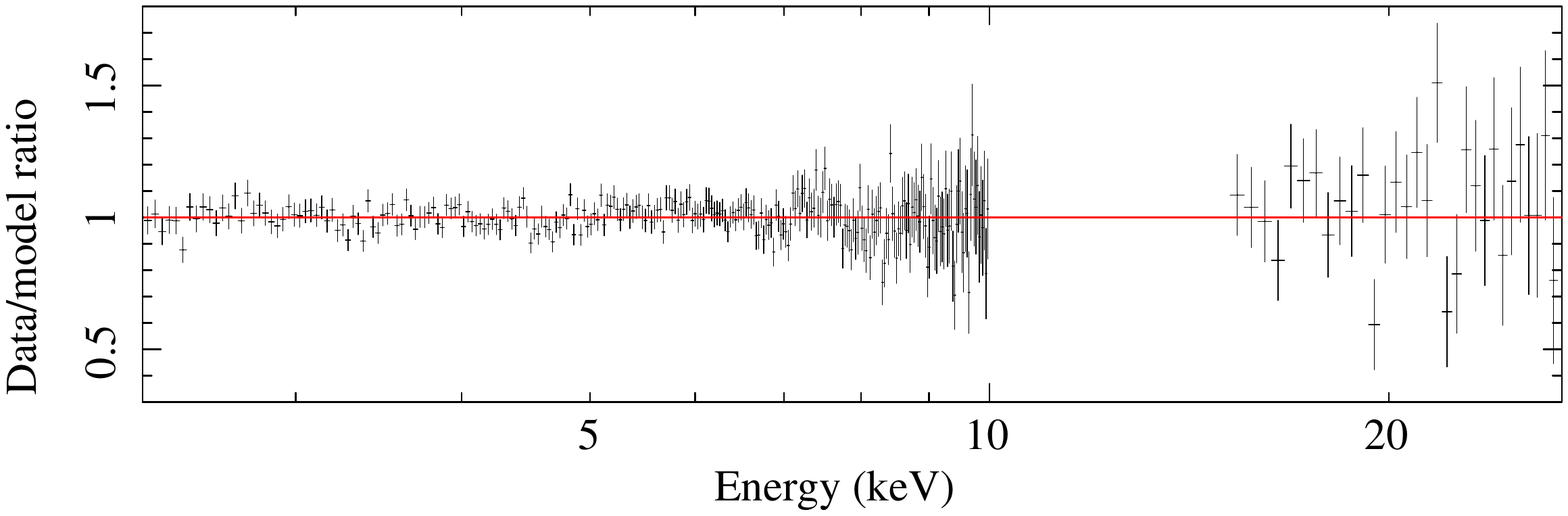}
    \end{minipage}
    \begin{minipage}[c]{0.5\textwidth}
        \includegraphics[trim=0 193 0 80,clip,width=\textwidth,angle=0]{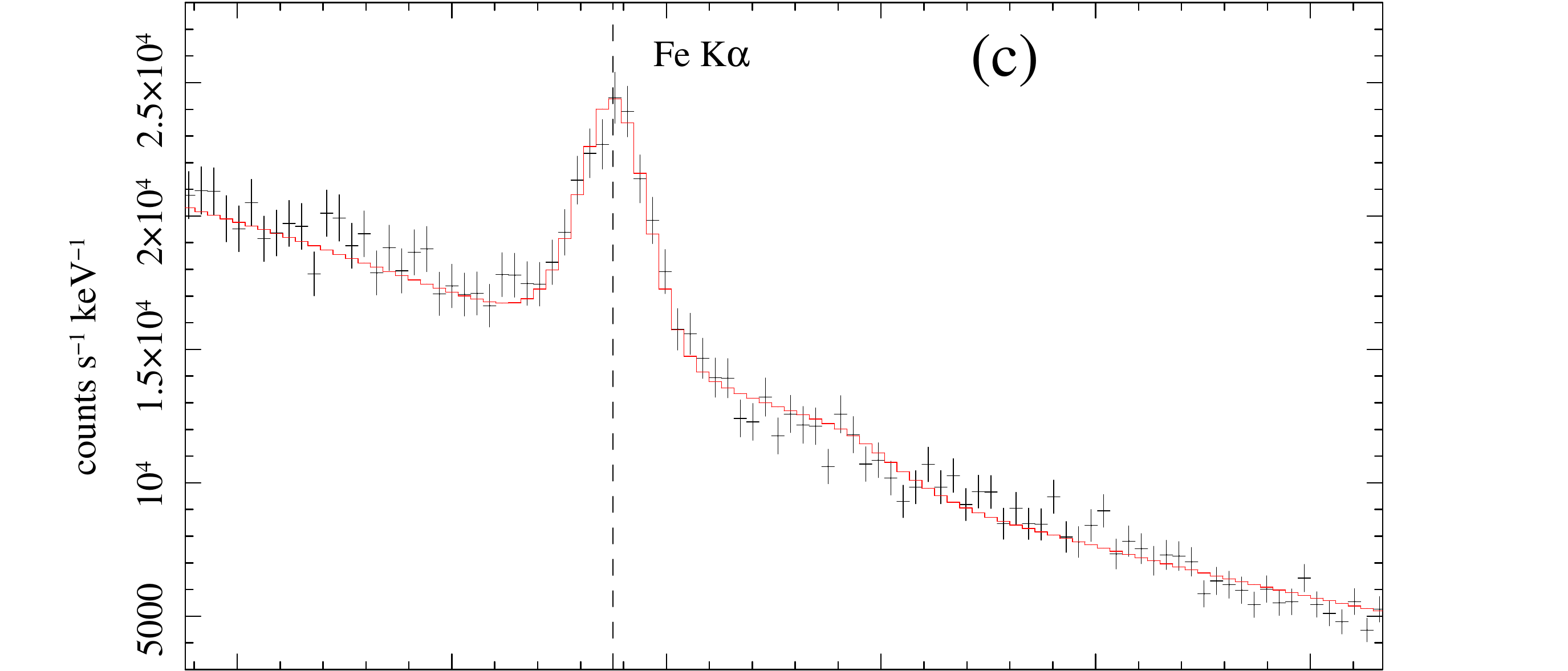}
      \begin{minipage}[c]{\textwidth}
        \includegraphics[trim=0 30 0 360,clip,width=1.\textwidth,angle=0]{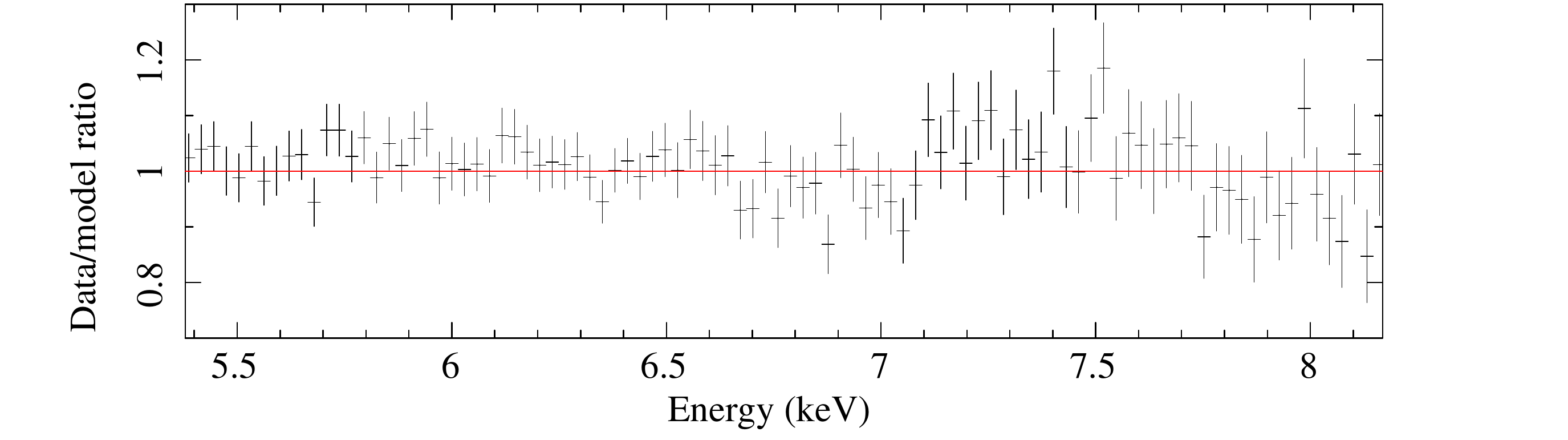}
      \end{minipage}
    \end{minipage}

    \caption{\footnotesize NGC3227 703022050 \label{fig-n3227-50}}
\end{figure*}

\begin{figure*}[t!]
    \begin{minipage}[c]{0.5\textwidth}
      \includegraphics[trim=0 50 0 -200,clip,width=1.\textwidth,angle=0]{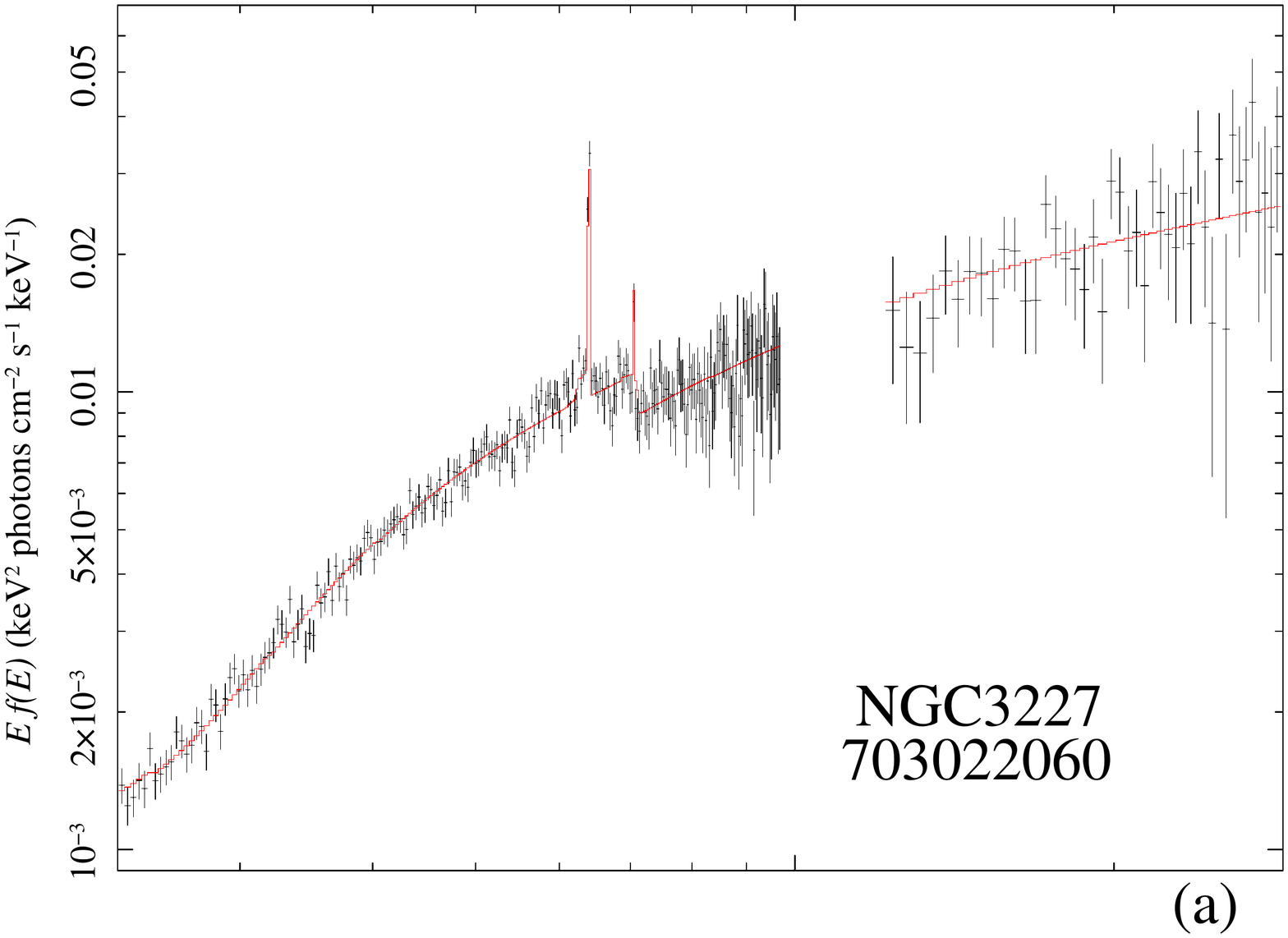}
    \end{minipage}
    \begin{minipage}[c]{0.5\textwidth}\vspace{-0pt}
      \includegraphics[trim=0 30 0 50,clip,width=1.\textwidth,angle=0]{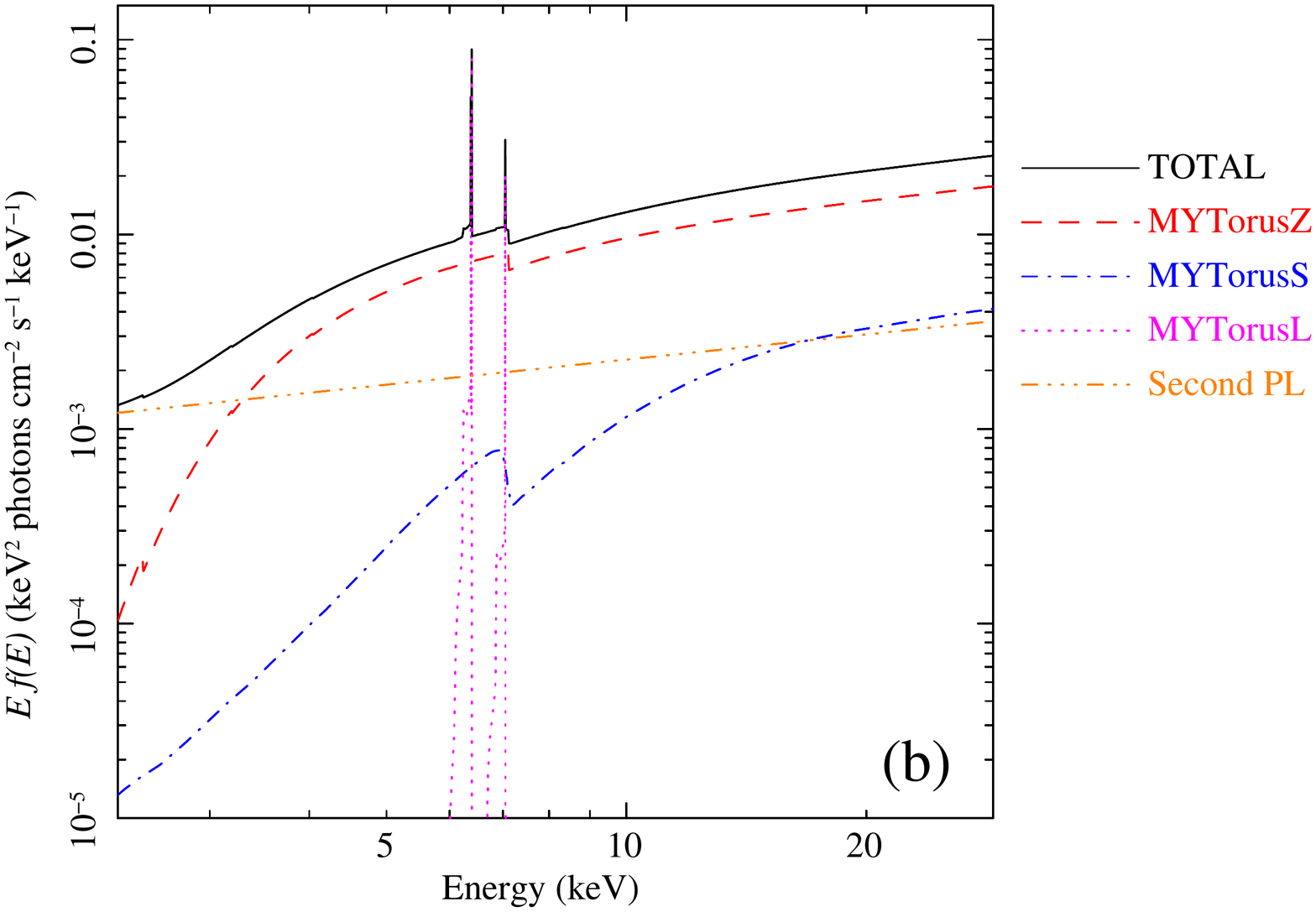}
      \vspace{-0pt}
    \end{minipage}\\
    
    \begin{minipage}[c]{0.5\textwidth}
        \includegraphics[trim=0 -300 0 360,clip,width=1.\textwidth,angle=0]{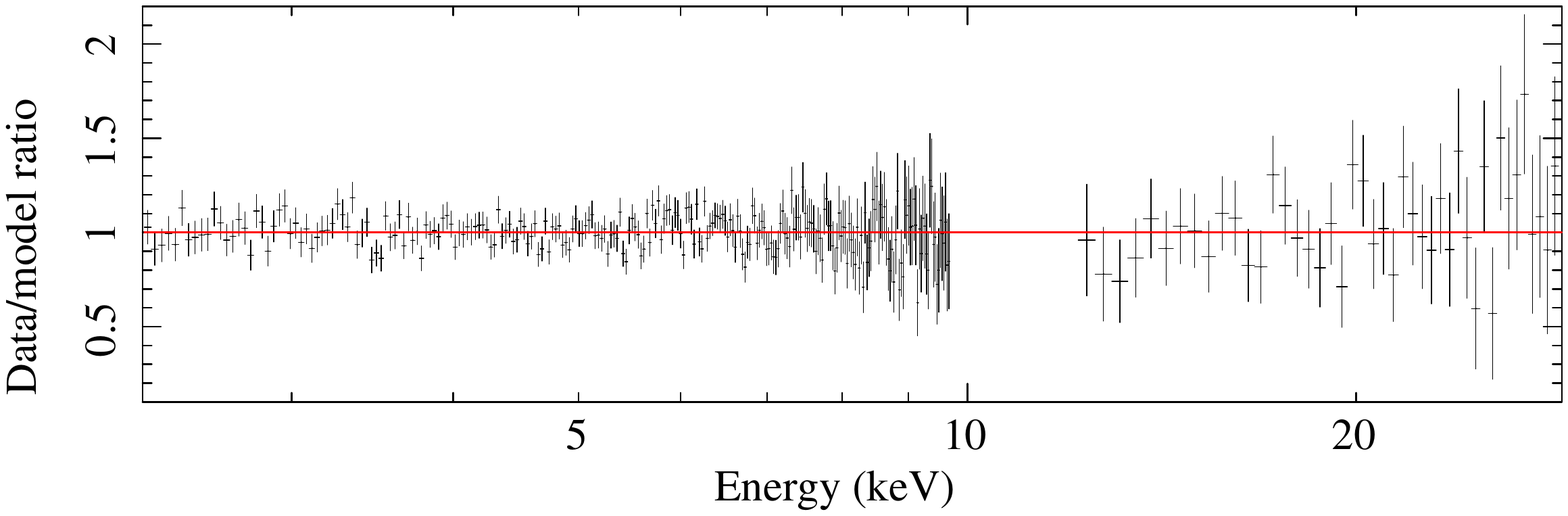}
    \end{minipage}
    \begin{minipage}[c]{0.5\textwidth}
        \includegraphics[trim=0 193 0 80,clip,width=\textwidth,angle=0]{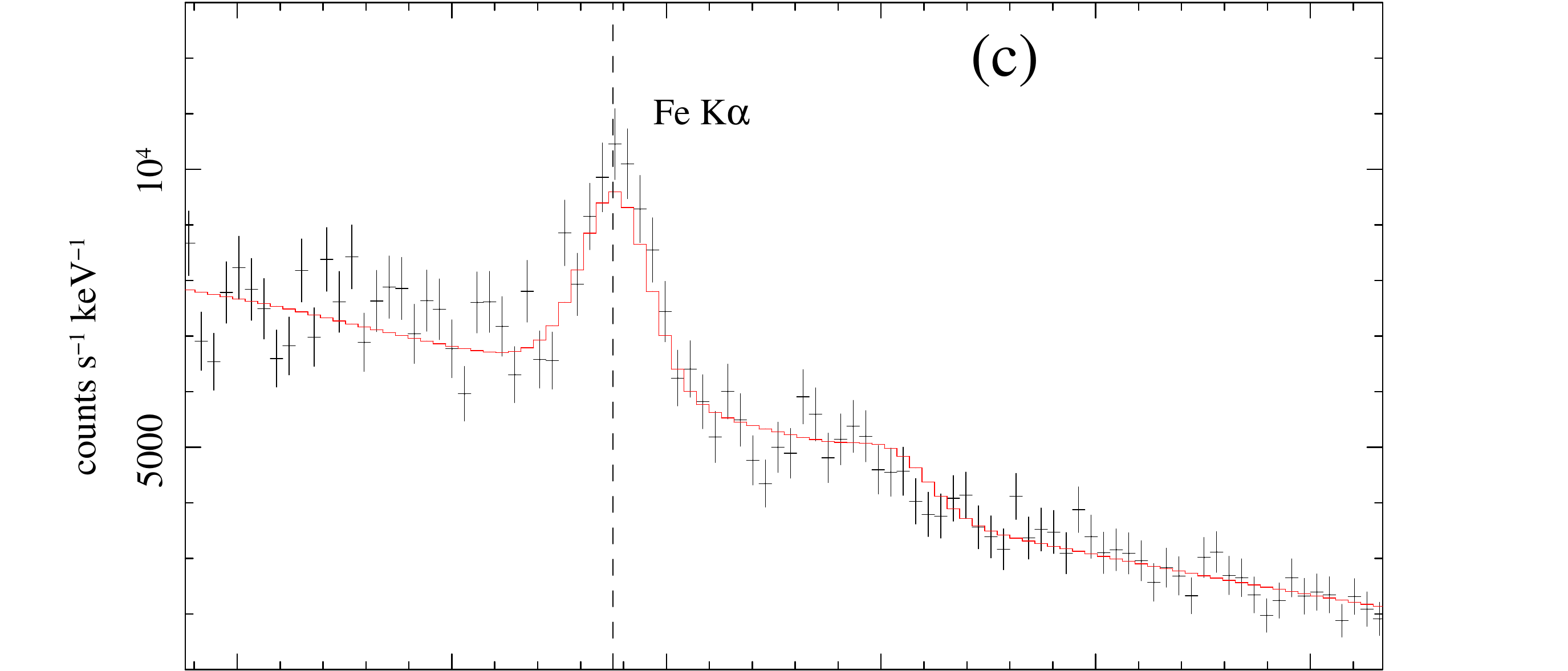}
      \begin{minipage}[c]{\textwidth}
        \includegraphics[trim=0 30 0 360,clip,width=1.\textwidth,angle=0]{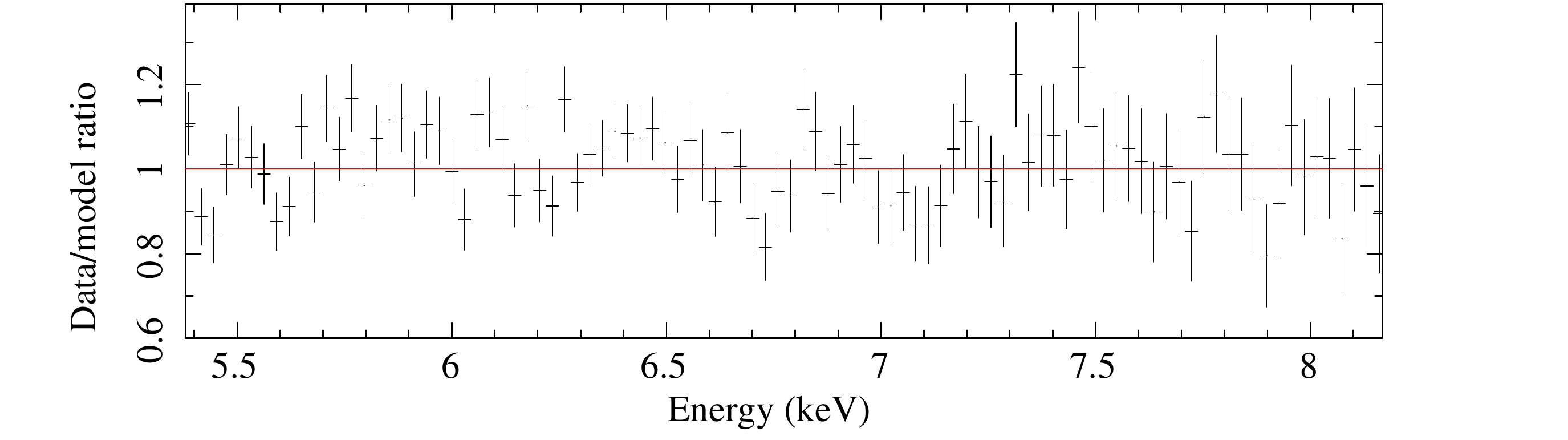}
      \end{minipage}
    \end{minipage}

    \caption{\footnotesize NGC3227 703022060 \label{fig-n3227-60}}
\end{figure*}

\begin{figure*}[t!]
    \begin{minipage}[c]{0.5\textwidth}
      \includegraphics[trim=0 50 0 -200,clip,width=1.\textwidth,angle=0]{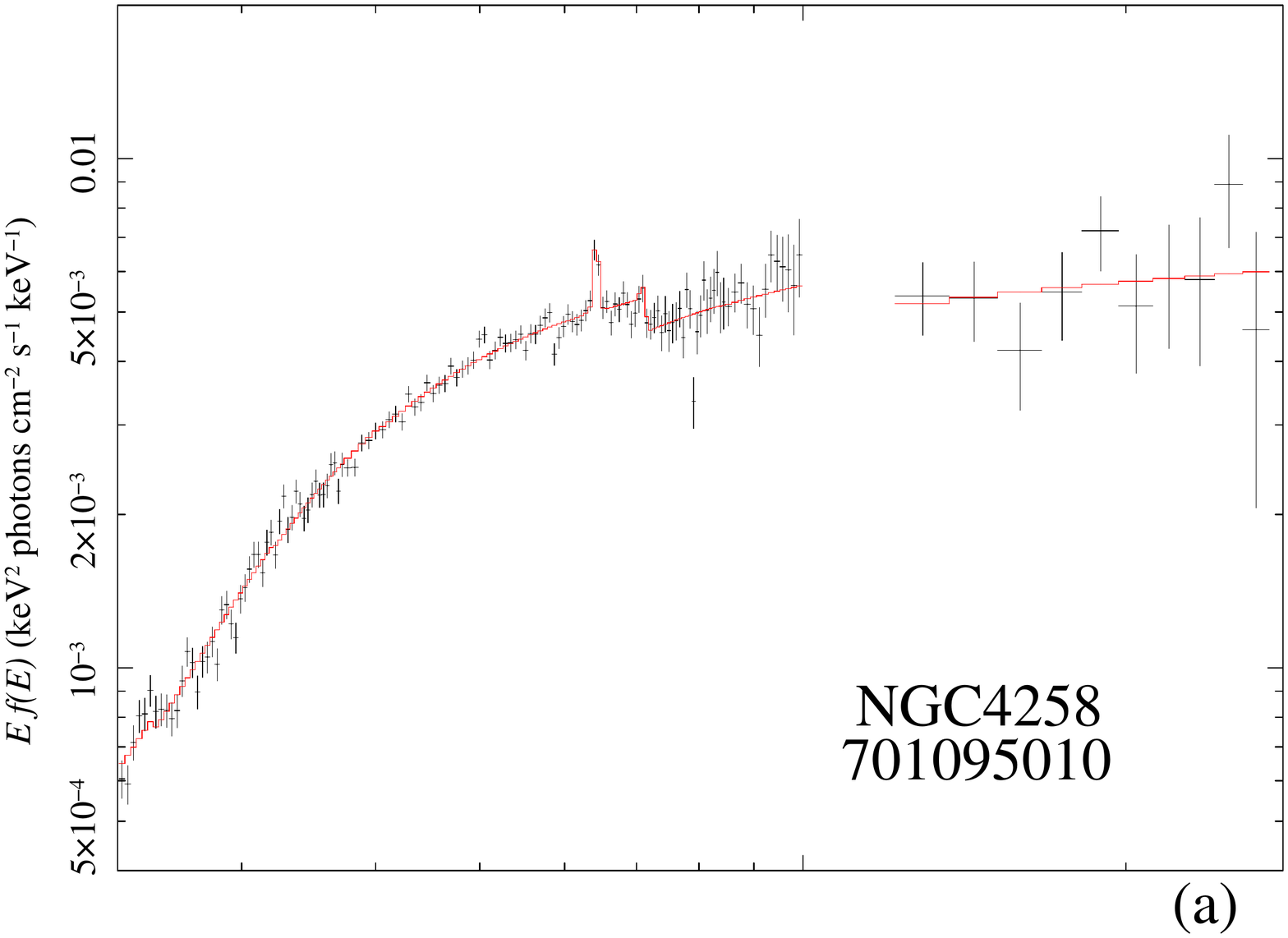}
    \end{minipage}
    \begin{minipage}[c]{0.5\textwidth}\vspace{-0pt}
      \includegraphics[trim=0 30 0 50,clip,width=1.\textwidth,angle=0]{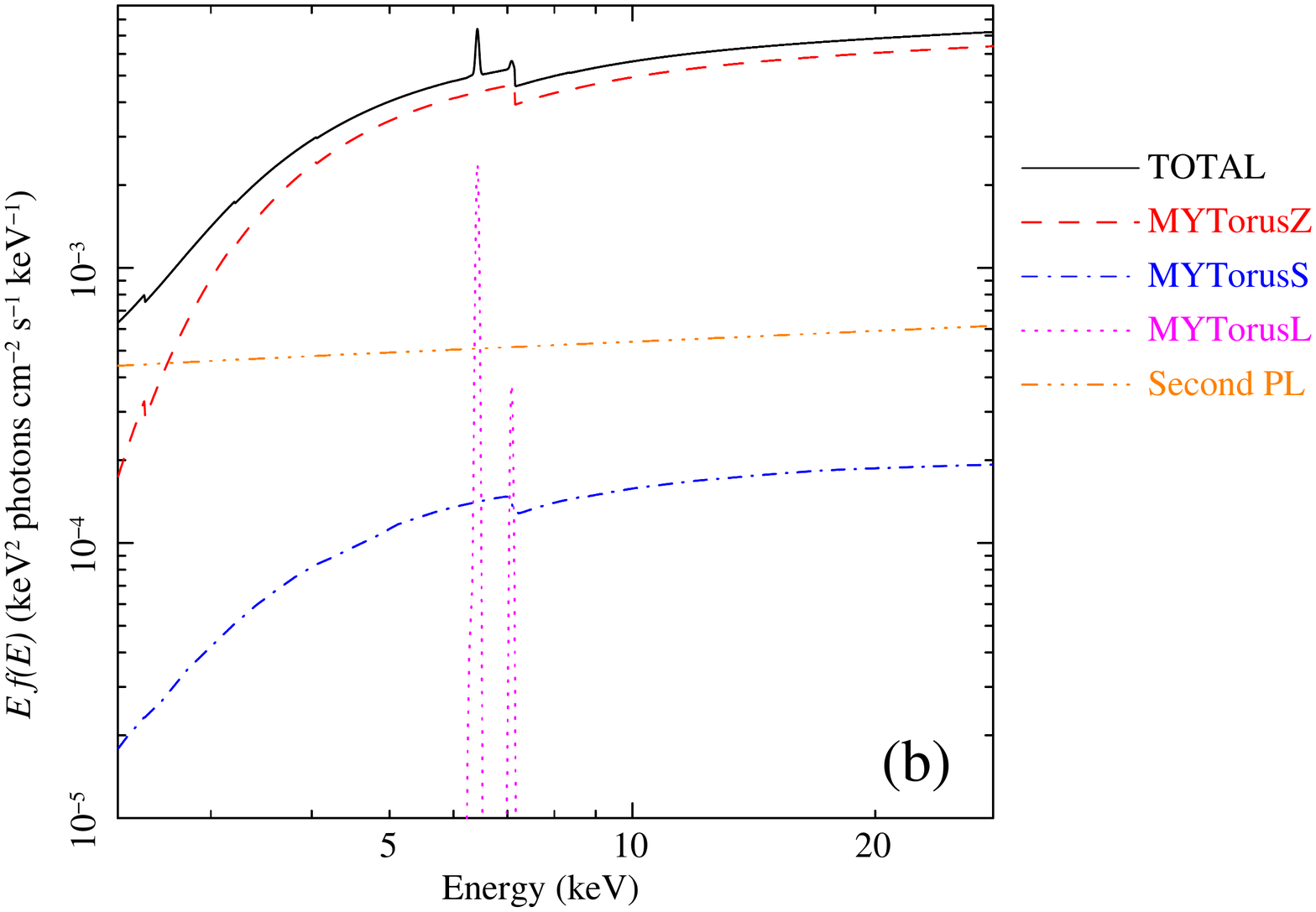}
      \vspace{-0pt}
    \end{minipage}\\
    
    \begin{minipage}[c]{0.5\textwidth}
        \includegraphics[trim=0 -300 0 360,clip,width=1.\textwidth,angle=0]{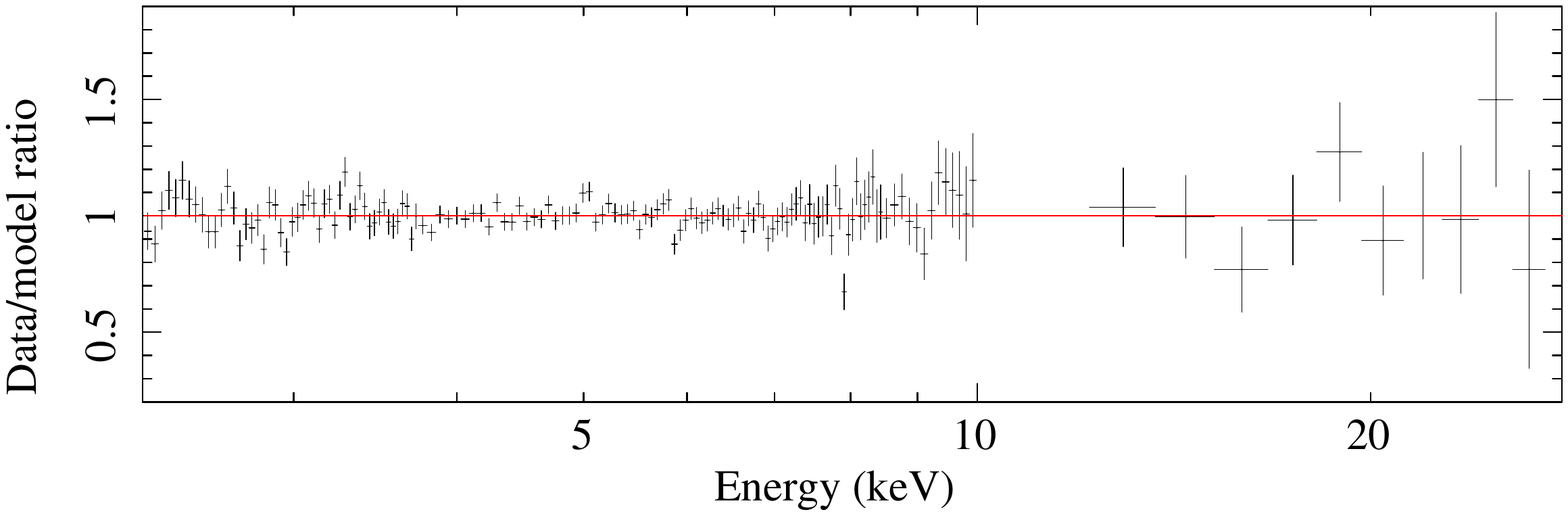}
    \end{minipage}
    \begin{minipage}[c]{0.5\textwidth}
        \includegraphics[trim=0 193 0 80,clip,width=\textwidth,angle=0]{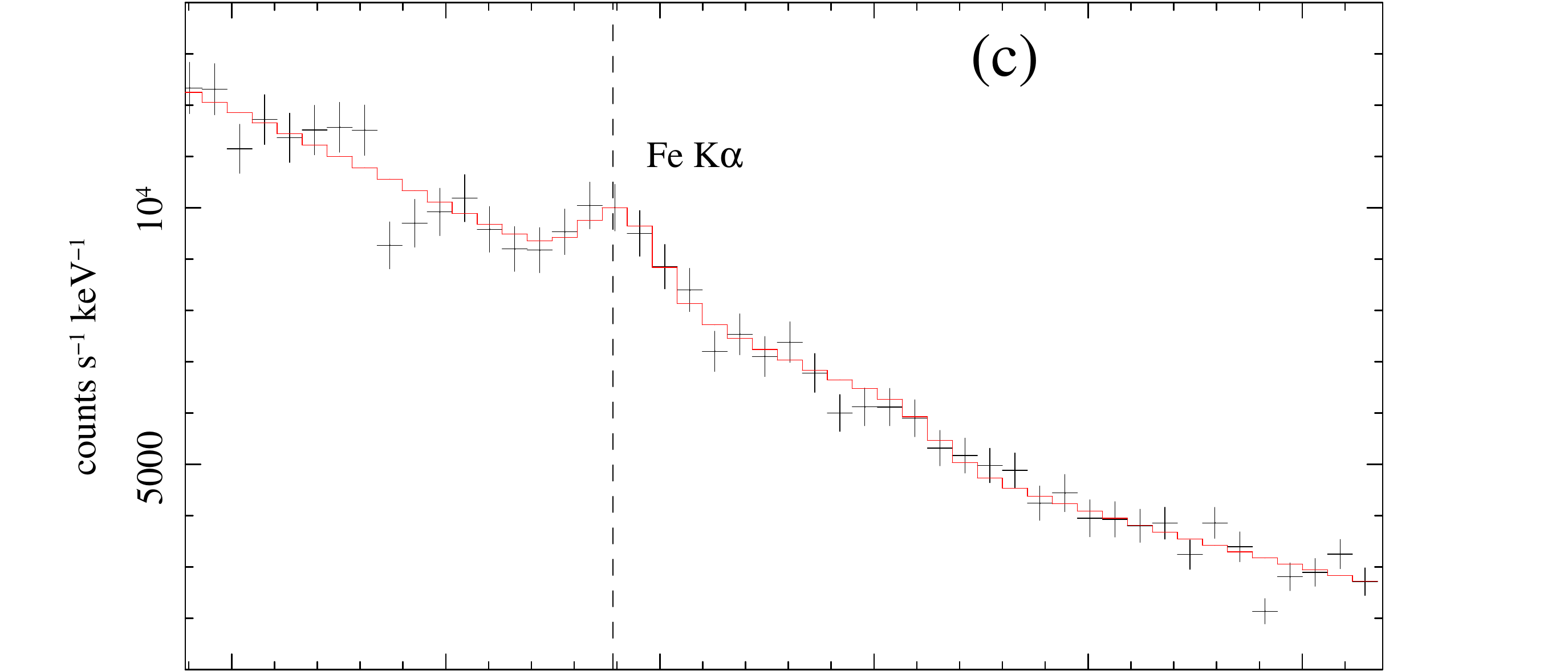}
      \begin{minipage}[c]{\textwidth}
        \includegraphics[trim=0 30 0 360,clip,width=1.\textwidth,angle=0]{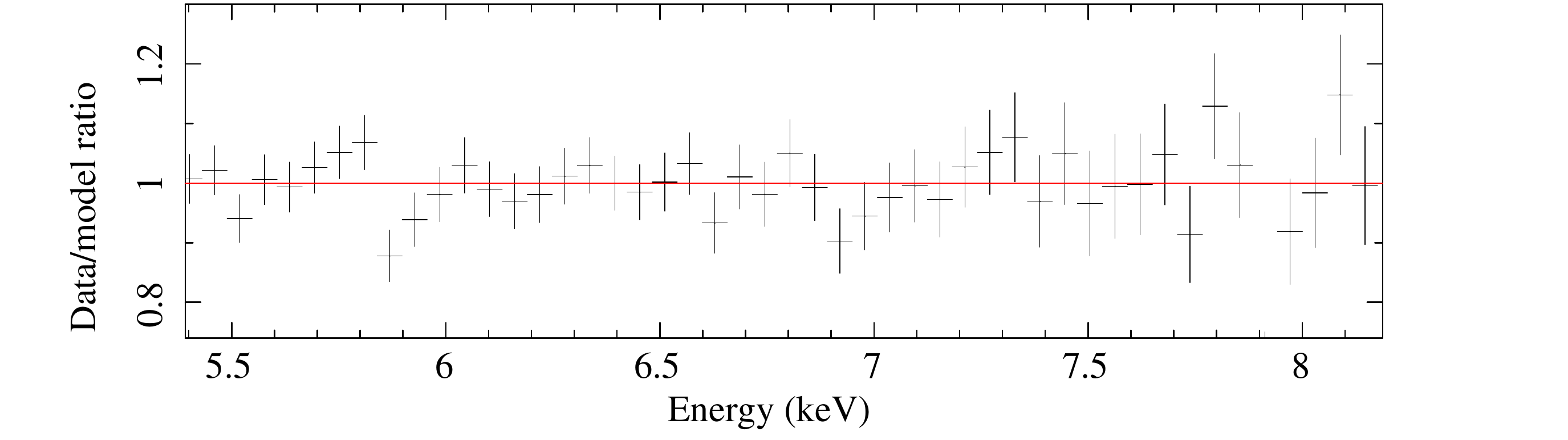}
      \end{minipage}
    \end{minipage}

    \caption{\footnotesize NGC4258 701095010 \label{fig-n4258}}
\end{figure*}

\begin{figure*}[t!]
    \begin{minipage}[c]{0.5\textwidth}
      \includegraphics[trim=0 50 0 -200,clip,width=1.\textwidth,angle=0]{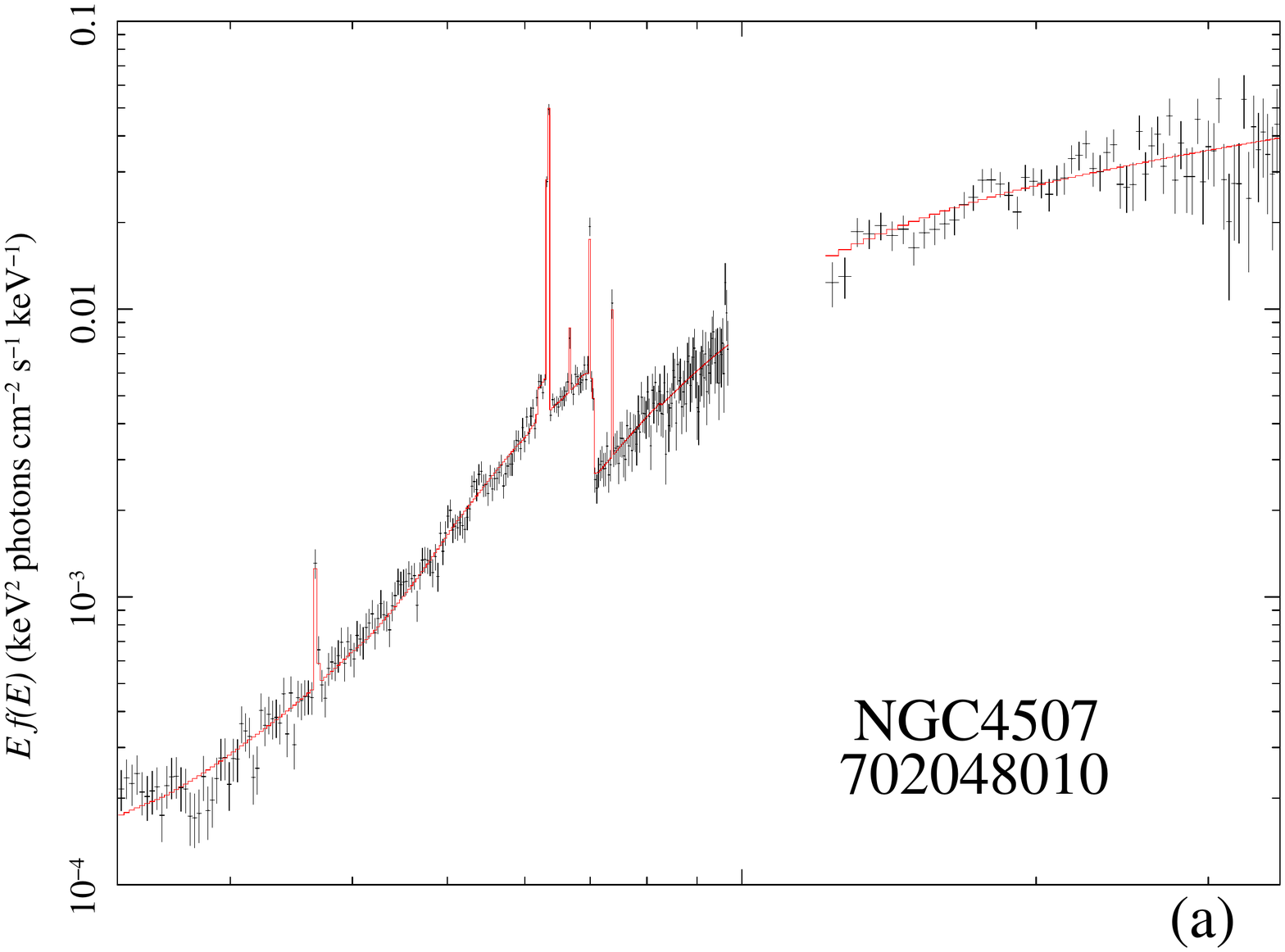}
    \end{minipage}
    \begin{minipage}[c]{0.5\textwidth}\vspace{-0pt}
      \includegraphics[trim=0 30 0 50,clip,width=1.\textwidth,angle=0]{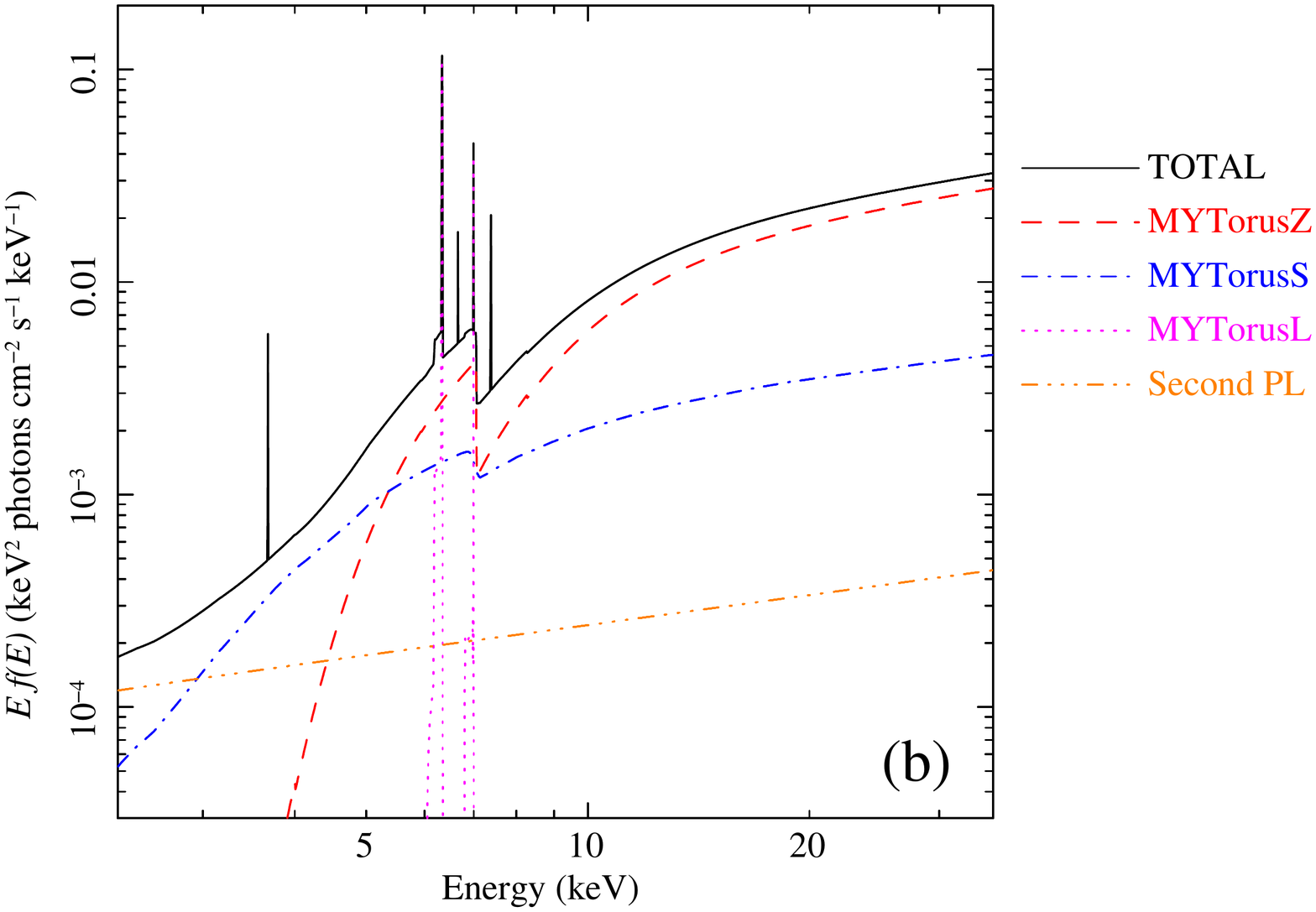}
      \vspace{-0pt}
    \end{minipage}\\
    
    \begin{minipage}[c]{0.5\textwidth}
        \includegraphics[trim=0 -300 0 360,clip,width=1.\textwidth,angle=0]{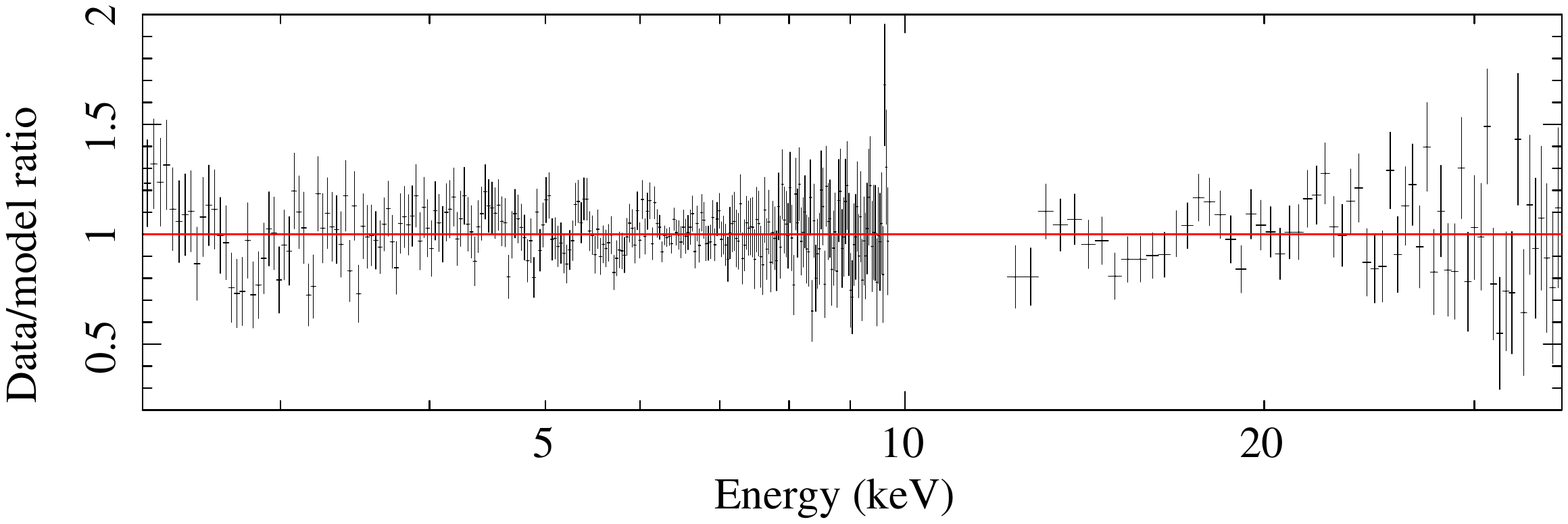}
    \end{minipage}
    \begin{minipage}[c]{0.5\textwidth}
        \includegraphics[trim=0 193 0 80,clip,width=\textwidth,angle=0]{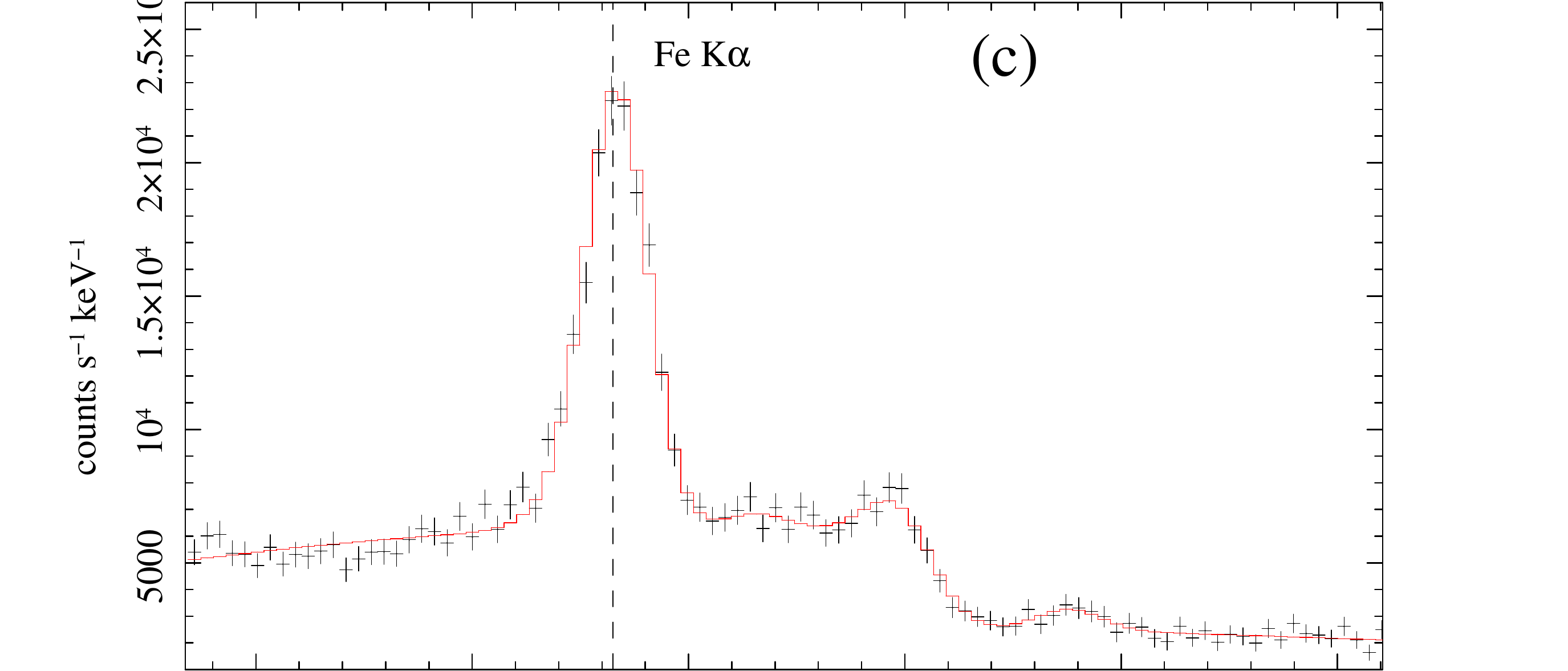}
      \begin{minipage}[c]{\textwidth}
        \includegraphics[trim=0 30 0 360,clip,width=1.\textwidth,angle=0]{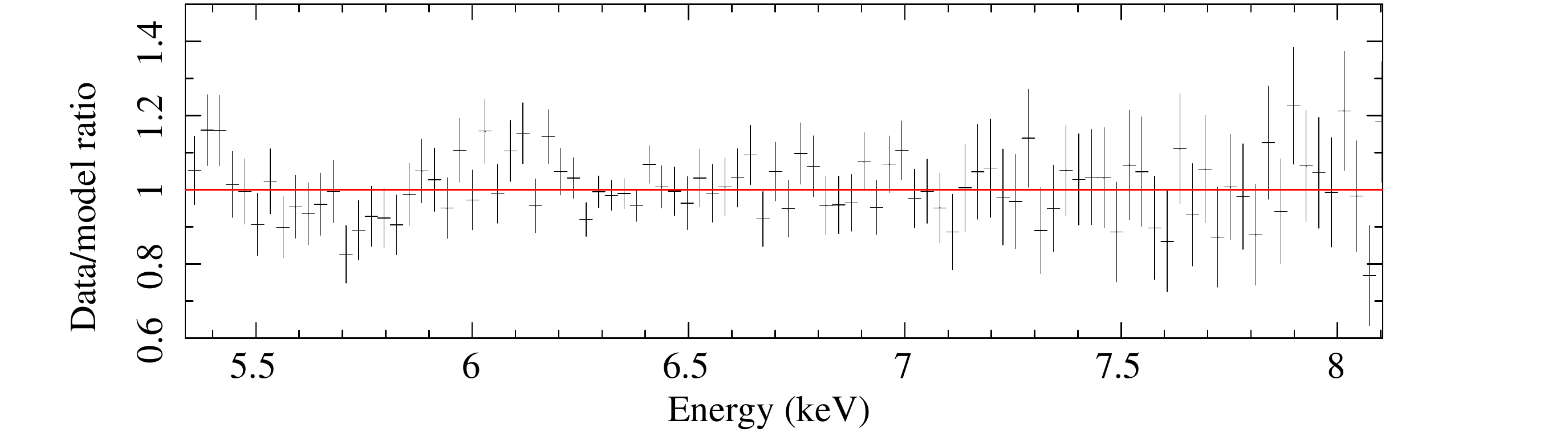}
      \end{minipage}
    \end{minipage}

    \caption{\footnotesize NGC4507 702048010 \label{fig-n4507}}
\end{figure*}

\begin{figure*}[t!]
    \begin{minipage}[c]{0.5\textwidth}
      \includegraphics[trim=0 50 0 -200,clip,width=1.\textwidth,angle=0]{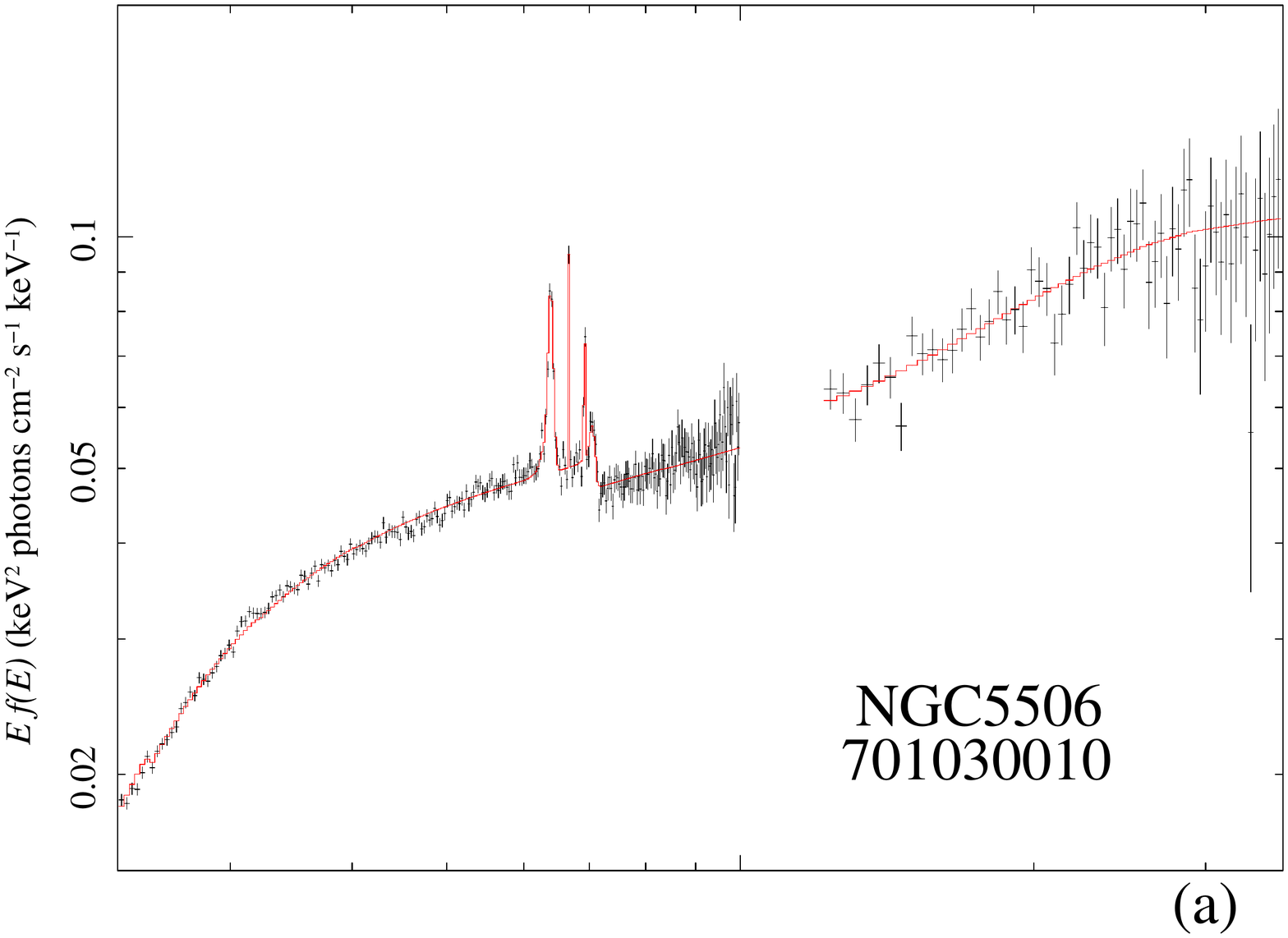}
    \end{minipage}
    \begin{minipage}[c]{0.5\textwidth}\vspace{-0pt}
      \includegraphics[trim=0 30 0 50,clip,width=1.\textwidth,angle=0]{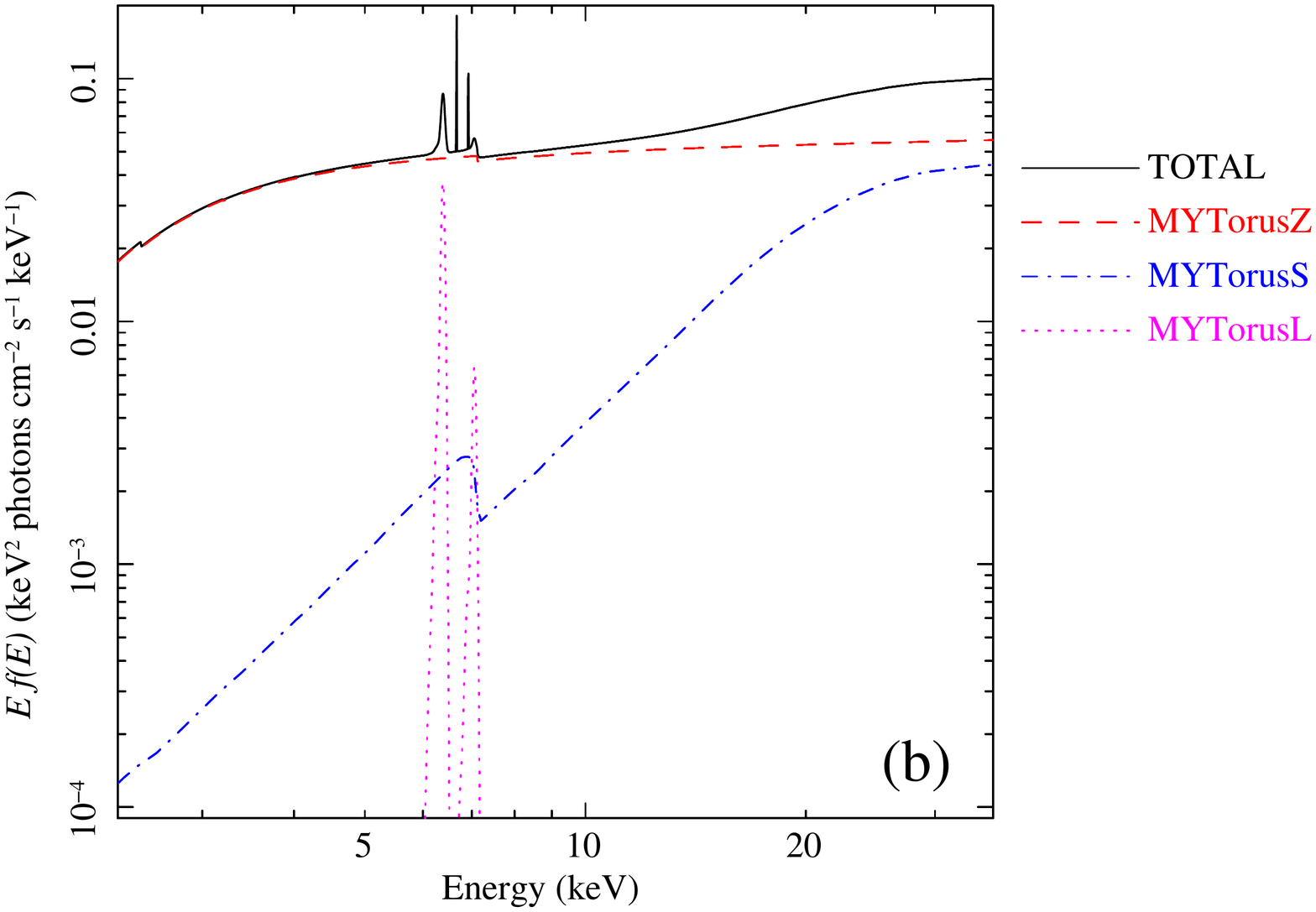}
      \vspace{-0pt}
    \end{minipage}\\
    
    \begin{minipage}[c]{0.5\textwidth}
        \includegraphics[trim=0 -300 0 360,clip,width=1.\textwidth,angle=0]{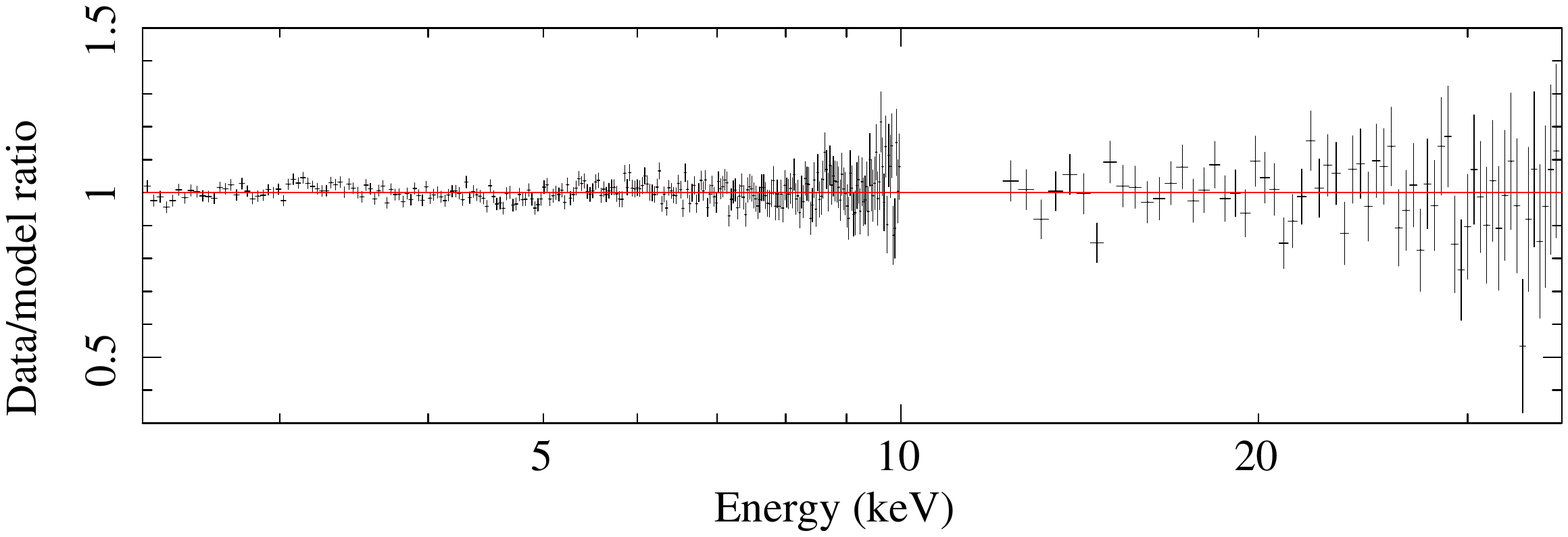}
    \end{minipage}
    \begin{minipage}[c]{0.5\textwidth}
        \includegraphics[trim=0 193 0 80,clip,width=\textwidth,angle=0]{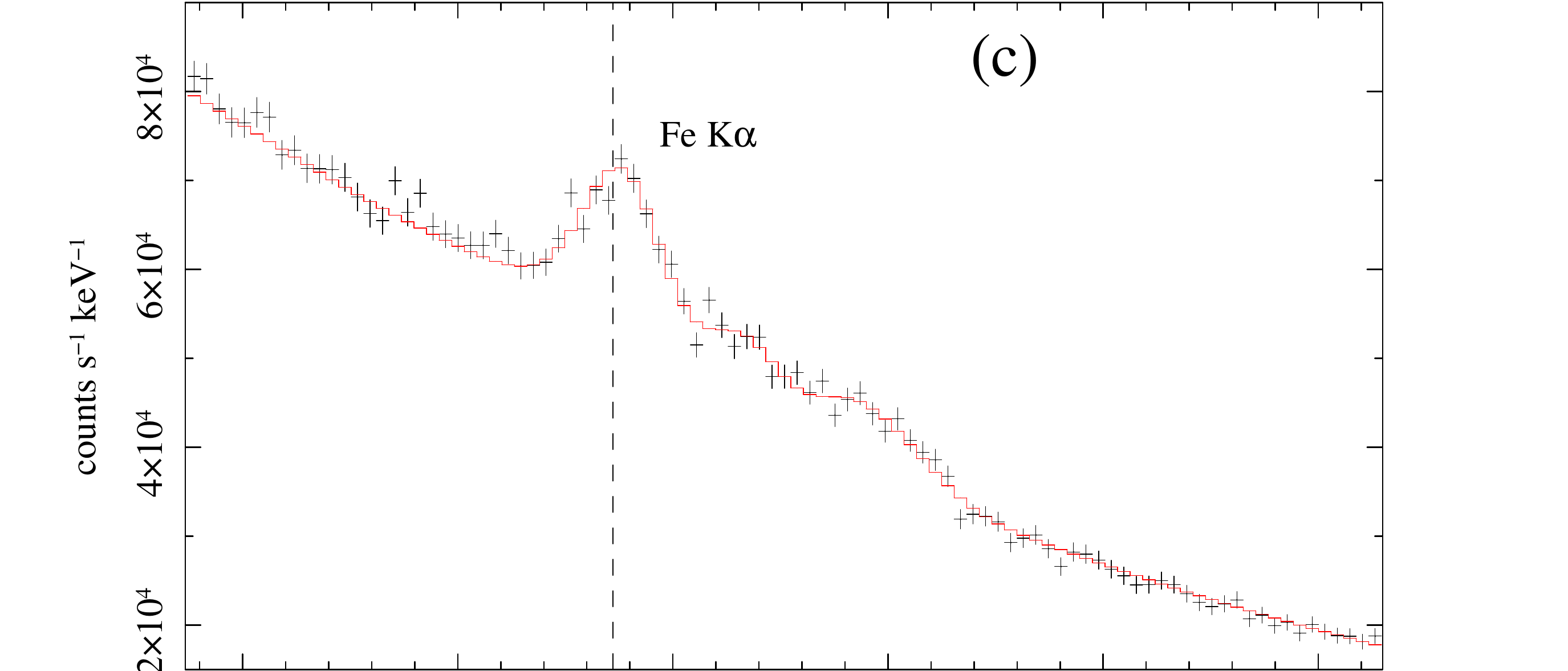}
      \begin{minipage}[c]{\textwidth}
        \includegraphics[trim=0 30 0 360,clip,width=1.\textwidth,angle=0]{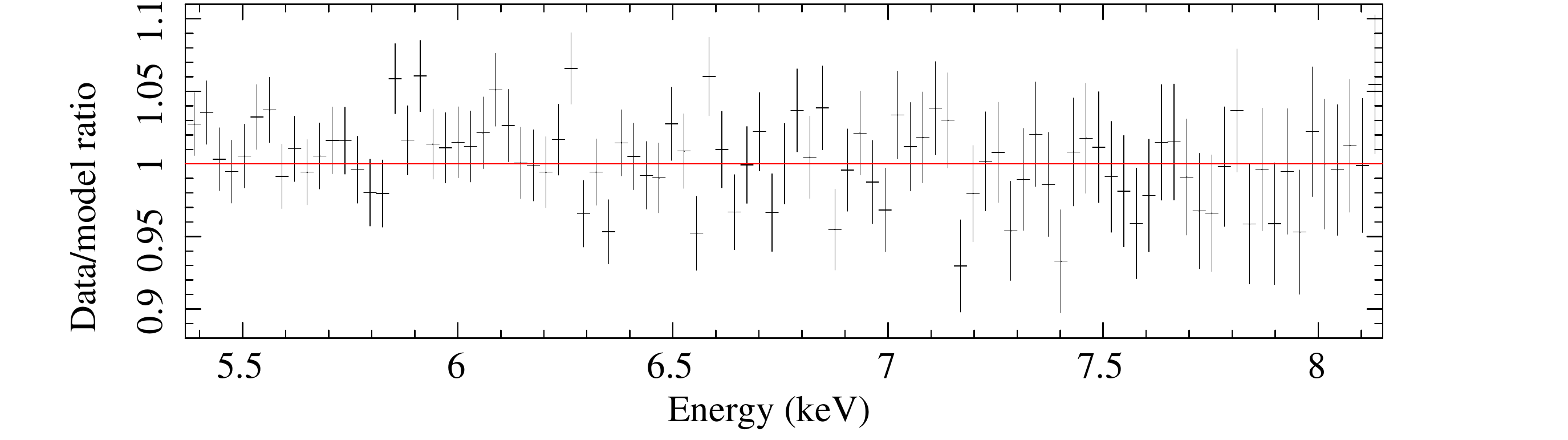}
      \end{minipage}
    \end{minipage}

    \caption{\footnotesize NGC5506 701030010 \label{fig-n5506-10}}
\end{figure*}

\begin{figure*}[t!]
    \begin{minipage}[c]{0.5\textwidth}
      \includegraphics[trim=0 50 0 -200,clip,width=1.\textwidth,angle=0]{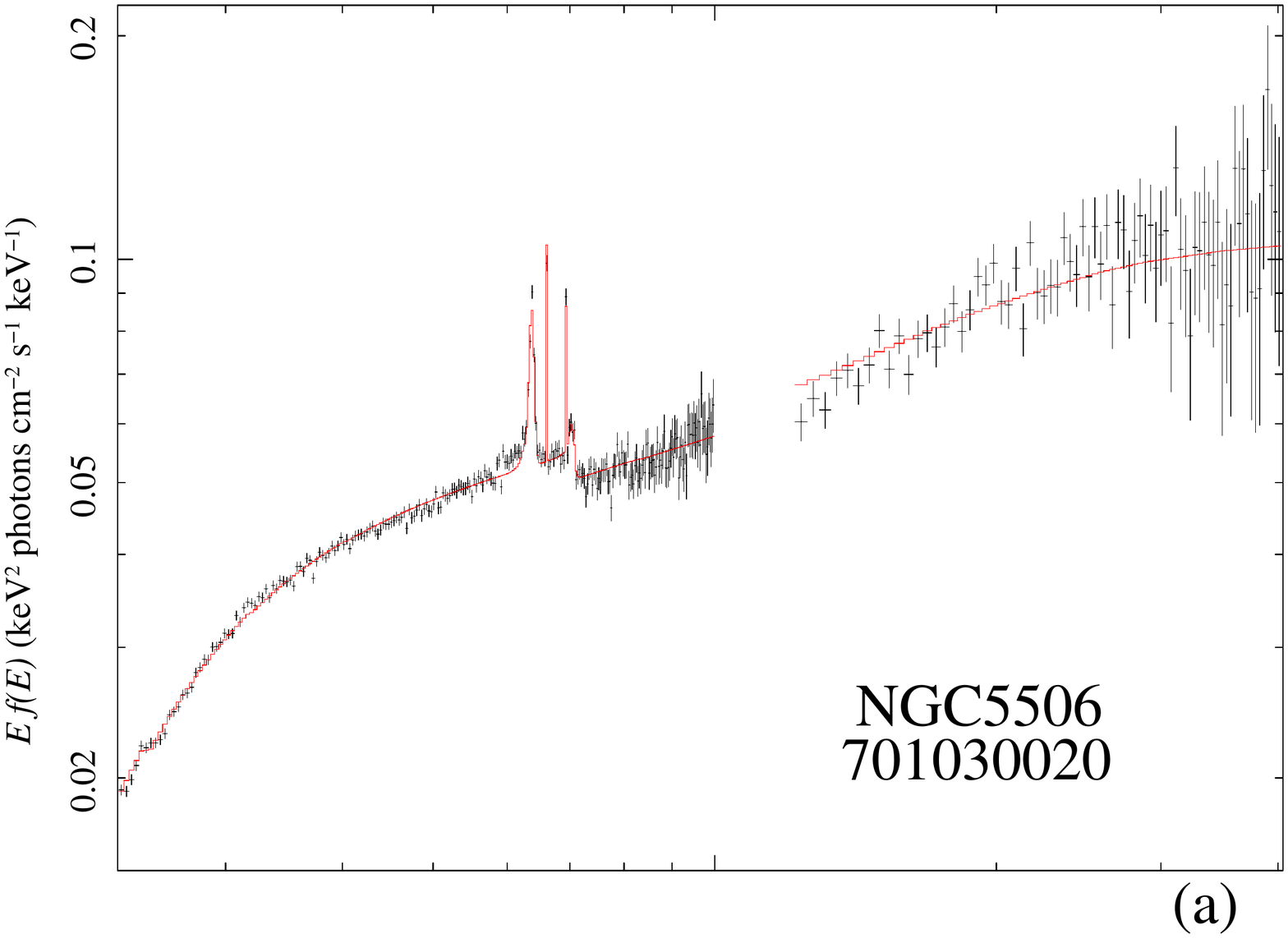}
    \end{minipage}
    \begin{minipage}[c]{0.5\textwidth}\vspace{-0pt}
      \includegraphics[trim=0 30 0 50,clip,width=1.\textwidth,angle=0]{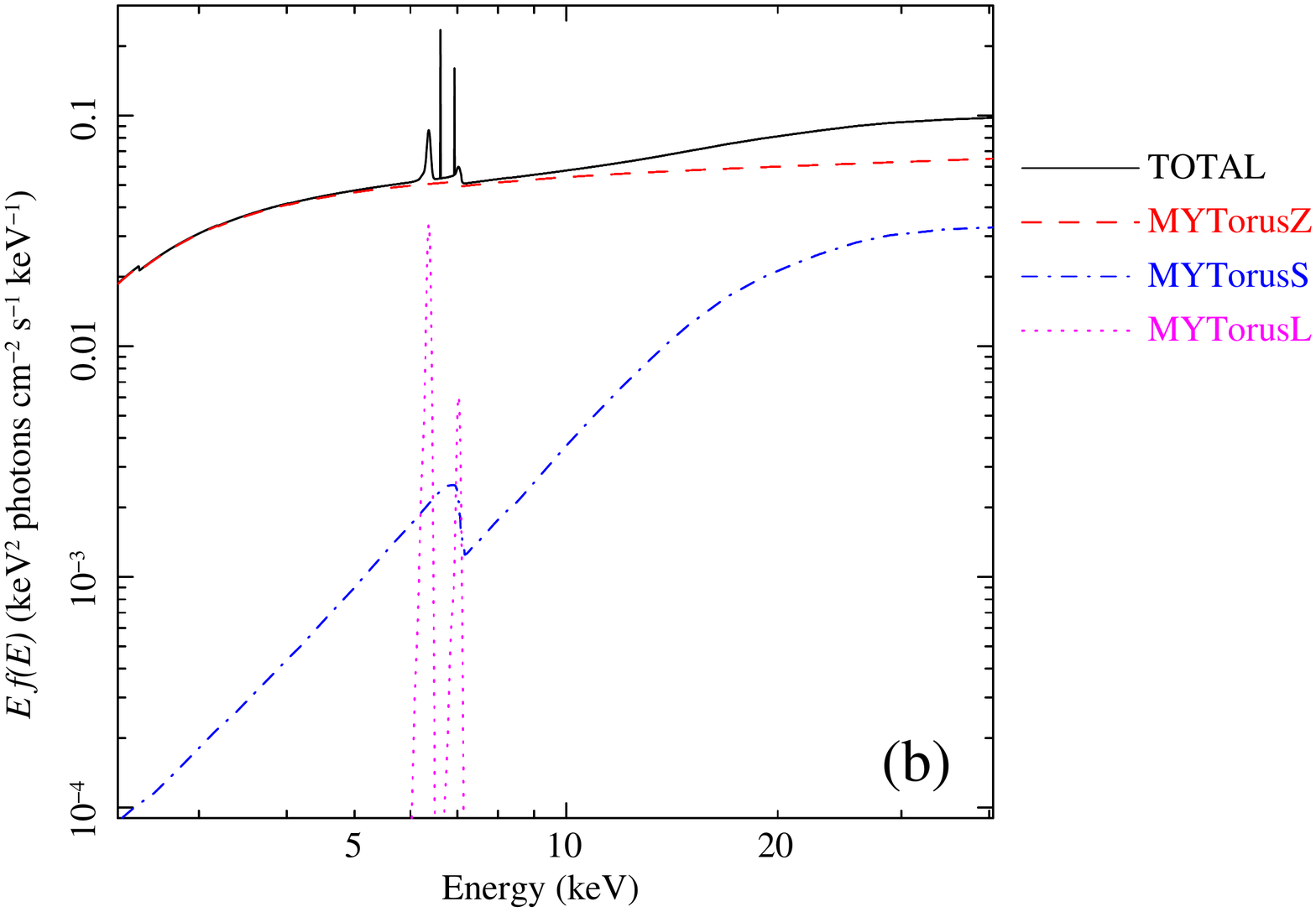}
      \vspace{-0pt}
    \end{minipage}\\
    
    \begin{minipage}[c]{0.5\textwidth}
        \includegraphics[trim=0 -300 0 360,clip,width=1.\textwidth,angle=0]{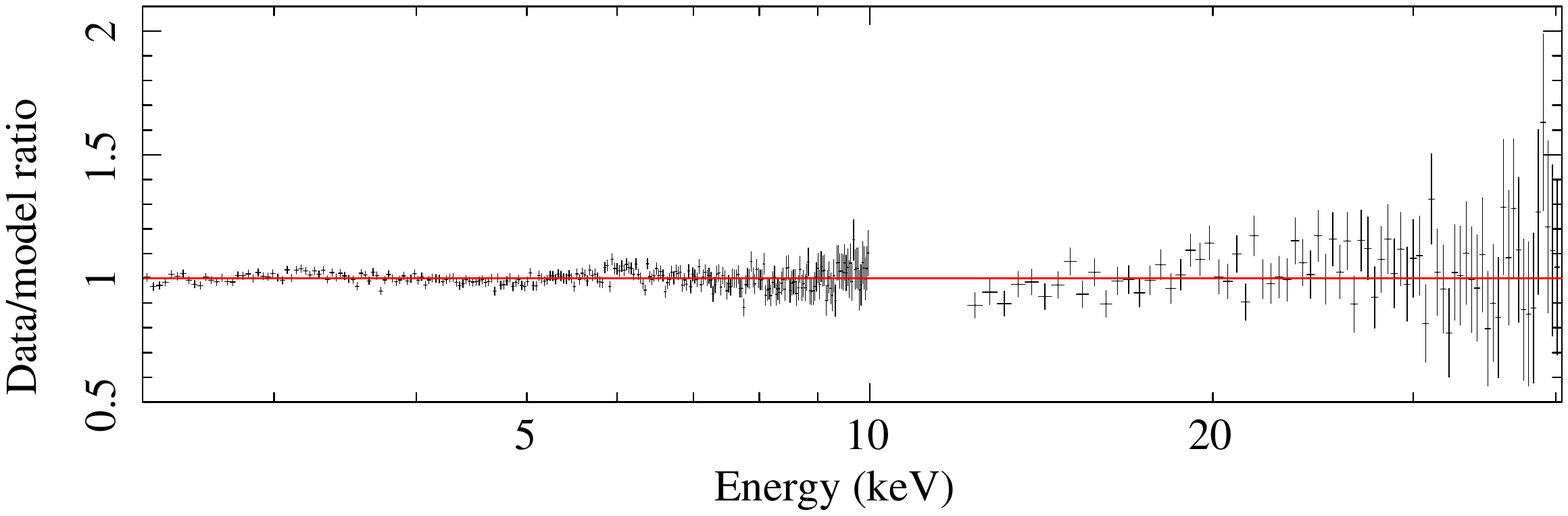}
    \end{minipage}
    \begin{minipage}[c]{0.5\textwidth}
        \includegraphics[trim=0 193 0 80,clip,width=\textwidth,angle=0]{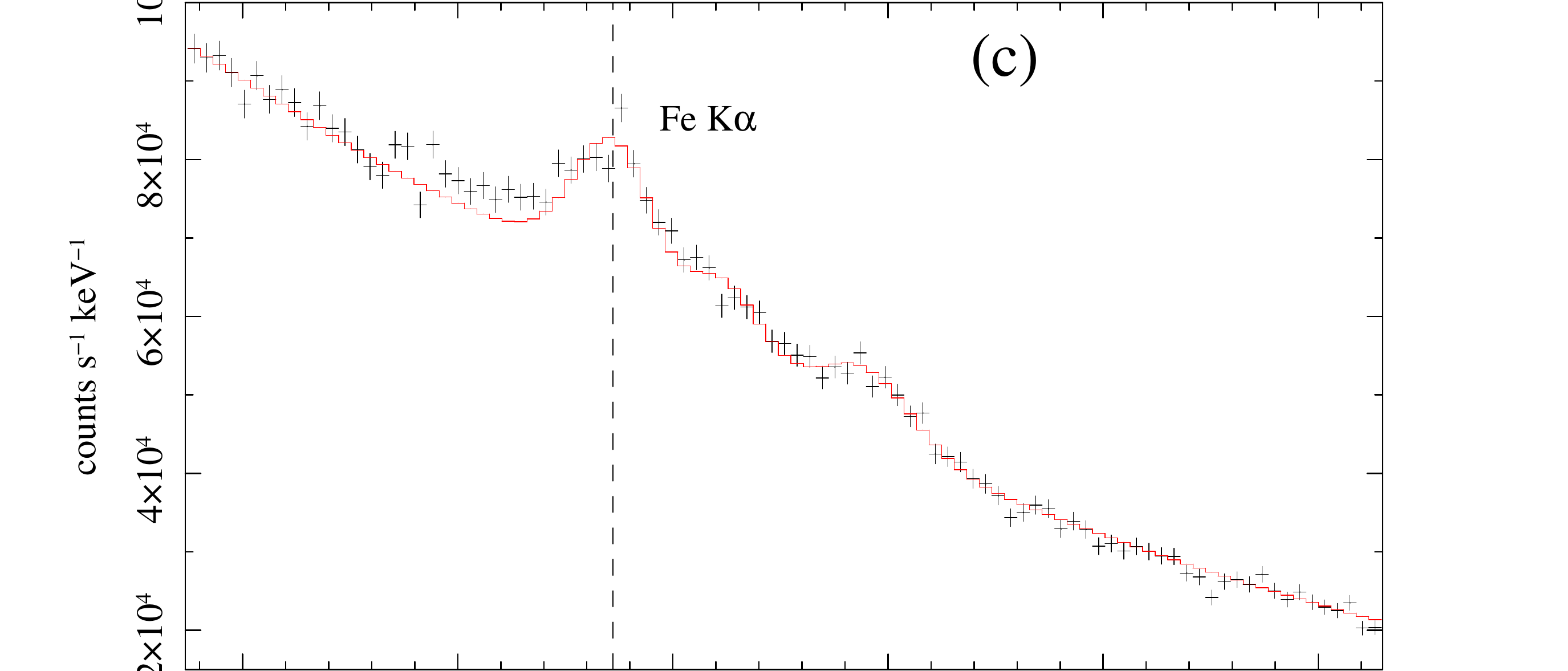}
      \begin{minipage}[c]{\textwidth}
        \includegraphics[trim=0 30 0 360,clip,width=1.\textwidth,angle=0]{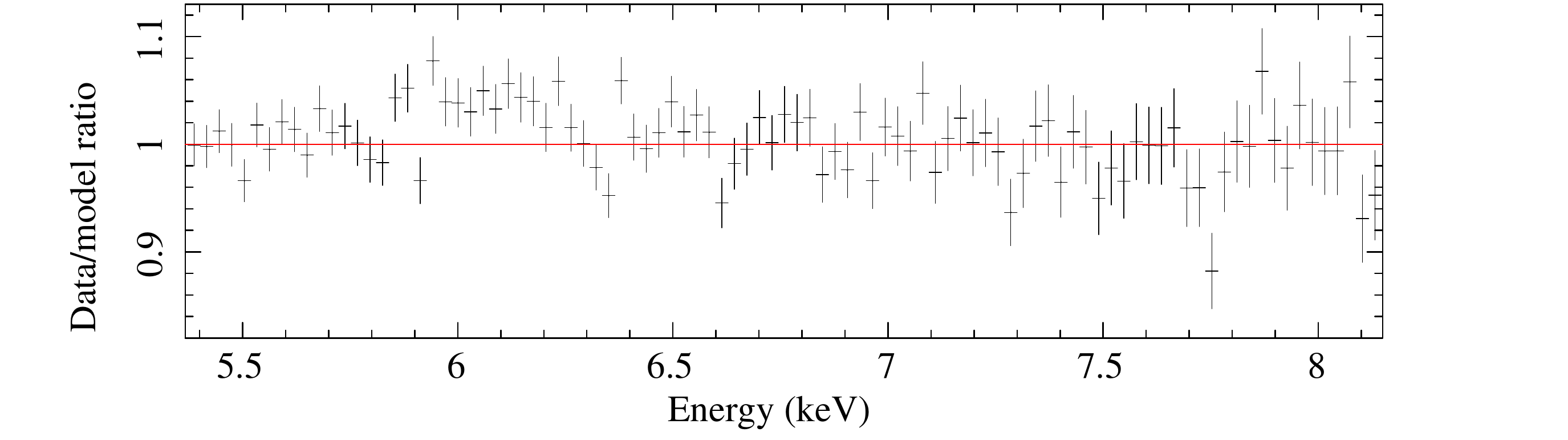}
      \end{minipage}
    \end{minipage}

    \caption{\footnotesize NGC5506 701030020 \label{fig-n5506-20}}
\end{figure*}

\begin{figure*}[t!]
    \begin{minipage}[c]{0.5\textwidth}
      \includegraphics[trim=0 50 0 -200,clip,width=1.\textwidth,angle=0]{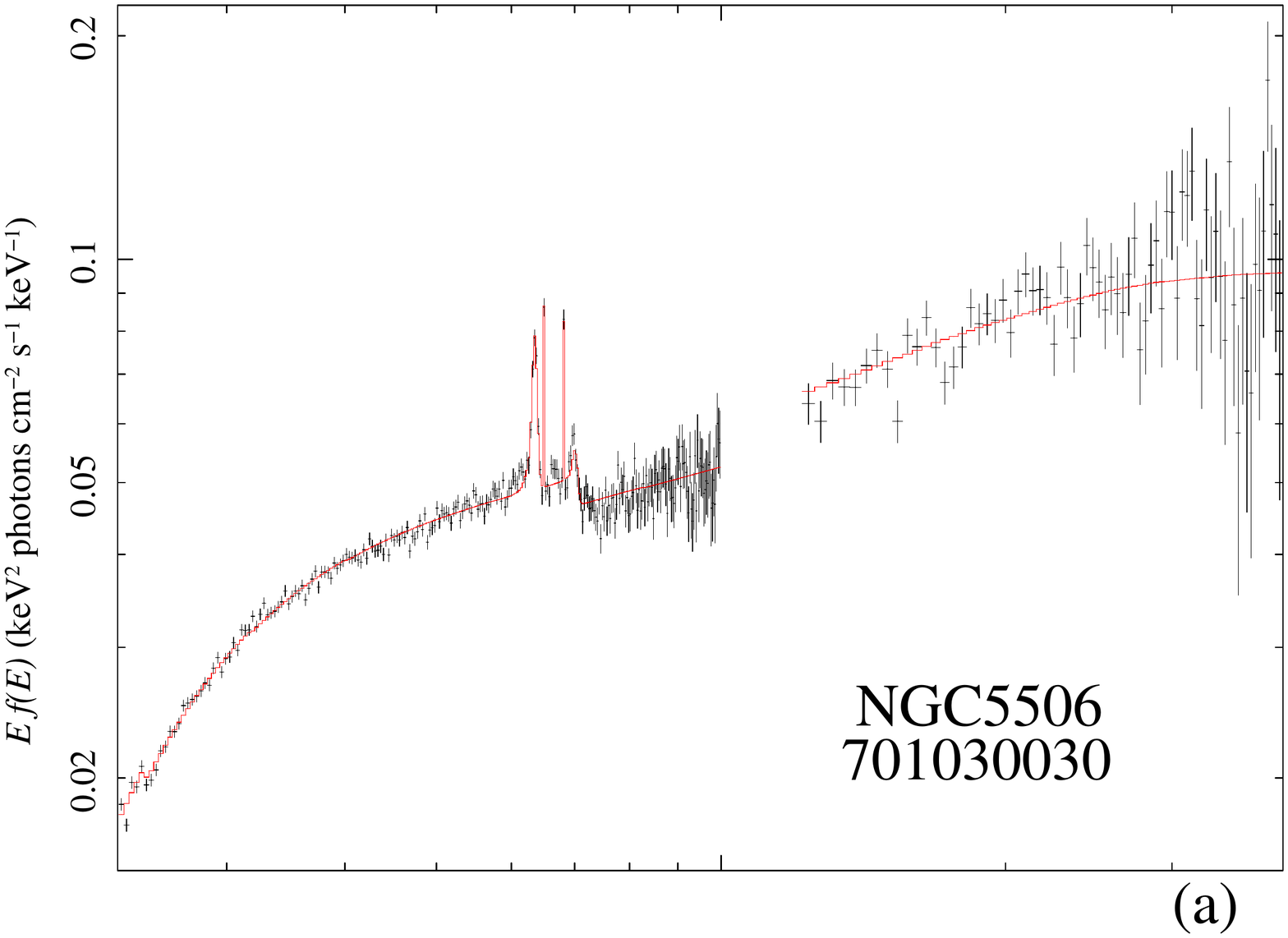}
    \end{minipage}
    \begin{minipage}[c]{0.5\textwidth}\vspace{-0pt}
      \includegraphics[trim=0 30 0 50,clip,width=1.\textwidth,angle=0]{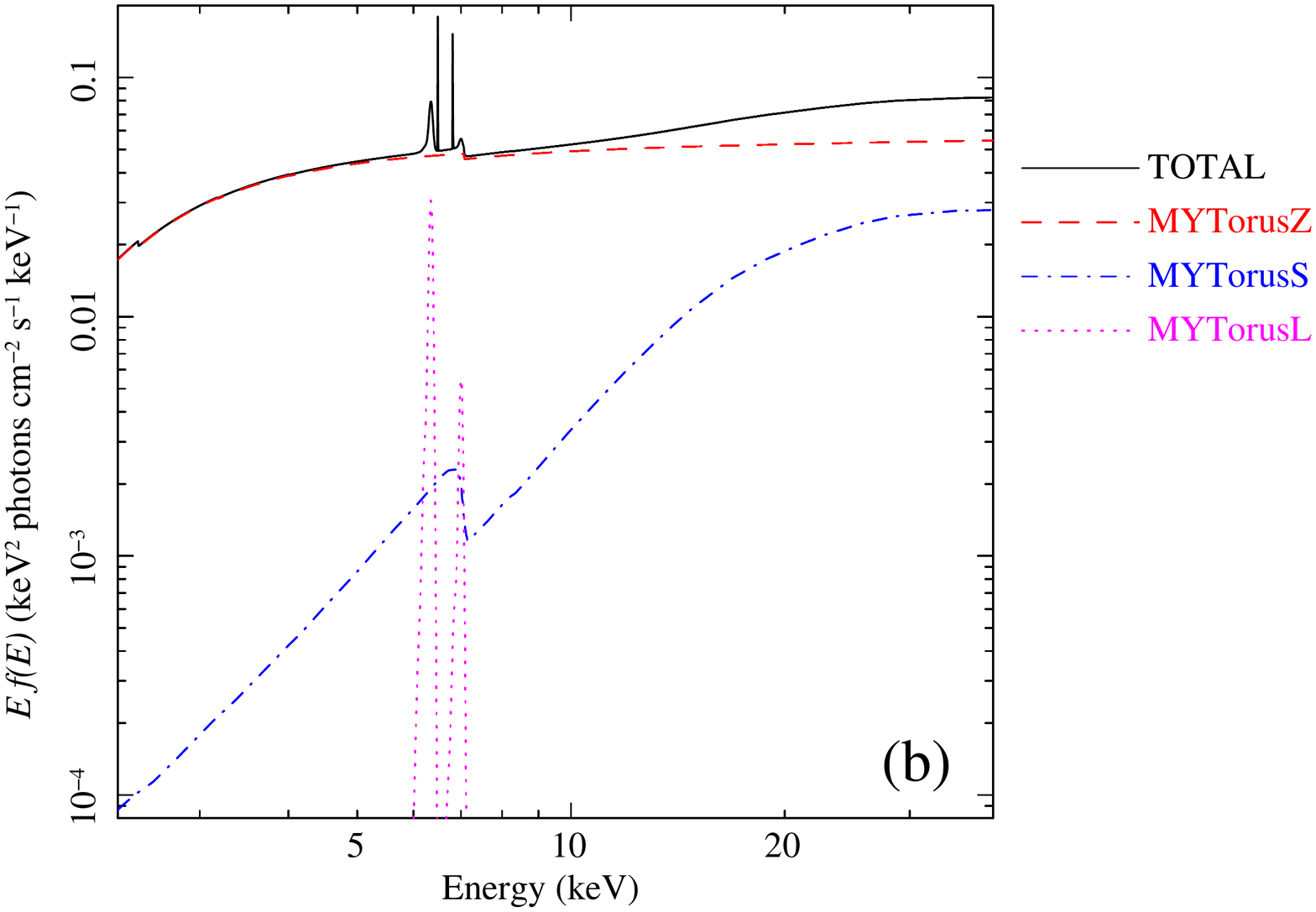}
      \vspace{-0pt}
    \end{minipage}\\
    
    \begin{minipage}[c]{0.5\textwidth}
        \includegraphics[trim=0 -300 0 360,clip,width=1.\textwidth,angle=0]{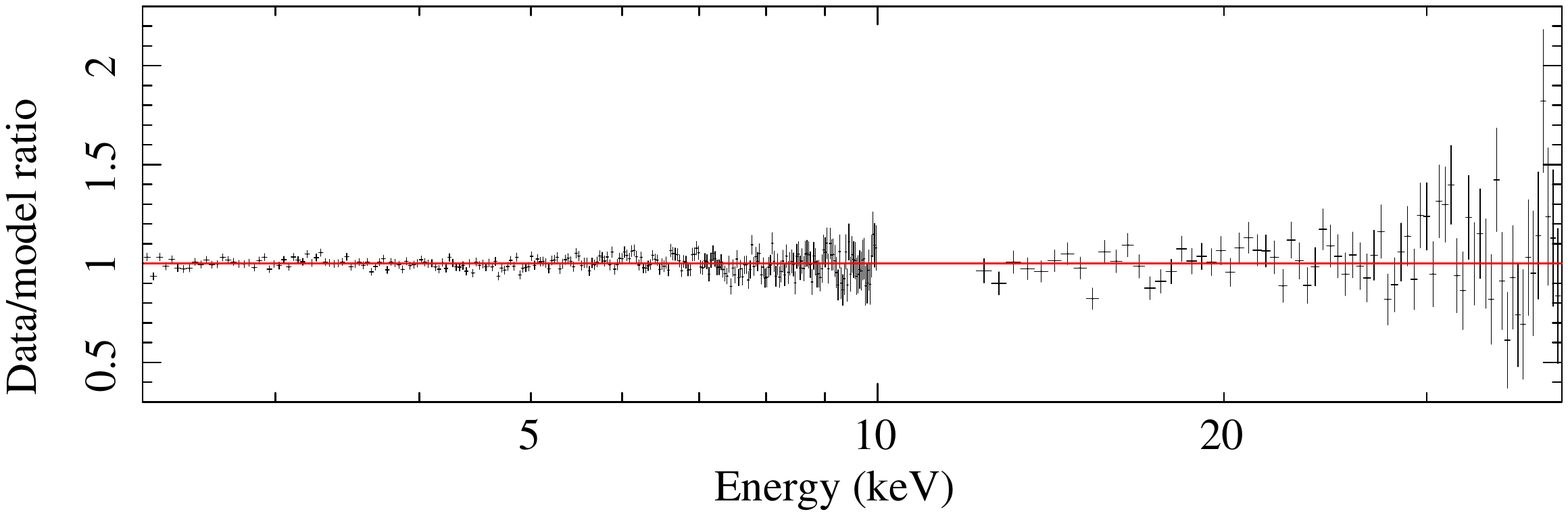}
    \end{minipage}
    \begin{minipage}[c]{0.5\textwidth}
        \includegraphics[trim=0 193 0 80,clip,width=\textwidth,angle=0]{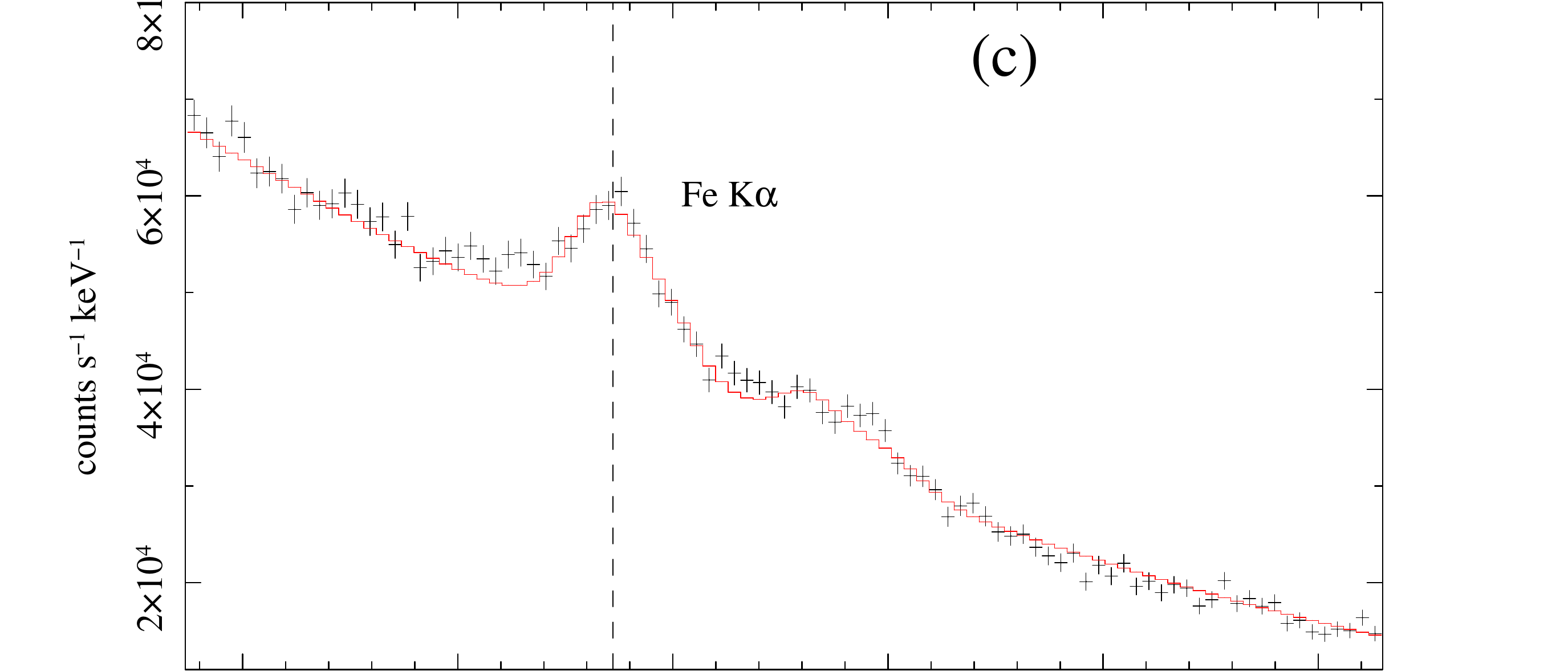}
      \begin{minipage}[c]{\textwidth}
        \includegraphics[trim=0 30 0 360,clip,width=1.\textwidth,angle=0]{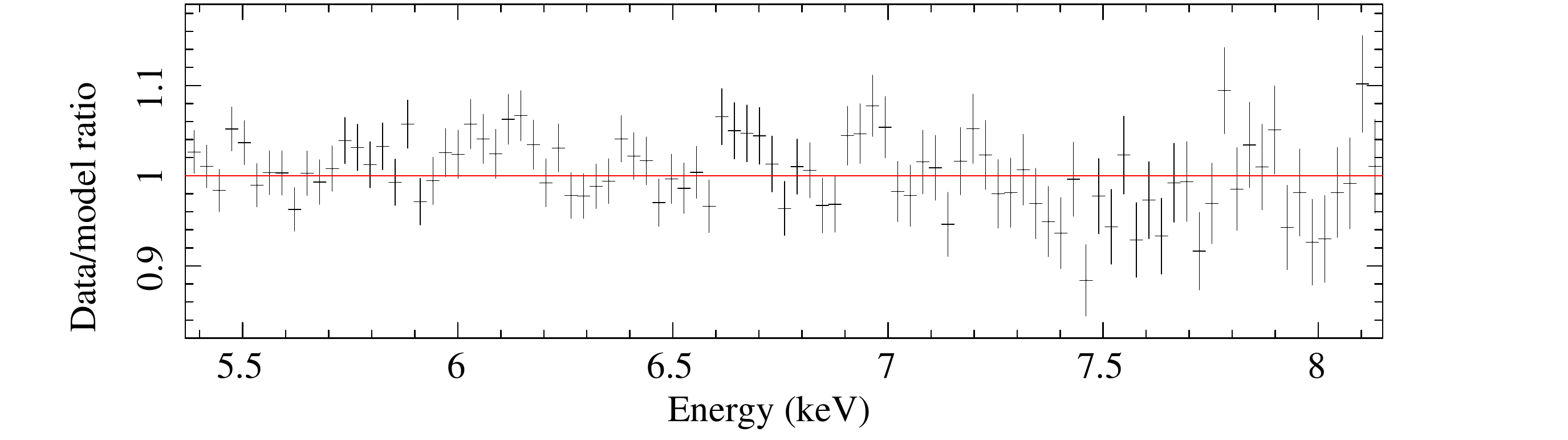}
      \end{minipage}
    \end{minipage}

    \caption{\footnotesize NGC5506 701030030 \label{fig-n5506-30}}
\end{figure*}

%

%
%



\bibliographystyle{likeapj}
\bibliography{masterbib}

\end{document}